\documentstyle[epsf,onecolumn]{jpsj_sissa}

\newcommand{\pr}[3]{Phys.\ Rev. {\bf #1} (#2) #3}
\newcommand{\prb}[3]{Phys.\ Rev.\ B {\bf #1} (#2) #3}
\newcommand{\prl}[3]{Phys.\ Rev.\ Lett. {\bf #1} (#2) #3}
\newcommand{\rmp}[3]{Rev.\ Mod.\ Phys. {\bf #1} (#2) #3}
\newcommand{\jpsj}[3]{J.\ Phys.\ Soc.\ Jpn. {\bf #1} (#2) #3}
\newcommand{\ptp}[3]{Prog.\ Theor.\ Phys. {\bf #1} (#2) #3}
\newcommand{\jpcs}[3]{J.\ Phys.\ Chem.\ Sol. {\bf #1} (#2) #3}
\newcommand{\epjb}[3]{Eur.\ Phys.\ J. B {\bf #1} (#2) #3}
\newcommand{\jpf}[3]{J.\ Phys.\ (France) {\bf #1} (#2) #3}

\newcommand{\eq}[1]{eq. (\ref{#1})}
\newcommand{\eqd}[2]{eqs. (\ref{#1}) and (\ref{#2})}
\newcommand{\eqt}[3]{eqs. (\ref{#1}), (\ref{#2}) and (\ref{#3})}

\newcommand{\del}{\partial}
\newcommand{\Tr}{{\rm Tr}}
\newcommand{\Tc}{T_{\rm c}}
\newcommand{\TN}{T_{\rm N}}
\newcommand{\rmc}{{\rm c}}
\newcommand{\rme}{{\rm e}}
\newcommand{\romap}{{\rm p}}
\newcommand{\rms}{{\rm s}}

\newcommand{\calL}{{\cal L}}
\newcommand{\calO}{{\cal O}}

\newcommand{\veps}{\varepsilon}
\newcommand{\bfveps}{\mib{\veps}}
\newcommand{\vepsz}{\veps^{(0)}}
\newcommand{\vepskin}{\veps_{\rm kin}}

\newcommand{\vepscor}[1]{\veps_{{\rm cor}#1}}
\newcommand{\bfphi}{\mib{\phi}}
\newcommand{\bfvarphi}{\mib{\varphi}}
\newcommand{\Ekin}{E_{\rm kin}}

\newcommand{\bfchi}{\mib{\chi}}
\newcommand{\DeltaMH}{\Delta^{\rm MH}}
\newcommand{\muz}{\mu^{(0)}}
\newcommand{\bfSigma}{\mib{\Sigma}}

\newcommand{\ho}{{\hat{\omega}}}
\newcommand{\delho}{\delta\ho}
\newcommand{\ave}[1]{{\langle#1\rangle}}
\newcommand{\com}[2]{\left[#2\right]_{#1}\!}
\newcommand{\up}{\uparrow}
\newcommand{\down}{\downarrow}

\newcommand{\Psid}{\Psi^\dagger}

\newcommand{\bfPsi}{{\bf\Psi}}
\newcommand{\bfPsid}{\bfPsi^\dagger}

\newcommand{\VO}[2]{V${}_{#1}$O${}_{#2}$}

\newcommand{\LSTiO}[4]{La${}_{#1}$Sr${}_{#2}$Ti${}_{#3}$O${}_{#4}$}
\newcommand{\LSVO}[4]{La${}_{#1}$Sr${}_{#2}$V${}_{#3}$O${}_{#4}$}

\newcommand{\BSCCO}[5]{Bi${}_{#1}$Sr${}_{#2}$Ca${}_{#3}$Cu${}_{#4}$O${}_{#5}$}

\title{Filling-Control Metal-Insulator Transition in the Hubbard Model Studied by the Operator Projection Method}
\def\runtitle{Filling-control Metal-Insulator Transition in the Hubbard Model Studied by the Operator Projection Method}

\author{Shigeki Onoda and Masatoshi Imada}
\def\runauthor{Shigeki Onoda and Masatoshi Imada}

\inst{Institute for Solid State Physics, University of Tokyo,
Kashiwanoha 5-1-5, Kashiwa, Chiba 277-8581, Japan}
\recdate{}

\abst{Thermodynamic and dynamical properties of filling-control metal-insulator transition (MIT) in the Hubbard model are studied by the operator projection method, especially in two dimensions. This is a non-perturbative analytic approach to many-body systems. The present theory incorporates the Mott-Hubbard, Brinkman-Rice and Slater pictures of the MIT into a unified framework, together with reproducing low-energy narrow band arising from spin-charge fluctuations. At half filling, single-particle spectra $A(\omega, k)$ show formation of two Hubbard bands and their antiferromagnetic shadows separated by a Mott gap in the plane of energy $\omega$ and momentum $k$ with lowering temperatures. These four bands produce splitting to two low-energy narrow bands and two SDW-like bands in the dispersion. Near half filling, the low-energy narrow band persists at low temperatures. This narrow band has a particularly weak dispersion and large weights around $(\pi, 0)$ and $(0, \pi)$ momenta. The velocity of these low-energy excitations is shown to vanish towards the MIT, indicating the mass divergence as in the Brinkman-Rice picture, but most prominently around $(\pi,0)$ and $(0,\pi)$ with strong momentum dependence. This reflects the suppression of the coherence near the MIT. Main structures in $A(\omega, k)$ show remarkable similarities to quantum Monte-Carlo results in two dimensions as well as in the one-dimensional Hubbard model. The charge compressibility appears to diverge with decreasing doping concentration in both one and two dimensions in agreement with the exact and quantum Monte-Carlo results. We also discuss implications of the flat dispersion formed near the Fermi level to the observations in high-$\Tc$ cuprate superconductors.}

\kword{metal-insulator transition, Hubbard model, projection operator}

\begin{document}
\sloppy
\maketitle

\section{Introduction}
\label{sec:Introduction}

The issue of the metal-insulator transitions (MIT) postulated by Mott~\cite{Mott} has opened extensive studies on the subject from both theoretical and experimental aspects~\cite{RMP_Imada}. Pioneering works by Hubbard~\cite{Hubbard1,Hubbard2,Hubbard3} and Kanamori~\cite{Kanamori63} played an active part in theoretical studies on strong electron correlations in narrow-band electron systems~\cite{Fulde}. In spite of a long history of studies on strongly correlated electrons~\cite{Fulde,RMP_Imada}, an appropriate and satisfactory theoretical description of the MIT remains controversial, even for the simple Hubbard model. Particularly, in one and two dimensions, fluctuations become significant near the MIT. This makes it difficult to reliably describe physical properties related to the MIT. This paper aims at developing a formalism of operator projection~\cite{Nakajima58,Zwanzig,Mori65} to construct a new non-perturbative analytic theory of many-body correlated electrons~\cite{OPM_jpsjletter01}. This method allows us to systematically improve the Hartree-Fock theory, perturbation theories, Hubbard approximations~\cite{Hubbard1,Hubbard3} and two-pole approximations~\cite{HarrisLange67,Roth69}. Using this new method, we give static and dynamical properties of the Hubbard model in one and two dimensions at and near half filling.

Early theories on transitions from metals to insulators in clean systems are divided into three pictures: (i) By means of the Green's function technique based on Heisenberg equations of motion~\cite{Zubarev60}, Hubbard described an MIT as the splitting of two Hubbard bands due to strong local Coulomb repulsion compared with the original bandwidth~\cite{Hubbard1, Hubbard3}. This description is based on the view provided by Mott~\cite{Mott}. The Hubbard I approximation was improved as two-pole approximations in a spirit of operator projections~\cite{HarrisLange67, Roth69}. These approximations are valid in the strongly correlated limits. However, they violate the Luttinger sum rule in paramagnetic metals. They do not reproduce the correct low-density limit given by Kanamori~\cite{Kanamori63} either. (ii) Using the Gutzwiller variational method~\cite{Gutzwiller65}, Brinkman and Rice predicted that at half filling, an MIT occurs at a finite critical interaction strength $U_{\rm c}$ due to the disappearance of Fermi-liquid quasiparticles accompanied by the divergence of the quasiparticle mass~\cite{BrinkmanRice70}. However, spin correlations as well as incoherent excitations appearing as the Hubbard bands are neglected in this theory. (iii) Early arguments on the Slater mechanism~\cite{Slater} of transitions from paramagnetic metals to antiferromagnetic (AF) insulators are based on the Hartree-Fock theory. In this mechanism, the folding of the Brillouin zone accompanied by the AF phase transition produces an energy gap in single-particle excitations. Gaussian fluctuations around the Hartree-Fock solution are treated by a random phase approximation (RPA). Self-consistent renormalization (SCR) theory developed by Moriya and coworkers modified the Hartree-Fock approximation and the RPA in the problem of itinerant magnetism~\cite{Moriya_Spin-Fluctuations}. However, these theories fail to reproduce Mott-insulating features which persist even above the Neel temperature $\TN$.

Experimentally, Mott-Hubbard MITs have been observed in a lot of compounds~\cite{Mott,RMP_Imada}, e.g., \VO{2}{3}, \LSTiO{1-x}{x}{}{3} and \LSVO{1-x}{x}{}{3}.
Especially, \VO{2}{3}-based compounds have provided a typical example for studying the MIT~\cite{McWhan71,Mott,RMP_Imada}. In a gross view, as the temperature and the bandwidth are varied, the compounds exhibit phase transitions among three different phases; antiferromagnetic insulating (AFI), paramagnetic metallic (PM) and paramagnetic insulating (PI) phases. The ground state is AFI under weak pressure and PM under strong pressure. The AFI phase has been considered to originate from the Slater mechanism. On the other hand, a strong renormalization of the quasiparticle mass occurs in the PM phase, as in the Brinkman-Rice picture. With increasing temperature, the AFI system undergoes an AFI-PM phase transition under strong pressure, while a PI phase appears above $\TN$ under weak pressure. Electronic properties in the PI phase are consistent with the Mott-Hubbard picture. This compound also undergoes a filling-control MIT with decreasing V concentration~\cite{Bao93}. In spite of a long history of the studies, a sufficient unified description of overall properties in the global phase diagram is still missing. \LSTiO{1-x}{x}{}{3} and \LSVO{1-x}{x}{}{3} are other examples where a filling-control MIT occurs accompanied by the AF phase transition and the quasiparticle mass enhancement. These are other examples which can not be described without overcoming the three separate scenarios of the MIT.

The discovery of high-$\Tc$ cuprate superconductors~\cite{BednortzMuller86} has promoted further intensive studies on the issue of the MIT with the aim of explaining their peculiar features observed near the Mott-insulating phase~\cite{Batlogg_SPRINGER,RMP_Imada}. A proper understanding and description of the high-$\Tc$ cuprate superconductivity remains still challenging, since it is difficult to correctly describe strong electron correlations that prominently emerge near the MIT. High-$\Tc$ cuprate superconductors have a quasi two-dimensional electronic structure. Anderson proposed that their effective model is given by the single-band Hubbard model for the Cu $3d_{x^2-y^2}$ orbital strongly hybridized with O $2p$ orbitals~\cite{Anderson87}. The undoped compounds are Mott insulators and exhibit the AF long-ranged order at low temperatures. Holes doped into these Mott insulators induce superconductivity at low temperatures. With further increasing holes or raising the temperature, they undergo a phase transition from the superconducting (SC) phase to a paramagnetic metal. Suppression of the quantum-mechanical coherence seems to reduce $\Tc$ with decreasing hole concentration below the optimum value. In contrast with the normal metals with a reduced Fermi temperature, the simple Brinkman-Rice picture does not describe the mechanism of this suppression of the coherence. This is because a pseudogap appears in low-energy single-particle, spin and charge excitations~\cite{Yasuoka,Berthier,ARPESrev,Ding96Nature,OnodaImada_AFMdSC}. AF spin excitations have a low-energy mode. Its characteristic energy decreases with decreasing $\Tc$ within the pseudogap. The competition of this mode with the singlet-pair correlations giving rise to the superconductivity appears to be the underlying origin of the pseudogap behavior. Pseudogap in charge excitations seems to continuously evolve into the SC gap which does not scale with $\Tc$. Towards the AFI phase, spectral weights outside the pseudogap must shift to high energies above the Mott gap. However, there seems no reliable theoretical framework that can describe such strong correlation effects near the MIT. For this purpose, it is crucial to develop a theoretical framework which gives a unified scheme containing the three aspects of the MIT, at least.

Therefore, construction of a unified theory which captures all the above three pictures of the MIT has attracted a general interest in the field of correlated electrons not only in the context of the specific subject of the MIT. In fact, any one of the above three pictures alone is not sufficient to describe the MIT even in the Hubbard model. Without a unified scheme, one can not reproduce the global phase diagrams of compounds that undergo an MIT. It is also difficult to clarify the mechanism of the high-$\Tc$ cuprate superconductivity.

The Brinkman-Rice solution is also derived as a nonmagnetic saddle point in a functional integral representation~\cite{KotliarRuckenstein86} based on the slave-boson formulation~\cite{Barnes76,Coleman84,ReadNewns83,Read85}, namely slave-boson mean-field (SBMF) theory. Including the Gaussian fluctuations around this nonmagnetic saddle point gives the Hubbard bands and incoherent features of single-particle excitations~\cite{CastellaniKotliarRaimondiGrilliWangRozenberg92, RaimondiCastellani93}. Short-ranged spin correlations, however, appear as higher-order loop corrections and were neglected in these theories. Instead, by considering both nonmagnetic and magnetic saddle points, the functional integral theory based on the slave-boson formulation gives an overall treatment of ferromagnetism, antiferromagnetism and MIT~\cite{KotliarRuckenstein86}. The results agree with those of Hartree-Fock theory in the weak-coupling regime. They are also consistent with Nagaoka's theorem~\cite{Nagaoka66} and Kanamori's results~\cite{Kanamori63} on ferromagnetism in the strong-coupling regime. For further improvements to include short-ranged spin correlations, however, it is required to overcome a difficulty associated with the uncontrolled treatment of the local constraints.

The dynamical mean-field theory (DMFT)~\cite{RMP_DMFT, MetznerVollhardt89} succeeded in incorporating the Hubbard and Brinkman-Rice pictures into one theoretical framework. This theory gives the exact Green's function in infinite dimensions. It predicts that at half filling, a resonance appears at the chemical potential inside the Hubbard gap below a critical interaction strength $U_{\rm c}$. The resonance has a constant intensity and a decreasing bandwidth as $U$ increases.  This agrees with the Brinkman-Rice picture~\cite{BrinkmanRice70}. The DMFT has also been applied to the finite-dimensional Hubbard model and other correlated models~\cite{RMP_DMFT}. However, it suffers from the following two related oversimplifications. Though this theory gives correct dynamical properties of the local self-energy part, it neglects its momentum dependence. In addition, the DMFT ignores the short-ranged two-particle correlations including spin correlations.

Some attempts have been made to improve the dynamical mean-field theory (DMFT). Recently, Jarrell and coworkers have developed the dynamical cluster approximation (DCA)~\cite{Hettler98,Hettler00,Maier00epjb,MoukouriHuscroftJarrell,HuscroftJarrellMaierMoukouriTahvildarzadeh99,MoukouriJarrell00}. In the DCA, the cluster problem with $N_{\rm c}(>1)$ sites is solved. In the case of $N_{\rm c}=1$, the DCA gives the original formulation of the DMFT~\cite{RMP_DMFT}. Then, $N_{\rm c}$ independent momentum regions appear in calculating the self-energy part. For the two-dimensional Hubbard model at half filling, the DCA succeeded in reproducing a Mott insulator with sharp peaks at the low-energy edges of the Hubbard bands energetically separated by an antiferromagnetic (AF) pseudogap at finite temperatures. At weak-to-intermediate couplings, this prediction contrasts with the paramagnetic solution of the DMFT. The DMFT predicts that the resonance emerges at the lowest energy~\cite{RMP_DMFT}. In the strong-coupling regime, the DCA and DMFT give similar Mott-insulating features. In the DCA, however, one has to solve the cluster problem. Although it captures some aspect of strong momentum dependence of the self-energy part, this gives a severe limitation due to a finite cluster size $N_{\rm c}$. This makes it difficult to consider the thermodynamic limit. It is also difficult to study more complicated systems with many degrees of freedom. Therefore, it is desired to build another analytic theory for studying correlated many-electron systems so that neither the system size problem nor the cluster size problem occurs.

From a viewpoint of the Slater picture, by using the diagrammatic method and/or Luttinger-Ward functional~\cite{AGD,Baym62,KadanoffBaym}, microscopic treatments to modify the RPA have been made~\cite{Baym62, KadanoffBaym, FLEX, Bickers91, TPSC, TPSC2, DeiszHessSerene96, SCRPA, SCRPA2}. They involve conserving approximations~\cite{Baym62,KadanoffBaym}, the fluctuation-exchange approximation~\cite{FLEX}, paramagnon theories, parquet approaches~\cite{Bickers91} and the two-particle self-consistent method~\cite{TPSC,TPSC2}. They can describe Fermi-liquid states in the weak-correlation regime~\cite{FLEX}, while they fail to predict Mott-insulating features~\cite{DeiszHessSerene96}, except the two-particle self-consistent method~\cite{TPSC,TPSC2}. The two-particle self-consistent method reproduces the AF pseudogap in the half-filled square-lattice Hubbard model at low temperatures~\cite{TPSC,TPSC2}. However, the Green's function and the self-energy part are not self-consistently determined in this method.

Apart from analytic theories, the quantum Monte-Carlo (QMC) simulation is a powerful tool to study strongly correlated models if the negative sign problem is not serious~\cite{RMP_Imada}. The static properties of the MIT in the two-dimensional Hubbard model were investigated by QMC calculations~\cite{FurukawaImada91, FurukawaImada92, FurukawaImada93}. The compressibility, thus the charge susceptibility was shown to diverge towards the MIT~\cite{FurukawaImada91}. QMC calculations, supplemented with an approximate analytic continuation by the maximum entropy method, have given some characteristic dynamical properties of the MIT in two dimensions~\cite{BulutScalapinoWhite94, HankePreuss95, GroberZacherEder98, AssaadImada99, GroberEderHanke00}: Nearly separated Hubbard bands develop near half filling. The Luttinger sum rule which is a criterion of the validity of the Fermi liquid in two- and three-dimensional metals at $T=0$ appears to be violated at least at $T\ge J=4t^2/U$ possibly due to thermal effects~\cite{BulutScalapinoWhite94}. This reflects that the Fermi temperature is suppressed near the MIT. Near half filling, a sharp peak of the local density of states develops near the Fermi level with decreasing temperature~\cite{BulutScalapinoWhite94}. Furthermore, at low temperatures $T<J$, extra two bands appear in addition to the Hubbard bands. A strongly weak dispersion around the $(\pi,0)$ and $(0,\pi)$ momenta has been found at zero temperature~\cite{AssaadImada99}. These features inferred from QMC calculations agree with the results of exact diagonalization~\cite{DaggottoOrtolaniScalapino92, RMP_Daggotto}. The strong momentum dependence of quasiparticle excitations is a crucial feature to be reproduced beyond the Mott-Hubbard, Brinkman-Rice and Slater pictures.

We next consider one-dimensional cases. Usual one-dimensional metals are classified into the Tomonaga-Luttinger (T-L) liquid~\cite{Tomonaga50,Luttinger63,Haldane81}. Spinons and holons play a role of elementary excitations in the T-L liquid. The Bethe ansatz wave function~\cite{LiebWu68} for the Hubbard model in one dimension suggests that in the case of $U=\infty$, spinons and holons are completely decoupled, and thereby the spin-charge separation is realized at all energies~\cite{OgataShiba90}. However, available spectral properties of the one-dimensional Hubbard model that can be extracted from the wave function are severely restricted to the case of half filling~\cite{MTakahashi70}, one hole doped into the Mott insulator~\cite{KawakamiOkiji89}, the strong-coupling limit~\cite{Schulz90, SorellaParola92}, the close vicinity of the Mott insulator~\cite{Schulz90} or $\omega=0$~\cite{ShastrySutherland90, KawakamiYang90}. For finite values of $U$, a mixing of spin and charge collective degrees of freedom occurs. In this case, information involved in the Bethe ansatz wave function~\cite{LiebWu68} is not sufficient for obtaining one-electron excitation spectra. Instead, QMC calculations supplemented with the maximum entropy method~\cite{PreussMuramatsuLindenDieterichAssaadHanke94} have given one-electron spectral properties at half filling which agree with those obtained by the SBMF approximation~\cite{KotliarRuckenstein86}.
For the one-dimensional $t$-$J$ model, exact diagonalization studies indicate the presence of spinon and holon branches in one-electron excitations~\cite{TohyamaMaekawa96,KimShenMotoyamaEisakiUchidaTohyamaMaekawa97} that are expected from the arguments for $J=0$~\cite{OgataShiba90,SorellaParola92}: One-electron excitations are formed by the convolution of spinons and holons. The bandwidth is given by $\pi J$ for spinons and $4t$ for holons. Then, they compose two branches in one-electron excitations mainly characterized by spinons and holons. At half filling, the holons become massive and this produces a Mott gap in the one-electron excitations. When the spin-charge separation occurs prominently, the spinon branch in the lower Hubbard band at half filling is found only for $|k|<\pi/2$. Then, the complete shadow structure is missing in one-electron excitations. This is allowed in one dimension, because even at $T=0$, there is no AF long-ranged order and the translational symmetry is never broken. For the Hubbard model with a finite value of $U$, however, a coupling between spinons and holons makes it difficult to discuss properties of one-electron excitation spectra.

In spite of these studies mentioned above, a sufficient theoretical framework that gives a unified treatment of Hubbard, Brinkman-Rice and Slater pictures together with strong momentum dependence of quasiparticle excitations and reproduces key features obtained in numerical simulations is still missing. Particularly, it remains unsolved how electronic spectra change from metals to Mott insulators. This is an important subject in the investigation of the MIT. To study this subject, in the rest of this section, we introduce a promising theoretical formulation of the operator projection method (OPM) employed in the previous~\cite{OPM_jpsjletter01} and the present papers. The OPM is a non-perturbative analytic approach which systematically improves the Hartree-Fock theory, the Dyson-equation formalism of the Green's function, theories based on spin fluctuations~\cite{Baym62, KadanoffBaym, FLEX, Bickers91, TPSC, TPSC2, DeiszHessSerene96}, Hubbard approximations~\cite{Hubbard1, Hubbard3} and two-pole approximations~\cite{Roth69, Nolting72, BeenenEdwards95}. The OPM also reproduces a sharp peak near the Fermi level in doped cases, in agreement with the results of the dynamical mean-field theory~\cite{RMP_DMFT}. The velocity that characterizes the low-energy single-particle excitations is shown to vanish towards the MIT particularly around $(\pi,0)$ and $(0,\pi)$. This suggests that the quasiparticle mass diverges towards the MIT, though the enhancement of the damping rate makes it difficult to define quasiparticles. This mass divergence is to some extent consistent with the Brinkman-Rice picture, although its strong momentum dependence is beyond the Brinkman-Rice picture. Moreover, the OPM reproduces a sharp peak of the density of states at each edge of the Mott gap that leads to the divergence of the compressibility towards half filling. This contrasts with the Brinkman-Rice and dynamical mean-field theory. Therefore, the method opens a way to construct a unified scheme beyond the three different pictures. Furthermore, our theory given in the previous~\cite{OPM_jpsjletter01} and the present papers does not suffer from the problem on the system size, unlike the dynamical cluster approximation~\cite{MoukouriHuscroftJarrell}.

The projection technique in the operator space to calculate correlation functions in the linear response theory~\cite{Kubo} was developed by Nakajima~\cite{Nakajima58}, Zwanzig~\cite{Zwanzig} and Mori~\cite{Mori65}. This is a non-perturbative approach based on Heisenberg equations of motion~\cite{Zubarev60}. It leads to a continued-fraction expansion of a correlation function. The expansion is uniquely determined and can also be described by the series of Dyson equations. This continued-fraction expression also uniquely determines a corresponding moment expansion. Though an exact proof is absent, the continued-fraction expansion is empirically known to converge more rapidly than series expansions and to have a larger convergence region. Any theory involving an equation of motion is called an equation-of-motion approach. In this sense, the operator projection theory belongs to a family of equation-of-motion approaches~\cite{Hubbard1, Hubbard2, Hubbard3, Roth69, Nolting72, BeenenEdwards95, MatsumotoMancini97, Plakida, MatsumotoSaikawaMancini96}. Operator projection theories are, however, more sophisticated in that it produces a systematic way of performing a continued-fraction expansion in energy. The OPM opens a possibility to systematically extract the lower-energy spectral properties projecting out higher-energy degrees of freedom in a self-consistent fashion.

The formulation of the OPM is explicitly given in \S~\ref{sec:Formulation}. Here, we summarize the connection of the OPM with other known theoretical frameworks.

If one performs the projection procedure for the single-particle operators up to the first order, it gives the Hartree-Fock theory, in the symmetry-broken phases as well as symmetry-unbroken phases, the free-electron properties. At this stage, the higher-order dynamics remains as the self-energy part in the so-called Dyson equation for the Green's function. This gives a starting point of the perturbation theory, and thus conserving approximations~\cite{Baym62, KadanoffBaym}, the fluctuation-exchange approximation~\cite{FLEX}, paramagnon theories, parquet approaches~\cite{Bickers91} and the two-particle self-consistent method~\cite{TPSC, TPSC2}. As we have already mentioned, these theories fail to predict Mott-insulating features in a self-consistent manner. Here, we note that this is because only the zeroth- and first-order moments of the Green's function are exact at this stage. To study strong correlation effects that prominently emerge near the MIT, a natural step is to find a way of treating the self-energy part in a systematic fashion beyond the Gaussian approximations. Since the perturbation theory works for the self-energy part more effectively than for the Green's function itself~\cite{AGD}, it seems natural to expect that considering the Dyson equation for the self-energy part improves its estimate. The projection technique is useful for this purpose.

The second-order projection filters out the local nature of the Hubbard model. Here, the atomic properties and the superexchange term are involved. In the Hubbard approximations~\cite{Hubbard1, Hubbard2, Hubbard3}, this procedure was partly taken: Hubbard obtained the exact second-order moments of the Green's function but not the third-order moments. His attempts of making a self-consistent theory do not constitute the systematic projection procedure. Roth improved the Hubbard I approximation along the concept of the operator projection theory~\cite{Roth69}. This introduces the superexchange term into the self-energy part which gives rise to a correct treatment of the third-order moments. However, the higher-order dynamics appearing in the Dyson equation of the self-energy part, which is hereafter called the second-order self-energy part, was ignored. Then, his results for the Green's function have only two poles (two-pole approximation). Some attempts to take account of higher-order dynamics were made by Matsumoto and Mancini~\cite{MatsumotoMancini97}. In their method called the composite operator method, the higher-order dynamics were included only within the two-site level in a local picture. However, it is not known how reliable the composite operator method is in the presence of short-ranged fluctuations. It is crucial to incorporate short-ranged AF spin correlations near continuous magnetic phase transitions. These short-ranged correlations can be included in the operator projection method (OPM) by several choices of decoupling approximations for evaluating the second-order self-energy part through convolutions of the single-particle Green's function and two-particle susceptibilities~\cite{OPM_jpsjletter01}.

The results obtained with the present method are given in \S~\ref{sec:Results}.
The present OPM reproduces the following single-particle properties in two dimensions: (I) The filling-control MIT takes place at half filling. A Mott gap opens and completely separates the two Hubbard bands. (II) At low temperatures, strong AF spin fluctuations produce an AF shadow in each Hubbard band, yielding a four-band structure. This structure in energy and momentum is regarded as a superposition of two low-energy narrow bands and two SDW-like bands. (III) Upon hole (electron) doping, the narrow bands appear to naturally merge into the top of the lower (upper) Hubbard band. Then, the low-energy band persists near half filling. This narrow band has a particularly weak dispersion and dominant weights in the momentum regions around $(\pi, 0)$ and $(0, \pi)$. (IV) Due to these momentum regions, a sharp peak of the local density of states grows at the Fermi level. With increasing temperature, the sharp peak diminishes due to thermal effects. (V) The Fermi velocity characterizing the low-energy narrow band is found to vanish as $\ave{n}\to1$, reflecting a suppressed Fermi degeneracy near the MIT. This suggests that the quasiparticle mass diverges towards the MIT. This is consistent with the Brinkman-Rice picture. However, our results show that an enhancement of the damping rate of quasiparticles makes it difficult to define quasiparticles. It also shows characteristic and strong momentum dependence of quasiparticles excitations beyond the Brinkman-Rice picture. Such enhanced momentum dependence may generate momentum-dependent electron differentiations in the low-energy level of the energy hierarchy formation. (VI) For the one-dimensional half-filled Hubbard model, the OPM also reproduces a Mott insulator at an intermediate coupling. The Hubbard bands separated by a Mott gap appear. At lower temperatures, their AF shadows develop similarly to the two-dimensional case, in spite of the absence of the AF long-ranged order. The single-particle excitations in one dimension consist of the convolutions of the spinon and holon excitations which couple each other in the case of finite values of $U$. The present results suggest that at an intermediate coupling, the mixing of spinons and holons is too large to interpret the single-particle excitations in terms of the simple picture based on the completely decoupled spinons and holons~\cite{KimShenMotoyamaEisakiUchidaTohyamaMaekawa97}. These features obtained by the OPM are consistent with QMC results~\cite{BulutScalapinoWhite94,HankePreuss95,GroberZacherEder98,AssaadImada99,GroberEderHanke00,PreussMuramatsuLindenDieterichAssaadHanke94}, as we will discuss in this paper.

The Mott insulators are incompressible, while the metals are compressible. This is reflected in the charge susceptibility; in the Mott insulator, the charge susceptibility vanishes. On the other hand, approaching the MIT in the metallic phase, the compressibility, thus the charge susceptibility diverges towards the MIT~\cite{RMP_Imada}. The present method also correctly reproduces this remarkable property.

It is remarkable that the present theory captures the three essential features of the MIT as given in the (i) Mott-Hubbard, (ii) Brinkman-Rice, and (iii) Slater pictures. In fact, this theory reproduces the Mott-Hubbard picture at moderate temperatures near the MIT. It also reproduces the enhancements of the quasiparticle mass and damping rate near the MIT. This reflects the suppression of the Fermi degeneracy near the MIT. This reminds us of the Brinkman-Rice picture. The growth of AF spin fluctuations at low temperatures yields AF shadows of the Hubbard bands in the single-particle dispersions. In \S~\ref{sec:TotalView}, we will give our view of the MIT in the two-dimensional Hubbard model. All the three aspects are crucial to correctly describe the MIT. We also note that previous equation-of-motion approaches have not taken account of short-ranged AF spin correlations.

In \S~\ref{sec:ARPES}, we compare the present results with experimental observations in high-$\Tc$ cuprate superconductors. Similarity of the narrow band obtained here to that observed in high-$\Tc$ cuprate superconductors implies that to discuss the superconductivity, it is required to include strong electron correlations related to the MIT.

\S~\ref{sec:Summary} is devoted to the summary of this paper. We also discuss problems left for future study.

\section{Formulation of operator projection method}
\label{sec:Formulation}

Below, we follow the general formalism of the OPM which has been briefly discussed in the previous letter~\cite{OPM_jpsjletter01}.
We take the Hubbard Hamiltonian with an electron transfer $t_{\mibs{x},\mibs{x}'}$ from an atomic site $\mib{x}'$ to $\mib{x}$ and the local Coulomb repulsion $U$;
\begin{equation}
H\equiv-\sum_{\mibs{x},\mibs{x}',s}t_{\mibs{x},\mibs{x}'}c^\dagger_{\mibs{x}s}c_{\mibs{x}'s}+U\sum_{\mibs{x}}n_{\mibs{x}\up}n_{\mibs{x}\down},
\label{eq:Formulation:H}
\end{equation}
with $n_{\mibs{x}s}\equiv c^\dagger_{\mibs{x}s}c_{\mibs{x}s}$. For simplicity, we restrict ourselves to the case of $t_{\mibs{x},\mibs{x}'}=t_{\mibs{x}',\mibs{x}}$. Applications to more general situations are straightforward.
We define $\bar{H}\equiv H-\mu\sum_{\mibs{x}s}n_{\mibs{x}s}$ with the chemical potential $\mu$.
We perform the projection procedure for the electron creation and annihilation operators at an atomic site $\mib{x}$ with a spin index $s$, $c^\dagger_{\mibs{x},s}$ and $c_{\mibs{x},s}$.
It is useful to define the fermionic $4N$-component vector operator $\bfPsi$, which is composed of
\begin{eqnarray}
\Psi_{\mibs{x}}&\equiv&\left(\frac{\rho_0+\rho_3}{2}\sigma_0-i\frac{\rho_0-\rho_3}{2}\sigma_2\right){}^t(c_{\mibs{x}\up},c_{\mibs{x}\down},c^\dagger_{\mibs{x}\up},c^\dagger_{\mibs{x}\down})
\nonumber\\
&=&{}^t(c_{\mibs{x}\up},c_{\mibs{x}\down},-c^\dagger_{\mibs{x}\down},c^\dagger_{\mibs{x}\up}).
\label{eq:Formulation:Psi_x}
\end{eqnarray}
in the real-space representation. $N$ represents the number of the atoms. $\bfPsid$ is defined as the Hermitian conjugate operator of $\bfPsi$. In the real-space representation, it is written as
\begin{equation}
\Psi^\dagger_{\mibs{x}}=(c^\dagger_{\mibs{x}\up},c^\dagger_{\mibs{x}\down},-c_{\mibs{x}\down},c_{\mibs{x}\up}).
\label{eq:Psid_x}
\end{equation}
$\sigma_0$ ($\rho_0$) and $\mib{\sigma}$ ($\mib{\rho}$) are the identity and Pauli matrices, respectively, which operate to the spin (charge) space of vector operators as $\Psi$ from the left-hand side and those as $\Psid$ from the right-hand side.

We define the commutator of an operator $A$ with $\bar{H}$ as $\ho A\equiv[A,\bar{H}]_-$ and the thermal average of $A$ at temperature $T$ as $\ave{A}\equiv\Tr[e^{-\bar{H}/T}A]/Z$ with the partition function $Z\equiv\Tr[e^{-\bar{H}/T}]$. The response function of $\bfPsi$ is introduced in the $4N\times4N$ matrix representation as $\mib{K}_{\Psi,\Psid}(t)$. Its $(a, a')\otimes(\mib{x},\mib{x}')$ component is defined by
\begin{equation}
K_{\Psi,\Psid}^{aa'}(t;\mib{x},\mib{x}')
\equiv-i\ave{\com{+}{\Psi_{\mibs{x}}^a(t),(\Psid){}_{\mibs{x}'}^{a'}}}.
\label{eq:Formulation:K_Psi,Psid}
\end{equation}
Here, the Heisenberg operator for an operator $A$ has been introduced as%
\begin{equation}
A(t)\equiv e^{i\bar{H}t}A e^{-i\bar{H}t}.
\end{equation}
The fermion anticommutation relations for $c_{\mibs{x},s}$ and $c^\dagger_{\mibs{x}',s'}$ leads to
\begin{equation}
\mib{K}_{\Psi,\Psid}(0)=-i\mib{I}
\label{eq:Formulation:K0_Psi,Psid}
\end{equation}
with the $4N\times4N$ identity matrix $\mib{I}$. The Fourier transform of \eq{eq:Formulation:K_Psi,Psid} gives the susceptibility $\bfchi_{\Psi,\Psid}(\omega)$, which is nothing but the Green's function $\mib{G}(\omega)$ in the $4N\times4N$ matrix representation. In more general, for arbitrary $4N$-component vector operators $\bfphi$ and $\bfvarphi$, their response function and susceptibilities can be defined in the $4N\times4N$ matrix representation as $\mib{K}_{\phi,\varphi^\dagger}(t)$ and $\bfchi_{\phi,\varphi^\dagger}(\omega)$, respectively. Throughout this paper, the $(a,a')\otimes(\mib{x},\mib{x}')$ component of the response function $\mib{K}_{\phi,\varphi^\dagger}(t)$ is defined by
\begin{equation}
K^{aa'}_{\phi,\varphi^\dagger}(t)\equiv-i\ave{\com{\pm}{\phi^a_{\mibs{x}}(t),(\varphi^\dagger){}^{a'}_{\mibs{x}'}}},
\label{eq:Formulation:K_phi,varphi}
\end{equation}
with the commutator and the anticommutator when $\mib{\phi}$ and $\mib{\varphi}$ are operators with even and odd number of particles, respectively. The susceptibility $\bfchi_{\phi,\varphi^\dagger}(\omega)$ is given by the Fourier transform of $\mib{K}_{\phi,\varphi^\dagger}(t)$;
\begin{equation}
\bfchi_{\phi,\varphi^\dagger}(\omega)=\int_0^\infty\!dt\,e^{i\omega t}\mib{K}_{\phi,\varphi^\dagger}(t).
\label{eq:Formulation:chi_phi,varphi}
\end{equation}

\subsection{First-order projection}
\label{subsec:1OP}

As is well known~\cite{OPM_jpsjletter01,Fulde}, the first-order projection reproduces the Hartree-Fock theory. It also allows us to construct the Dyson equation for the Green's function which is obtained from the perturbation theory~\cite{AGD}. At this order of projection, only the zeroth- and first-order moments of the Green's functions appear explicitly. In the following, we give detailed arguments of the projection.

The projection procedure is given by
\begin{subequations}
\begin{eqnarray}
\ho\bfPsi&=&\bfveps^{(11)}\bfPsi+\delho\bfPsi,
\slabel{eq:1OP:ho-Psi}\\
\bfveps^{(11)}&\equiv&\mib{K}_{\ho\Psi,\Psid}(0)\mib{K}^{-1}_{\Psi,\Psid}(0),
\slabel{eq:1OP:veps11}\\
\delho\bfPsi&\equiv&(1-P_1)\ho\bfPsi.
\slabel{eq:1OP:delho-Psi}
\end{eqnarray}
\label{eq:1OP:project1-Psi}
\end{subequations}
$P_1$ is defined as the first-order projection operator which operates to an arbitrary fermionic $4N$-component vector operator $\bfphi$ as
\begin{equation}
P_1\bfphi=\mib{K}_{\phi,\Psid}(0)\mib{K}^{-1}_{\Psi,\Psid}(0)\bfPsi.
\label{eq:1OP:P_1}
\end{equation}
$P_1$ satisfies the property of the projection operator,
\begin{equation}
P_1^2=P_1.
\label{eq:1OP:P_1^2}
\end{equation}
We note that the new operator $\delho\bfPsi$ satisfies
\begin{equation}
\mib{K}_{\Psi,(\delho\Psi)^\dagger}(0)=\mib{K}_{\delho\Psi,\Psid}(0)=0
\label{eq:1OP:K_Psi,delho-Psid}
\end{equation}
by definition. It allows us to express the single-particle Green's function $\mib{G}(\omega)=\bfchi_{\Psi,\Psid}(\omega)$ in the form of a Dyson equation as
\begin{eqnarray}
\mib{G}(\omega)
&=&i\left[\mib{G}^{(0)}{}^{-1}(\omega)-\bfSigma_1(\omega)\right]^{-1}
\mib{K}_{\Psi,\Psid}(0),
\label{eq:1OP:Dyson-G}\\
\mib{G}^{(0)}(\omega)&=&\left[\omega\mib{I}-\bfveps^{(11)}\right]^{-1},
\label{eq:1OP:chi_Psi^0}\\
\bfSigma_1(\omega)
&=&-i\bfchi_{\delho\Psi,(\delho\Psi)^\dagger}^{\rm irr}(\omega)
\mib{K}_{\Psi,\Psid}^{-1}(0).
\label{eq:1OP:Dyson-chi_Psi}
\end{eqnarray}
The irreducible part of $\bfchi_{\delho\Psi,(\delho\Psi)^\dagger}(\omega)$
with respect to $\mib{G}^{(0)}(\omega)$ has been introduced;
\begin{eqnarray}
\lefteqn{\bfchi_{\delho\Psi,(\delho\Psi)^\dagger}^{\rm irr}(\omega)}
\nonumber\\
&=&\left[\bfchi_{\delho\Psi,(\delho\Psi)^\dagger}^{-1}(\omega)
-i\mib{K}_{\Psi,\Psid}^{-1}(0)\mib{G}^{(0)}(\omega)\right]^{-1}.
\label{eq:1OP:chi_delho-Psi_delho-Psid}
\end{eqnarray}
For the Hubbard model, one obtains
\begin{subequations}
\begin{eqnarray}
\veps^{(11)}_{\mibs{x},\mibs{x}'}
&=&\vepsz_{\mibs{x},\mibs{x}'}\rho_3\sigma_0-U\delta_{\mibs{x},\mibs{x}'}
\left(\ave{\mib{S}_{\mibs{x}}}\rho_0\mib{\sigma}
-\frac{\ave{\Delta^{\bar{i}}_{\mibs{x}}}}{\sqrt{2}}\rho_{\bar{i}}\sigma_0\right),\nonumber\\
\label{eq:1OP:veps11_x}\\
\vepsz_{\mibs{x},\mibs{x}'}&=&-t_{\mibs{x},\mibs{x}'}
-\left(\mu-\frac{U}{2}\ave{n_{\mibs{x}}}\right)\delta_{\mibs{x},\mibs{x}'}.
\label{eq:1OP:vepsz_x}
\end{eqnarray}
\end{subequations}
Hereafter, the summation over $\bar{i}=1$ and $2$ should be taken. We have introduced the local charge, spin and pairing operators at the atomic site $\mib{x}$ as
\begin{subequations}
\begin{eqnarray}
n_{\mibs{x}}&\equiv&n_{\mibs{x}\up}+n_{\mibs{x}\down},
\label{eq:1OP:n_x}\\
\mib{S}_{\mibs{x}}&\equiv&c^\dagger_{\mibs{x}s}\mib{\sigma}_{ss'}c_{\mibs{x}s'}/2,
\label{eq:1OP:S_x}\\
\Delta^{1}_{\mibs{x}}&\equiv&\frac{1}{\sqrt{2}}(c_{\mibs{x}\up}c_{\mibs{x}\down}
+c^\dagger_{\mibs{x}\down}c^\dagger_{\mibs{x}\up}),
\label{eq:1OP:Deltass1}\\
\Delta^{2}_{\mibs{x}}&\equiv&\frac{i}{\sqrt{2}}(c_{\mibs{x}\up}c_{\mibs{x}\down}
-c^\dagger_{\mibs{x}\down}c^\dagger_{\mibs{x}\up}).
\label{eq:1OP:Deltass2}
\end{eqnarray}
\end{subequations}
The superconducting state with the isotropic $s$-wave pairing symmetry can be excluded in the repulsive Hubbard model ($U\ge0$), i.e., $\ave{\Delta^{i}_{\mibs{x}}}=0$. The expectation value $\ave{\mib{S}_{\mibs{x}}}$, if it does not vanish, is chosen to break the symmetry in the $z$-axis, i.e., $\ave{S^{\pm}_{\mibs{x}}}=0$ without loss of generality. Then, \eq{eq:1OP:veps11_x} and $\delho\bfPsi$ are reduced to
\begin{subequations}
\begin{eqnarray}
\veps^{(11)}_{\mibs{x},\mibs{x}'}
&=&\vepsz_{\mibs{x},\mibs{x}'}\rho_3\sigma_0
-U\ave{S^z_{\mibs{x}}}\delta_{\mibs{x},\mibs{x}'}\rho_0\sigma_3,
\label{eq:1OP:veps11_x_2}\\
\delho\Psi_{\mibs{x}}&=&U\left(\frac{1}{2}\delta n_{\mibs{x}}\rho_3\sigma_0
-\delta S^z_{\mibs{x}}\rho_0\sigma_3\right)\Psi_{\mibs{x}},
\label{eq:1OP:delhho-Psi_x}
\end{eqnarray}
\label{eq:1OP:project1-Psi_x_2}
\end{subequations}
with $\delta n_{\mibs{x}}=n_{\mibs{x}}-\ave{n_{\mibs{x}}}$ and
$\delta S^z_{\mibs{x}}=S^z_{\mibs{x}}-\ave{S^z_{\mibs{x}}}$.

If $\bfSigma_1(\omega)$ is ignored, then the present theory is reduced to the Hartree-Fock theory. The Hartree-Fock theory captures some of the many-body effects in the weak-coupling limit, while for the intermediate-to-strong couplings, the theory breaks down. In fact, in the strong-coupling regime, the functional integral representation of the action~\cite{Hertz76} gives an expansion in a small parameter $t/U$ around the Hartree-Fock solution. However, the Hartree-Fock theory fails to reproduce correct dynamical properties. In order to improve the Hartree-Fock theory, one must include the self-energy part corrections $\bfSigma_1(\omega)$.

For this purpose, one can adopt perturbation expansions on a single-particle basis, one-loop approximations, long-time or short-time approximations. However, these approximations do not give the energy- and momentum-dependent self-energy part in a correct manner. Particularly, if $\bfSigma_1(\omega)$ is self-consistently determined by means of simple framework such as perturbation expansion or one-loop approximations~\cite{Baym62,KadanoffBaym,FLEX,Bickers91}, the Mott insulator at half filling can not be reached~\cite{TPSC,DeiszHessSerene96}. This is because arguments along this line do not satisfy the Pauli principle and fail to give a correct high-energy behavior of the self-energy part~\cite{TPSC}.

The Pauli principle is a crucial aspect in reproducing a Mott insulator in the Hubbard model, since any deviation allows electrons (holes) to move through atomic sites already occupied by another electron (hole) with the same spin without any repulsion. Any incorrect treatment of the Pauli principle also violates the local moment sum rule~\cite{TPSC}. Only the zeroth- and first-order moments of the Green's function are correct in these theories, while the Pauli principle $n_{\mibs{x}s}^2=n_{\mibs{x}s}$ enters a theory in the evaluation of the second- and higher-order moments. We note that the Pauli principle can be exactly satisfied in each order of the equations of motion.

Another key factor to reproduce a Mott insulator is to correctly estimate the double occupancy. The double occupancy directly determines the local moment of the spin. In contrast with the self-consistent second-order perturbation theory, the dynamical mean-field theory supplemented with the iterative second-order perturbation theory reproduces the Mott insulator~\cite{RMP_DMFT}. This is also because the Pauli principle and the double occupancy are correctly treated in solving the impurity problem.

\subsection{Second-order projection}
\label{subsec:2OP}

As is well known, the perturbation theory works more effectively for the self-energy part in the form of the Dyson equation rather than for the Green's function itself. Therefore, it is natural to consider the Dyson equation for the self-energy part. The operator projection theory is useful for this purpose. The second-order projection reproduces the two-pole approximation to the Green's function. It also allows us to construct the Dyson equation for the self-energy part.

The second-order projection is given by
\begin{subequations}
\begin{eqnarray}
\ho\delho\bfPsi
&=&\bfveps^{(21)}\bfPsi+\bfveps^{(22)}\delho\bfPsi+\delho\delho\bfPsi,
\label{eq:2OP:ho-delho-Psi}\\
\bfveps^{(21)}&\equiv&\mib{K}_{\ho\delho\Psi,\Psid}(0)\mib{K}^{-1}_{\Psi,\Psid}(0),
\label{eq:2OP:veps21}\\
\bfveps^{(22)}&\equiv&\mib{K}_{\ho\delho\Psi,(\delho\Psi)^\dagger}(0)
\mib{K}^{-1}_{\delho\Psi,(\delho\Psi)^\dagger}(0),
\label{eq:2OP:veps22}\\
\delho\delho\bfPsi&\equiv&(1-P_2)\ho\delho\bfPsi.
\label{eq:2OP:delho-delho-Psi}
\end{eqnarray}
\label{eq:2OP:project2-Psi}
\end{subequations}
$P_2$ is defined as the second-order projection operator which operates to an arbitrary fermionic $4N$-component vector operator $\bfphi$ as
\begin{eqnarray}
P_2\bfphi&=&\mib{K}_{\phi,\Psi}(0)\mib{K}^{-1}_{\Psi,\Psid}(0)\bfPsi
\nonumber\\
&&{}+\mib{K}_{\phi,(\delho\Psi)^\dagger}(0)\mib{K}^{-1}_{\delho\Psi,(\delho\Psi)^\dagger}(0)\delho\bfPsi.
\label{eq:2OP:P_2}
\end{eqnarray}
It satisfies the condition
\begin{equation}
P_2^2=P_2.
\label{eq:2OP:P_2^2}
\end{equation}
From \eq{eq:2OP:project2-Psi}, one obtains the irreducible self-energy part
\begin{subequations}
\begin{eqnarray}
\hspace*{-20pt}\bfSigma_1(\omega)&=&\left[\bfSigma_1^{(0)}{}^{-1}(\omega)
-\bfSigma_2(\omega)\right]^{-1}\bfveps^{(21)},
\label{eq:2OP:Sigma_1}\\
\hspace*{-20pt}\bfSigma_1^{(0)}(\omega)&=&\left[\omega\mib{I}-\bfveps^{(22)}\right]^{-1},
\label{eq:2OP:Sigma_1^0}\\
\hspace*{-20pt}\bfSigma_2(\omega)
&=&-i\bfchi^{\rm irr}_{\delho\delho\Psi,(\delho\delho\Psi)^\dagger}(\omega)
\mib{K}^{-1}_{\delho\Psi,(\delho\Psi)^\dagger}(0).
\label{eq:2OP:Sigma_2}
\end{eqnarray}
\label{eq:2OP:Sigma_1,Sigma_2}
\end{subequations}
Here, we have introduced the irreducible part of $\bfchi_{\delho\delho\Psi,(\delho\delho\Psi)^\dagger}(\omega)$ with respect to $\mib{G}^{(0)}(\omega)$ and $\bfSigma_1(\omega)$ as
\begin{eqnarray}
\lefteqn{\bfchi_{\delho\delho\Psi,(\delho\delho\Psi)^\dagger}^{\rm irr}(\omega)}
\nonumber\\
&&=\left[\bfchi^{-1}_{\delho\delho\Psi,(\delho\delho\Psi)^\dagger}(\omega)
-i\mib{K}^{-1}_{\delho\Psi,(\delho\Psi)^\dagger}(0)\bfSigma_1^{(0)}(\omega)\right]^{-1}.
\label{eq:2OP:chi^irr_delho-Psi_delho-Psid}
\end{eqnarray}
For the repulsive Hubbard model,
\begin{subequations}
\begin{eqnarray}
\veps^{(21)}_{\mibs{x},\mibs{x}'}&=&U^2\delta_{\mibs{x},\mibs{x}'}M_{\mibs{x}},
\label{eq:2OP:veps21_x}\\
M_{\mibs{x}}&\equiv&\ave{\left(\frac{1}{2}\delta n_{\mibs{x}}\rho_3\sigma_0
-\delta S^z_{\mibs{x}}\rho_0\sigma_3\right)^2},
\label{eq:2OP:M_x}\\
\veps^{(22)}_{\mibs{x},\mibs{x}'}&=&-t^{(22)}_{\mibs{x},\mibs{x}'}\rho_3\sigma_0-\left(\mu_2\rho_3\sigma_0-\ave{S^z_x}\rho_0\sigma_3\right)
\delta_{\mibs{x},\mibs{x}'},
\label{eq:2OP:veps22_x}\\
t^{(22)}_{\mibs{x},\mibs{x}'}&\equiv&t_{\mibs{x},\mibs{x}'}
\langle\delta\calL_{\mibs{x}}\rho_3\delta\calL_{\mibs{x}'}\rho_3\rangle M_{\mibs{x}'}^{-1},
\label{eq:2OP:t^2_x}\\
\delta\calL_{\mibs{x}}&\equiv&\calL_{\mibs{x}}-\ave{\calL_{\mibs{x}}},
\label{eq:2OP:delta-calL_x}\\
\calL_{\mibs{x}}&\equiv&\frac{1}{2}\delta n_{\mibs{x}}\rho_3\sigma_0
-\delta\mib{S}_{\mibs{x}}\cdot\rho_0\mib{\sigma}
+\frac{1}{\sqrt{2}}\Delta^{\bar{i}}_{\mibs{x}}\rho_{\bar{i}}\sigma_0,
\label{eq:2OP:calL_x}\\
\mu_2&=&\muz_2\rho_0\sigma_0+\left[(1-\ave{n})\vepskin+\vepscor{2}\right]M_{\mibs{x}}^{-1},
\label{eq:2OP:mu2}\\
\muz_2&\equiv&\mu-U\left(1-\frac{1}{2}\ave{n_{\mibs{x}}}\right),
\label{eq:2OP:muz2}\\
\vepskin&=&\frac{-1}{N}\sum_{\mibs{k}}\!t_{\mibs{k}}\!
\left(\frac{1}{2}\ave{n_{0;\mibs{k}}}\rho_3\sigma_0
-\ave{S^z_{0;\mibs{k}}}\rho_0\sigma_3\right).
\label{eq:2OP:vepskin}
\end{eqnarray}
\label{eq:2OP:project2-Psi_x}
\end{subequations}
\noindent
with
\begin{subequations}
\begin{eqnarray}
n_{\mibs{q};\mibs{k}}&\equiv&\sum_sc^\dagger_{\mibs{k}-\mibs{q}/2,s}c_{\mibs{k}+\mibs{q}/2,s},
\label{eq:2OP:n_qk}\\
\mib{S}_{\mibs{q};\mibs{k}}&\equiv&\frac{1}{2}\sum_{ss'}c^\dagger_{\mibs{k}-\mibs{q}/2,s}\mib{\sigma}_{ss'}c_{\mibs{k}+\mibs{k}/2,s'}.
\label{eq:2OP:S_qk}
\end{eqnarray}
\end{subequations}
The kinetic energy per site summed over spins is expressed as
$\Ekin=\frac{1}{2}\Tr\vepskin(\rho_0+\rho_3)=
-\frac{1}{N}\sum_{\mibs{k},s}t_{\mibs{k}}\ave{c^\dagger_{\mibs{k}s}c_{\mibs{k}s}}$.
Below, a momentum independent energy shift $\vepscor{2}$
due to two-site correlated hopping terms which take the form
$t_{\mibs{x},\bar{\mibs{x}}}\ave{c^\dagger_{\mibs{x}s}c_{\bar{\mibs{x}}s'}
\delta n_{\mibs{x}-s}}$ in \eq{eq:2OP:veps22_x} are neglected. We note that in the particle-hole symmetric case, $\vepscor{2}$ vanishes. Then, in the case of $\ave{S^z_{\mibs{x}}}=0$, the remaining operator becomes
\begin{eqnarray}
\delho\delho\Psi_{\mibs{x}}&\approx&
\left(\delta_{\mibs{x},\bar{\mibs{x}}}(1-\ave{n})\vepskin M_{\mibs{x}}^{-1}
+t^{(22)}_{\mibs{x},\bar{\mibs{x}}}\right)\rho_3\delho\Psi_{\bar{\mibs{x}}}
\nonumber\\
&&{}-Ut_{\mibs{x},\bar{\mibs{x}}}\delta\calL_{\mibs{x}}\rho_3\Psi_{\bar{\mibs{x}}}.
\label{eq:2OP:delho-delho-Psi_x}
\end{eqnarray}

A general two-pole approximation by neglecting $\Sigma_{2\rme}(\omega)$ and its approximate treatment will be given in \S~\ref{subsec:Two-pole} and \S~\ref{subsec:Higher}, respectively.

\subsection{Neglecting $\bfSigma_2(\omega)$ --- Two-pole approximation}
\label{subsec:Two-pole}

If $\bfSigma_2(\omega)$ and the momentum-dependence of $\bfveps^{(22)}$ are neglected, the present formalism gives the Hubbard I approximation~\cite{Hubbard1}. The momentum dependence of $\bfveps^{(22)}$ appears through the equal-time short-ranged charge, spin and isotropic $s$-wave pairing correlations. This modifies the dispersions of the separated Hubbard bands. This yields the Roth's two-pole approximation~\cite{Roth69}.

The two-pole approximation has been found to give a good starting point in discussing main single-particle dispersions near the MIT. Two dispersions of the energetically separated Hubbard bands obtained with this approximation resemble those obtained by QMC calculations~\cite{Roth69, BeenenEdwards95}, while it does not reproduce the self-energy part at $\omega=0$ obtained by Kanamori in the low density limit. Two-pole approximations~\cite{HarrisLange67, Roth69, BeenenEdwards95} as well as Hubbard approximations~\cite{Hubbard1, Hubbard3} are also known to violate the Luttinger sum rule due to the formation of two distinct Hubbard bands. To overcome these problems, it is required to include the higher-order dynamics.

Here, we briefly discuss a general two-pole approximation. In the symmetry-unbroken phase, the electron Green's function with momentum $\mib{k}$ and spin $s$ reads
\begin{equation}
G_{\rme,s}(\omega,\mib{k})=\left[\omega-\vepsz_{\mibs{k}}
-\frac{\left(\frac{U}{2}\right)^2\ave{n}(2-\ave{n})}
{\omega-\tilde{\veps}_{\mibs{k}}}\right]^{-1}.
\label{eq:2OP:G_2^0}
\end{equation}
Here, the (1,1) components of \eqd{eq:2OP:veps22_x}{eq:2OP:t^2_x} and $\vepscor{2}$ have been introduced in the momentum space as
\begin{equation}
\tilde{\veps}_{\mibs{k}}=-\tilde{t}_{\mibs{k}}-\muz_2
-\frac{2(1-\ave{n})\Ekin+4E_{{\rm cor}2}}{\ave{n}(2-\ave{n})},
\label{eq:Two-pole:tildet_k}
\end{equation}
$\tilde{t}_{\mibs{k}}$ and $E_{{\rm cor}2}$, respectively. The above Green's function has two poles at $\omega=\omega^\pm_{\mibs{k}}$ with their residues $z^\pm_{\mibs{k}}$ where
\begin{eqnarray}
\omega^\pm_{\mibs{k}}&=&(\vepsz_{\mibs{k}}+\tilde{\veps}_{\mibs{k}}\pm\DeltaMH_{\mibs{k}})/2,
\label{eq:Two-pole:omega^pm_k}\\
z^\pm_{\mibs{k}}&=&(1\pm(\vepsz_{\mibs{k}}-\tilde{\veps}_{\mibs{k}})/\DeltaMH_{\mibs{k}})/2.
\label{eq:Two-pole:z^pm_k}
\end{eqnarray}
Here, we have introduced
\begin{equation}
\DeltaMH_{\mibs{k}}=\sqrt{U^2\ave{n}(2-\ave{n})+(\vepsz_{\mibs{k}}-\tilde{\veps}_{\mibs{k}})^2},
\label{eq:Two-pole:Delta_k}
\end{equation}
which generate the direct energy gap at the momentum $\mib{k}$ between the two poles lying in the upper and lower Hubbard bands. The momentum dependence diminishes as $U/t$ diverges. It is easily found that when the collective degrees of freedom are completely frozen, i.e., $\tilde{t}_{\mibs{k}}=0$, and the third term in \eq{eq:Two-pole:tildet_k} is neglected, then \eq{eq:2OP:G_2^0} recovers the Hubbard I solution.

Next, we consider the case where equal-time two-particle correlations with the momentum $\mib{Q}$ are significant. Then, through the growth of short-range correlations in \eq{eq:2OP:t^2_x}, $\tilde{t}_{\mibs{k}}$ behaves like $t_{\mibs{k}+\mibs{Q}}$ so that eq. (\ref{eq:2OP:G_2^0}) continuously connects two different phases that preserves and breaks the translational symmetry.
Actually, when the commensurate AF long-ranged order is stabilized at the ground state in the simple square-lattice Hubbard model, the nonzero ordered moment in this case creates new modes in single-particle excitation spectra. The new modes emerge due to the spontaneous breaking of the translational symmetry, namely the folding of the Brillouin zone. Therefore, when the higher-order dynamics is ignored, a discontinuous AF transition can occur, as it does in the Hartree-Fock approximation. However, this is not always valid. Upon continuous phase transition, this discontinuous feature should be removed by a gradual growth of the AF shadows. Any new mode leading to a continuous phase transition does not appear discontinuously at the phase transition, if and only if the condition given by $\tilde{t}_{\mibs{k}}=t_{\mibs{k}+\mibs{Q}}$ or equivalently
\begin{eqnarray}
\tilde{t}_{\mibs{x},\mibs{x}'}=\left\{\begin{array}{ll}
-t_{\mibs{x},\mibs{x}'}&\mbox{ for the different sublattices}
\nonumber\\
t_{\mibs{x},\mibs{x}'}&\mbox{ for the same sublattice}
\end{array}\right.
\label{eq:Two-pole:condition}
\end{eqnarray}
is satisfied at $T=0$ and $\tilde{t}_{\mibs{x},\mibs{x}'}$ changes as a continuous function of temperature. So far, few theories have reproduced such shadow-band effect even qualitatively within the one-loop approximations to $\bfSigma_1(\omega)$~\cite{TPSC,TPSC2}. The two-pole approximation can capture the shadows and continuously approach the AF phase from the paramagnetic phase. In this case, however, the two-pole approximation only gives the SDW bands similar to the Hartree-Fock results. Apart from the shadow features, the approximation does not introduce any additional feature beyond the Hartree-Fock approximation. In \S~\ref{sec:Results}, it will turn out that low-energy narrow bands playing a role of quasiparticles do emerge as significant additional features.

We note that as is clear from \eq{eq:2OP:t^2_x}, $\tilde{t}_{\mibs{k}}$ includes the superexchange term. This suggests that the momentum dependence of the self-energy part appears through the superexchange term. For further improvements of the theory, higher-order dynamics contained in $\bfSigma_2(\omega)$ has to be considered. The following arguments on the expectation value of the Hamiltonian illustrate this necessity.

In general, the expectation value of the Hamiltonian is given by
\begin{equation}
\ave{H}=T\!\sum_{\omega_n,\mibs{k},s}\!
\left[-t_{\mibs{k}}+\frac{U\ave{n}}{2}+\Sigma_{1\rme,s}(\omega_n,\mib{k})\right]
G_{\rme,s}(\omega_n,\mib{k}).
\label{eq:Two-pole:E}
\end{equation}
In the two-pole approximation without $\bfSigma_2(\omega)$, we obtain
\begin{eqnarray}
\ave{H}&=&U\sum_{\mibs{k},\pm}f(\omega^\pm_{\mibs{k}})
\left[z^\pm_{\mibs{k}}\left(-\frac{2t_{\mibs{k}}}{U}+\ave{n}\right)
\pm\frac{\ave{n}(2-\ave{n})}{2\DeltaMH_{\mibs{k}}/U}\right]
\nonumber\\
\label{eq:Two-pole:E_two-pole}
\end{eqnarray}
as a function of $\tilde{t}_{\mibs{k}}$. Here, we have introduced a fermionic Matsubara frequency $\omega_n=(2n+1)\pi T$ and the Fermi distribution function $f(\omega)=(e^{\omega/T}+1)^{-1}$. $\Sigma_{1\rme,s}(\omega,\mib{k})$ denotes the momentum-$\mib{k}$ spin-$s$ electron component of $\bfSigma_1(\omega)$. At $\ave{n}=1$, we obtain
\begin{equation}
\ave{H}=-\sum_{\mibs{k}}[t_{\mibs{k}}^2-(t_{\mibs{k}}+\tilde{t}_{\mibs{k}})^2/4]/U+O(t^3/U^2)
\label{eq:Two-pole:Eel1}
\end{equation}
when $U\gg T$, $t$. This clarifies that the superexchange term appears through \eq{eq:2OP:t^2_x}. However, there exist extra terms of order of $t^2/U$ in \eq{eq:Two-pole:Eel1} in addition to the superexchange term. Since the term of order of $t^2/U$ should be uniquely determined in terms of the strong-coupling expansion, the additional terms are to be cancelled with the terms from $\bfSigma_2(\omega)$.

\subsection{Higher-order dynamics}
\label{subsec:Higher}

Here, we consider the second-order self-energy part $\bfSigma_2(\omega)$ defined by \eq{eq:2OP:Sigma_2}. In principle, one can continue the projection procedure to derive the Dyson equation for $\bfSigma_2(\omega)$. The results of the third-order projection will be published elsewhere. In this paper, we employ an approximation to obtain $\bfSigma_2(\omega)$. In the symmetry-unbroken phase, its off-diagonal components vanish and it does not depend on the spin index. Then, the spin indices can be dropped out in the following notations and we have only to discuss the spin-independent electron component expressed as
\begin{equation}
\Sigma_{2\rme}(\omega,\mib{k})=\frac{4}{\ave{n}(2-\ave{n})}\chi_{\delho\delho\Psi,(\delho\delho\Psi)^\dagger}^{{\rm irr} 11}(\omega,\mib{k}).
\end{equation}

Several approximate methods are available to calculate $\Sigma_{2\rme}(\omega,\mib{k})$: the perturbation theory, one-loop approximation with two-particle susceptibilities obtained by other methods such as the self-consistent RPA~\cite{SCRPA}, the two-particle self-consistent method~\cite{TPSC,TPSC2} or the extension of the Roth's method~\cite{Roth69} by Mancini {\it et al.}~\cite{ManciniMarraMatsumoto95} which exploits equations of motion of the response functions of $\Psi_{\mibs{x}}$ with respect to $n_{\mibs{x}s}\Psi_{\mibs{x}'}$. Here, at least, the following two points are required in calculating $\Sigma_{2\rme}(\omega,\mib{k})$ along the present approach: One is that they should give a correct estimate of the moment of $\Sigma_{2\rme}(\omega,\mib{k})$, thus the fourth-order moment of the Green's function. The other is that the short-ranged correlations which are neglected in previous equation-of-motion approaches~\cite{Hubbard1, Hubbard2, Hubbard3, Roth69, BeenenEdwards95, MatsumotoMancini97} should be taken into consideration.

The perturbation theory gives an approximately good estimate of the moment of $\Sigma_{2\rme}(\omega,\mib{k})$. However, it ignores the short-ranged correlations. Since one of our main aims is to take account of both the short-ranged correlations and the Mott-Hubbard picture due to the strong local repulsion, the perturbation theory is not sufficient. A possible prescription to take account of the both is to decouple the three-body correlation function $\Sigma_{2\rme}(\omega,\mib{k})$ into the convolution of the single-particle Green's function and two-particle dynamical susceptibilities.

We note that $\bfSigma_2(\omega)$ has been evaluated within a two-site level
by means of a different decoupling scheme~\cite{MatsumotoMancini97}. However, short-ranged correlations are ignored there. If the AF short-ranged spin correlations are taken into account, the AF shadow bands of the Hubbard bands should appear, as we will obtain below.

First, we give our decoupling scheme. In the symmetry-unbroken phase, the spin-independent electron component of the second-order electronic self-energy part is evaluated by means of the following decoupling approximation as
\begin{subequations}
\begin{eqnarray}
\Sigma_{2\rme}(\omega_n,\mib{k})&=&\frac{1}{\ave{n}(2-\ave{n})}\sum_{\calO=n,\mibs{S},\Delta}\hspace*{-4pt}\Sigma_{2\rme \calO}(\omega_n,\mib{k}),
\label{eq:higher:Sigma_2e}\\
\Sigma_{2\rme \calO}(\omega_n,\mib{k})&=&\frac{T}{N}\sum_{m,\mibs{q}}\Lambda_{\calO}(\mib{k};\mib{q})^2G_\rme(\omega_n-\Omega_m,\mib{k}-\mib{q})
\nonumber\\
&&{}\times\chi_{\calO,\calO^\dagger}(\Omega_m,\mib{q})
\nonumber\\
&&{}\mbox{ for $\calO=n$ and $\mib{S}$},
\label{eq:higher:Sigma_2enS_decoupling}\\
\Sigma_{2\rme \Delta}(\omega_n,\mib{k})&=&-\frac{T}{N}\sum_{m,\mibs{q}}\Lambda_\Delta(\mib{k};\mib{q})^2G_\rme(\Omega_m-\omega_n,\mib{q}-\mib{k})
\nonumber\\
&&{}\times\chi_{\Delta,\Delta^\dagger}(\Omega_m,\mib{q}),
\label{eq:higher:Sigma_2eDelta_decoupling}
\end{eqnarray}
\label{eq:higher:Sigma_2e_decoupling}
\end{subequations}
where $\Omega_m=2\pi mT$ is a bosonic Matsubara frequency, $G_\rme$ is the spin-independent electron part of the Green's function and $\chi_{n,n}$, $\chi_{S^i,S^i}$ and $\chi_{\Delta,\Delta^\dagger}$ are the charge, spin and local-pair susceptibilities, respectively. The following vertices have also been introduced;
\begin{subequations}
\begin{eqnarray}
\Lambda_n(\mib{k};\mib{q})&=&-T_{\mibs{k};\mibs{q}},
\label{eq:higher:Lambda_n_kq}\\
\Lambda_{S^z}(\mib{k};\mib{q})&=&-2T_{\mibs{k};\mibs{q}},
\label{eq:higher:Lambda_S^z_kq}\\
\Lambda_{S^{x,y}}(\mib{k};\mib{q})&=&-2t_{\mibs{k}-\mibs{q}},
\label{eq:higher:Lambda_S^x,y_kq}\\
\Lambda_{\Delta}(\mib{k};\mib{q})&=&-2t_{\mibs{k}-\mibs{q}},
\label{eq:higher:Lamdba_Delta_kq}
\end{eqnarray}
\label{eq:higher:Lambda_kq}
\end{subequations}
with
\begin{equation}
T_{\mibs{k};\mibs{q}}=t_{\mibs{k}-\mibs{q}}-\tilde{t}_{\mibs{k}}
-\frac{2(1-\ave{n})}{\ave{n}(2-\ave{n})}\Ekin.
\label{eq:higher:T_k;q}
\end{equation}
Here, $\tilde{t}_{\mibs{k}}$ is the $(1,1)$ component of $t^{(22)}_{\mibs{k}}$. These equations are straightforwardly derived from \eqd{eq:2OP:Sigma_2}{eq:2OP:delho-delho-Psi_x}.

\begin{figure}[htb]
\begin{center}
\epsfxsize=7.cm
$$\epsffile{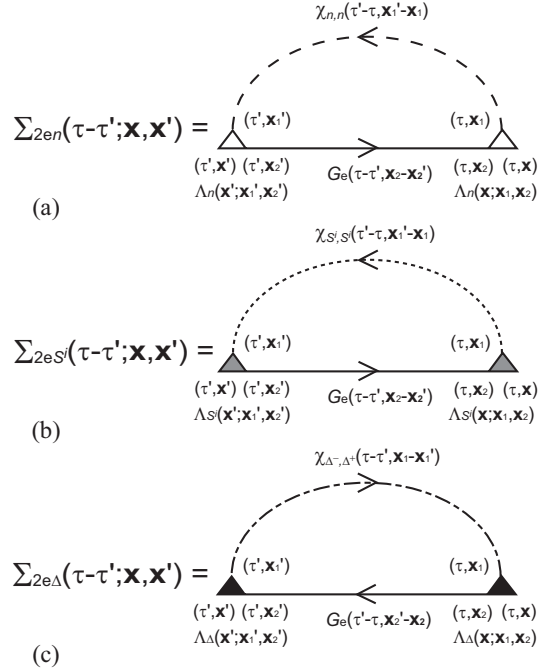}$$
\end{center}
\caption{Diagrammatic representations of the second-order self-energy part in our decoupling scheme. It can be decomposed into three parts shown in (a), (b) and (c) divided by $\ave{n}(2-\ave{n})$, as in \eq{eq:higher:Sigma_2e}. In (b), the superscript $i$ takes $x$, $y$ and $z$. Details are explained in the text.}
\label{fig:diagram}
\end{figure}

Figure~\ref{fig:diagram} shows the Feynman diagrams of the present decoupling approximation to the above five ($n$, $S^x$, $S^y$, $S^z$ and $\Delta$) components of $\Sigma_{2\rme}$ in the imaginary time and the real space. Here, the vertices have the following representations in the real space;
\begin{subequations}
\begin{eqnarray}
\Lambda_n(\mib{x};\mib{x}_1,\mib{x}_2)&=&\delta_{\mibs{x}_1,\mibs{x}_2}\left[\delta_{\mibs{x},\mibs{x}_1}\frac{2(1-\ave{n})\Ekin}{\ave{n}(2-\ave{n})}+t^{(22)}_{\mibs{x},\mibs{x}_1}\right]
\nonumber\\
&&{}-\delta_{\mibs{x},\mibs{x}_1}t_{\mibs{x}_1,\mibs{x}_2},
\label{eq:higher:Lambda_n_x}\\
\Lambda_{S^z}(\mib{x};\mib{x}_1,\mib{x}_2)&=&2\delta_{\mibs{x}_1,\mibs{x}_2}\left[\delta_{\mibs{x},\mibs{x}_1}\frac{2(1-\ave{n})\Ekin}{\ave{n}(2-\ave{n})}+t^{(22)}_{\mibs{x},\mibs{x}_1}\right]
\nonumber\\
&&{}-2\delta_{\mibs{x},\mibs{x}_1}t_{\mibs{x}_1,\mibs{x}_2},
\label{eq:higher:Lambda_S^z_x}\\
\Lambda_{S^{x,y}}(\mib{x};\mib{x}_1,\mib{x}_2)&=&-2\delta_{\mibs{x},\mibs{x}_1}t_{\mibs{x}_1,\mibs{x}_2},
\label{eq:higher:Lambda_S^x,y_x}\\
\Lambda_{\Delta}(\mib{x};\mib{x}_1,\mib{x}_2)&=&-2\delta_{\mibs{x},\mibs{x}_1}t_{\mibs{x}_1,\mibs{x}_2}.
\label{eq:higher:Lambda_Delta_x}
\end{eqnarray}
\end{subequations}

From \eq{eq:higher:Sigma_2e_decoupling}, it turns out that the second-order self-energy part takes the similar expression to the usual self-energy part, if one takes decoupling approximations to evaluate them. The difference between the self-energy part and the second-order self-energy part appears in their coupling constants. The coupling constant for a decoupled form of the self-energy part is $U$, while that for the second-order self-energy part is $\Lambda_{\calO}(\mib{k};\mib{q})$, of order of $t$.

The momentum dependences of $\Lambda_{\calO}(\mib{k};\mib{q})$ determine detailed contributions of two-particle correlations to $\Sigma_{2\rme}(\omega,\mib{k})$. To discuss these contributions, for simplicity, we consider the Hubbard model with the electron transfers restricted to the nearest neighbors. When AF spin correlations are so strong that the equal-time nearest-neighbor spin correlation has a large negative value, it happens that $\tilde{t}_{\mibs{k}}$ approximates $t_{\mibs{k}+\mibs{Q}}=-t_{\mibs{k}}$ with the staggered wave vector $\mib{Q}$. In such case, the first and second terms in \eq{eq:higher:T_k;q} nearly cancel each other for momentum $\mib{q}\sim\mib{Q}$. Particularly, at half filling, \eq{eq:higher:T_k;q} suggests that $T_{\mibs{k};\mibs{q}}$ nearly vanishes. Therefore, when AF spin correlations develop at half filling, the main contributions to $\Sigma_{2\rme}(\omega,\mib{k})$ are those from $\Sigma_{2\rme S^{x,y}}(\omega,\mib{k})$ and $\Sigma_{2\rme \Delta}(\omega,\mib{k})$. In the strong-coupling case, the double occupancy is strongly reduced. This suppresses the contribution from $\Sigma_{2\rme \Delta}(\omega,\mib{k})$. Finally, \eqt{eq:higher:Sigma_2enS_decoupling}{eq:higher:Lambda_kq}{eq:higher:T_k;q} suggest that for the momenta lying on the magnetic Brillouin zone boundary where $t_{\mibs{k}}$ and thus $t_{\mibs{k}-\mibs{Q}}$ vanish, all the contributions from the two-particle fluctuations with their momenta $\mib{q}=0$ and $\mib{Q}$ to $\Sigma_{2\rme}(\omega,\mib{k})$ disappear at half filling. Even away from half filling, only the charge and the longitudinal fluctuations dominantly contribute to $\Sigma_{2\rme}(\omega,\mib{k})$. On the other hand, away from the magnetic Brillouin zone boundary, $\Sigma_{2\rme}(\omega,\mib{k})$ is dominated by the contributions from the transverse AF spin fluctuations. Therefore, modifications of dispersions from the Hubbard bands stem from the spin excitation properties under strong AF correlations. In \S~\ref{sec:Results}, we actually see additional low-energy bands develop near half filling at low temperatures. We will discuss the issue in \S~\ref{sec:Results}.

Next, we must obtain the two-particle susceptibilities in an appropriate method. As we have already emphasized, effects of the short-ranged correlations must be incorporated into the present theory for a better description of the strongly correlated electrons. For this purpose, one should include the Stoner enhancement factor resulting from a summation of ladder or bubble diagrams and also obtained from the Gaussian approximation in the functional integral technique~\cite{Hertz76}. Although one can easily calculate the dynamical spin susceptibilities using the method developed by Mancini {\it et al.}~\cite{ManciniMarraMatsumoto95}, such enhancement factor is not taken into account in the approximation. Another requirement already noted above is that in the framework of the present OPM, the higher-order dynamics accounted by $\bfSigma_2(\omega)$ must reproduce its correct moment. The self-consistent RPA~\cite{SCRPA,SCRPA2} is based on the equation of motion of two-particle operators. Therefore, it seems that the approximation may be appropriate. The self-consistent RPA, however, does not satisfy the local moment sum rules explicitly given below. Instead of these approximate methods, the two-particle self-consistent (TPSC) method determines the irreducible vertices in the RPA-like forms of the two-particle susceptibilities so that the local-moment sum rules are fulfilled for the two-particle properties, not only charge and spin correlations~\cite{TPSC} but also local $s$-wave pairing correlations~\cite{TPSC2}. In this paper, we employ the TPSC susceptibilities as in the previous paper. If one employs the perturbation theory or the self-consistent RPA, they also reproduce a Mott insulator. However, they give weaker AF spin correlations than the TPSC method. 

Within the two-particle self-consistent approximation~\cite{TPSC,TPSC2}, the susceptibilities are calculated from
\begin{subequations}
\begin{eqnarray}
\chi_{n,n}(\Omega_m,\mib{k})&=&\frac{2\chi_{\rm ph}(\Omega_m,\mib{k})}{1+\Gamma_\rmc\chi_{\rm ph}(\Omega_m,\mib{k})},
\label{TPSC:chin}\\
\chi_{S^i,S^i}(\Omega_m,\mib{k})&=&\frac{2\chi_{\rm ph}(\Omega_m,\mib{k})}{1-\Gamma_s\chi_{\rm ph}(\Omega_m,\mib{k})},
\label{TPSC:chis}\\
\chi_{\Delta,\Delta^\dagger}(\Omega_m,\mib{k})&=&\frac{\chi_{\rm pp}(\Omega_m,\mib{k})}{1+\Gamma_\romap\chi_{\rm pp}(\Omega_m,\mib{k})},
\label{TPSC:chip}
\end{eqnarray}
\end{subequations}
with an approximation $\Gamma_\rms=4U\ave{n_{\mibs{x}\up}n_{\mibs{x}\down}}/\ave{n}^2$ and the self-consistency conditions for $\Gamma_\rms$, $\Gamma_\rmc$ and $\Gamma_\romap$ given by the local moment sum rules;
\begin{subequations}
\begin{eqnarray}
&&\hspace*{-20pt}\frac{T}{N}\sum_{m,\mibs{k}}\chi_{n,n}(\Omega_m,\mib{k})
=\ave{n}+2\ave{n_{\mibs{x}\up}n_{\mibs{x}\down}}-\ave{n}^2,
\label{eq:higher:sum_chic}\\
&&\hspace*{-20pt}\frac{T}{N}\sum_{m,\mibs{k}}\chi_{S^i,S^i}(\Omega_m,\mib{k})
=\ave{n}-2\ave{n_{\mibs{x}\up}n_{\mibs{x}\down}},
\label{eq:higher:sum_chis}\\
&&\hspace*{-20pt}\frac{T}{N}\sum_{m,\mibs{k}}\chi_{\Delta,\Delta^\dagger}(\Omega_m,\mib{k})
=\ave{n_{\mibs{x}\up}n_{\mibs{x}\down}}.
\label{eq:higher:sum_chip}
\end{eqnarray}
\end{subequations}
$\chi_{\rm ph}$ and $\chi_{\rm pp}$ are the particle-hole and particle-particle susceptibilities calculated from the bare Green's functions.

\section{Numerical results}
\label{sec:Results}

In the previous letter~\cite{OPM_jpsjletter01}, it was shown that the filling-control MIT is well described by the present method. In this section, we calculate the single-particle density of states, spectral functions, dispersions, momentum distributions and the chemical potential shift. Their dependences on $T$, $\ave{n}$ and/or $U$ are also discussed in detail.

The following results have been obtained with 2048 Matsubara frequencies in the $32\times32$ lattice for two-dimensional systems and in the $128$ sites chain for one-dimensional systems. For analytic continuations, we have used the Pade approximation~\cite{PadeAPPROX}.

\subsection{Mott-insulating properties at half filling in two dimensions}
\label{subsec:half-filling}

\begin{figure}[htb]
\begin{center}
\epsfxsize=8.cm
$$\epsffile{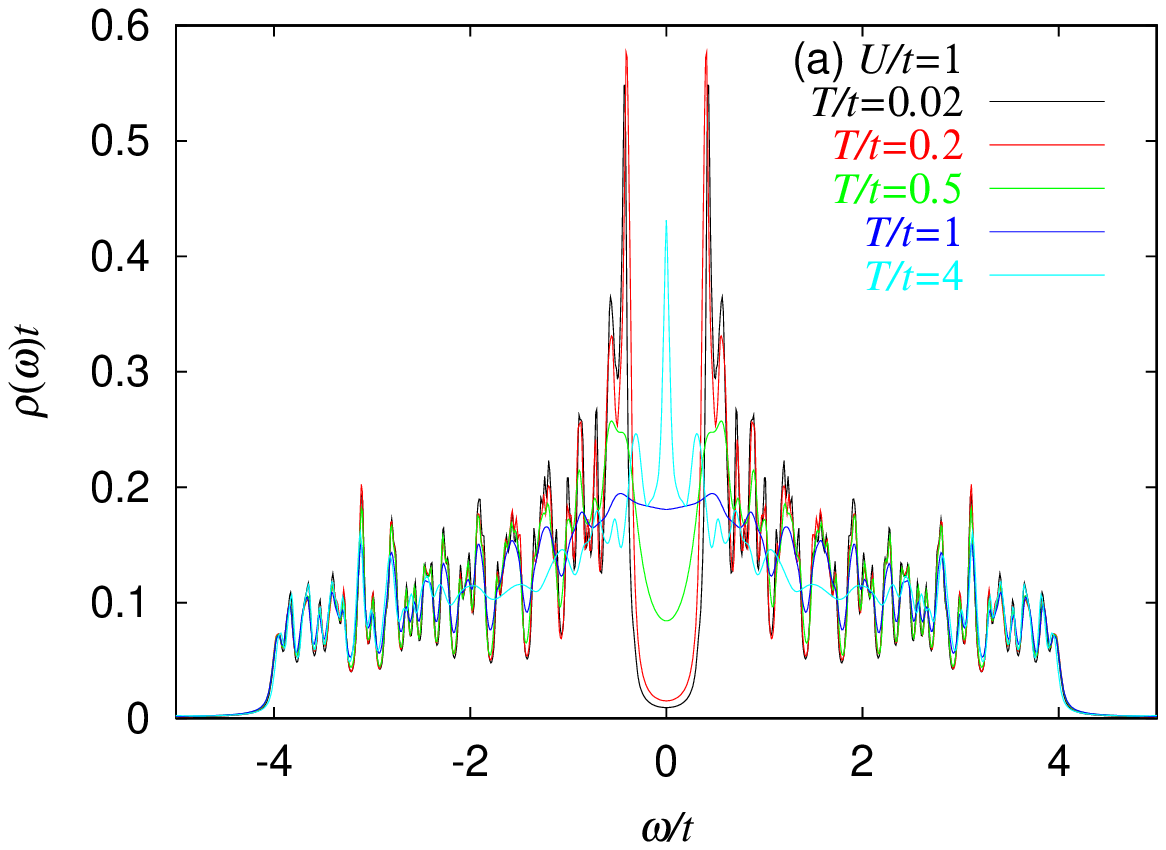}$$
\end{center}
\begin{center}
\epsfxsize=8.cm
$$\epsffile{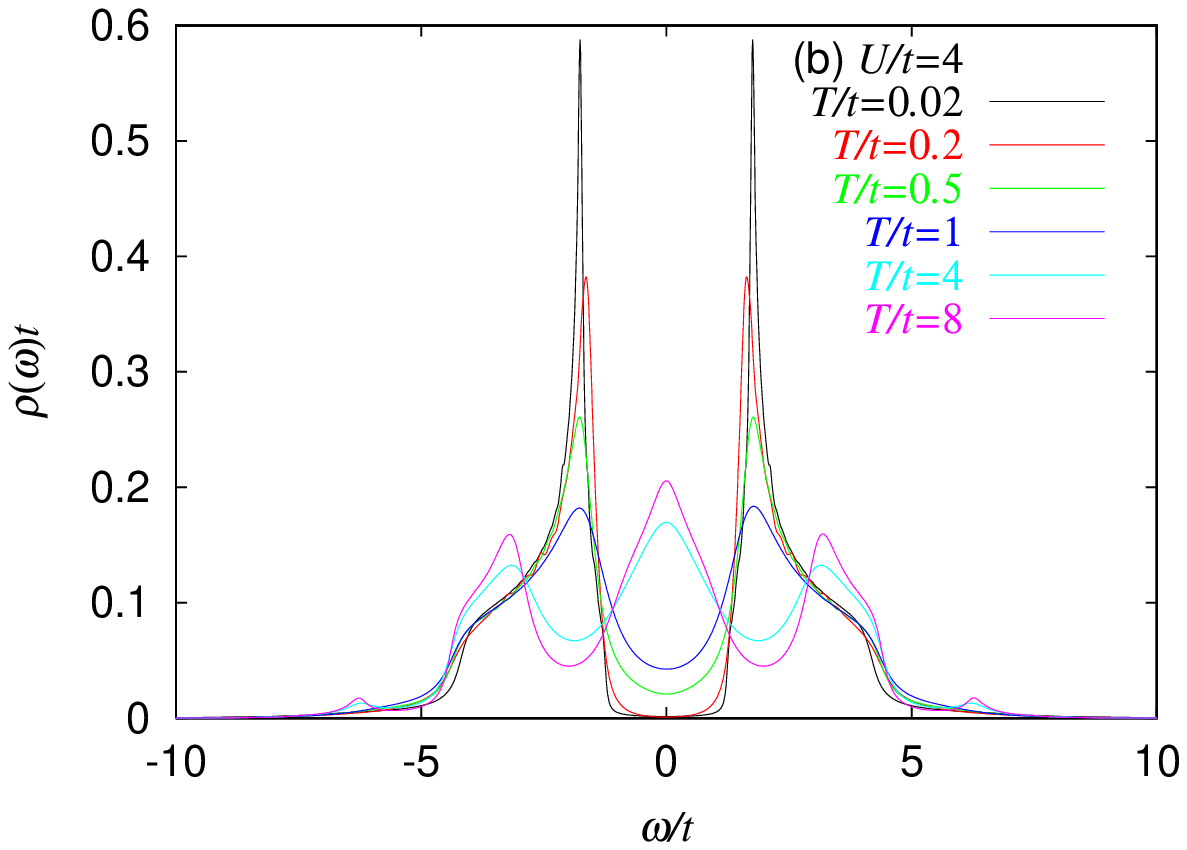}$$
\end{center}
\begin{center}
\epsfxsize=8.cm
$$\epsffile{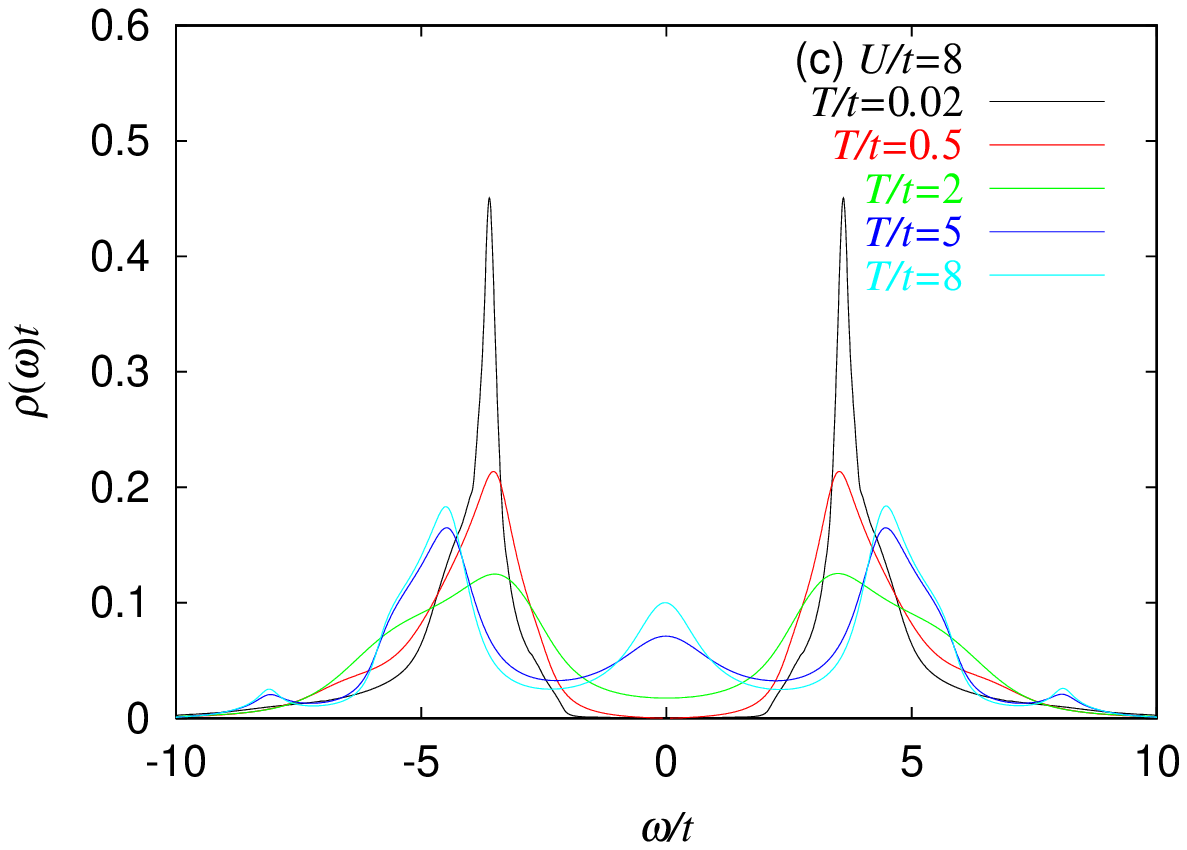}$$
\end{center}
\caption{The temperature dependence of the local density of states $\rho(\omega)$ for $U/t=1$ (a), $4$ (b) and $8$ (c) at half filling $\ave{n}=1$. For (a), $T/t=0.02$, $0.2$, $0.5$, $1$ and $4$. For (b), $T/t=0.02$, $0.2$, $0.5$, $1$, $4$ and $8$. For (c), $T/t=0.02$, $0.5$, $2$, $5$ and $8$. At lower temperatures, less spectral weights remain around $\omega=0$. The Mott insulating gap grows at low temperatures $T=0.02t$, while spectral weights gradually fill in the gap at high temperatures. It is remarkable that the Mott-Hubbard pseudogap remains even at $T=J\equiv 4t^2/U$ above which AFM spin correlations are not enhanced. On the other hand, at $T>U/2$, a low-energy incoherent band develops. Detailed discussions are given in the text.}
\label{fig:DOSel1t0U1,4,8}
\end{figure}
\begin{figure}[htb]
\begin{center}
\epsfxsize=8.cm
$$\epsffile{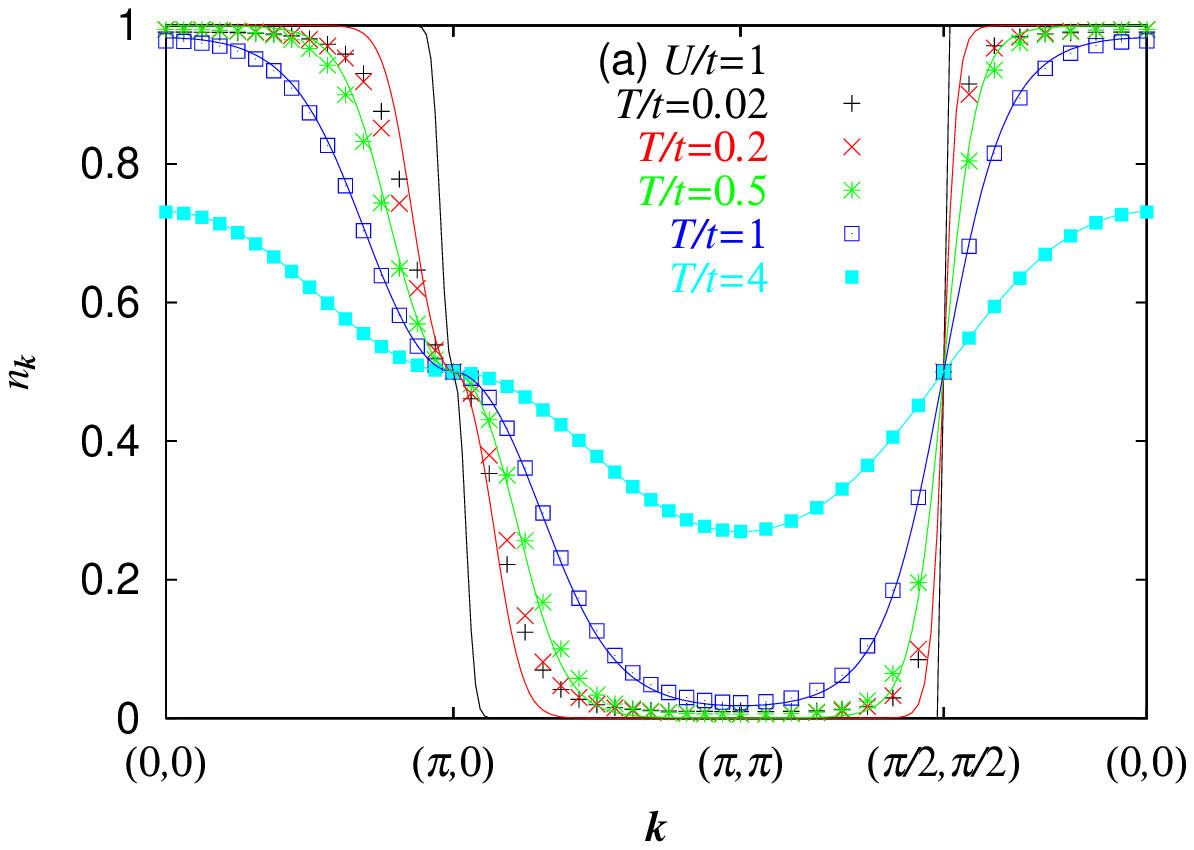}$$
\end{center}
\begin{center}
\epsfxsize=8.cm
$$\epsffile{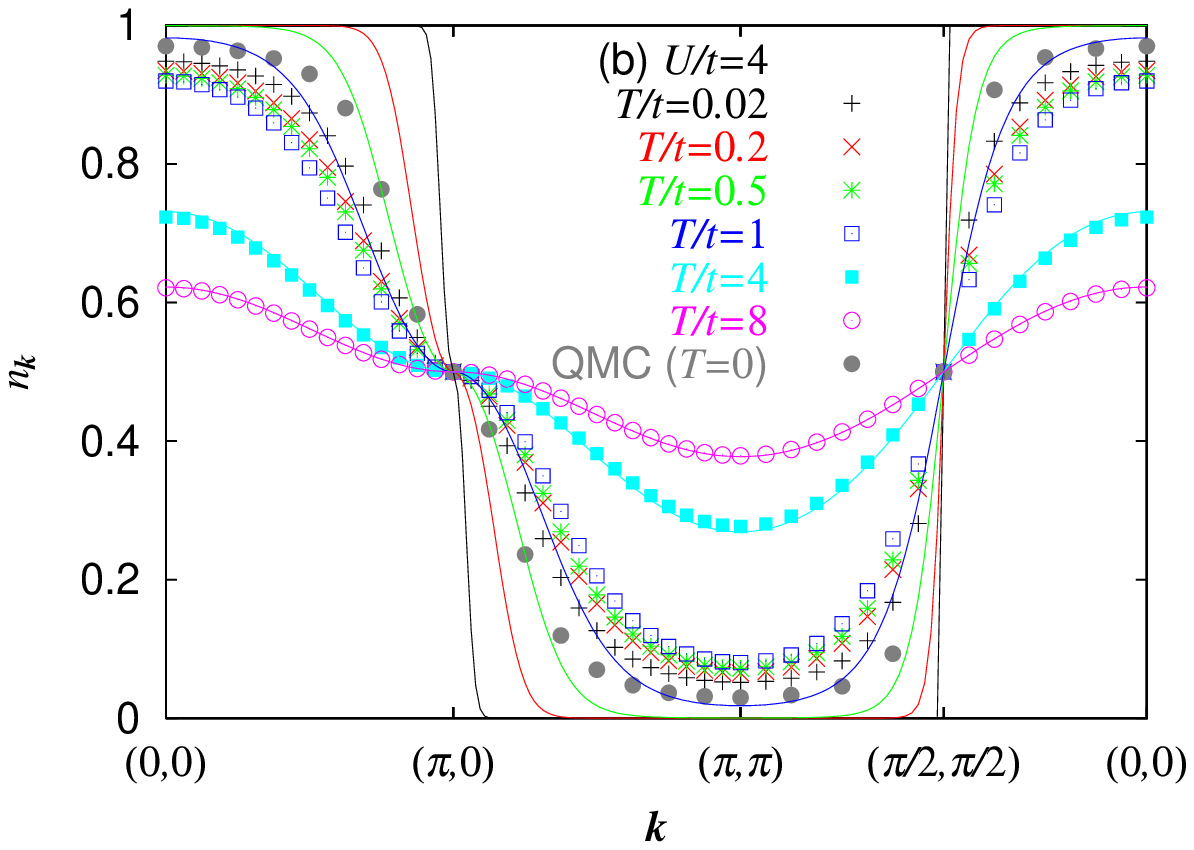}$$
\end{center}
\begin{center}
\epsfxsize=8.cm
$$\epsffile{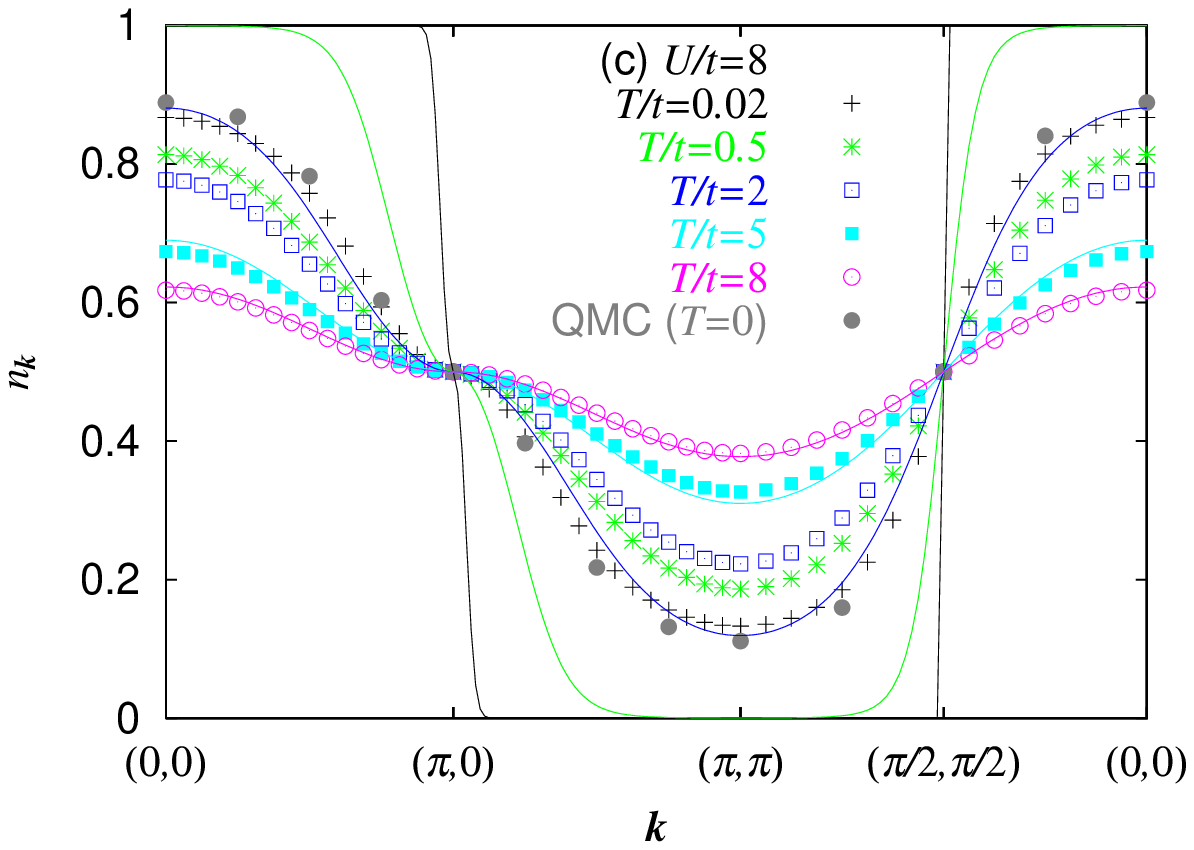}$$
\end{center}
\caption{Momentum distributions for $U/t=1$ (a), $4$ (b) and $U/t=8$ (c) at half filling $\ave{n}=1$ compared with the QMC results from ref.~\cite{AssaadImada99} (b) and ref.~\cite{Assaad} (c). For comparison, momentum distributions for $U=0$ and $\ave{n}=1$ at the same set of temperatures are plotted with solid lines in the corresponding colors in each panel. The results support that the system is a non-Fermi liquid in all these cases. Although Fig.~\ref{fig:DOSel1t0U1,4,8} shows that spectral weights develop around $\omega=0$ at $T>U/2$, these low-energy excitations are too strongly damped to produce a jump in $n_\mibs{k}$.}
\label{fig:momdisel1t0U1,4,8}
\end{figure}
\begin{figure}[htb]
\begin{center}
\epsfxsize=8.0cm
$$\epsffile{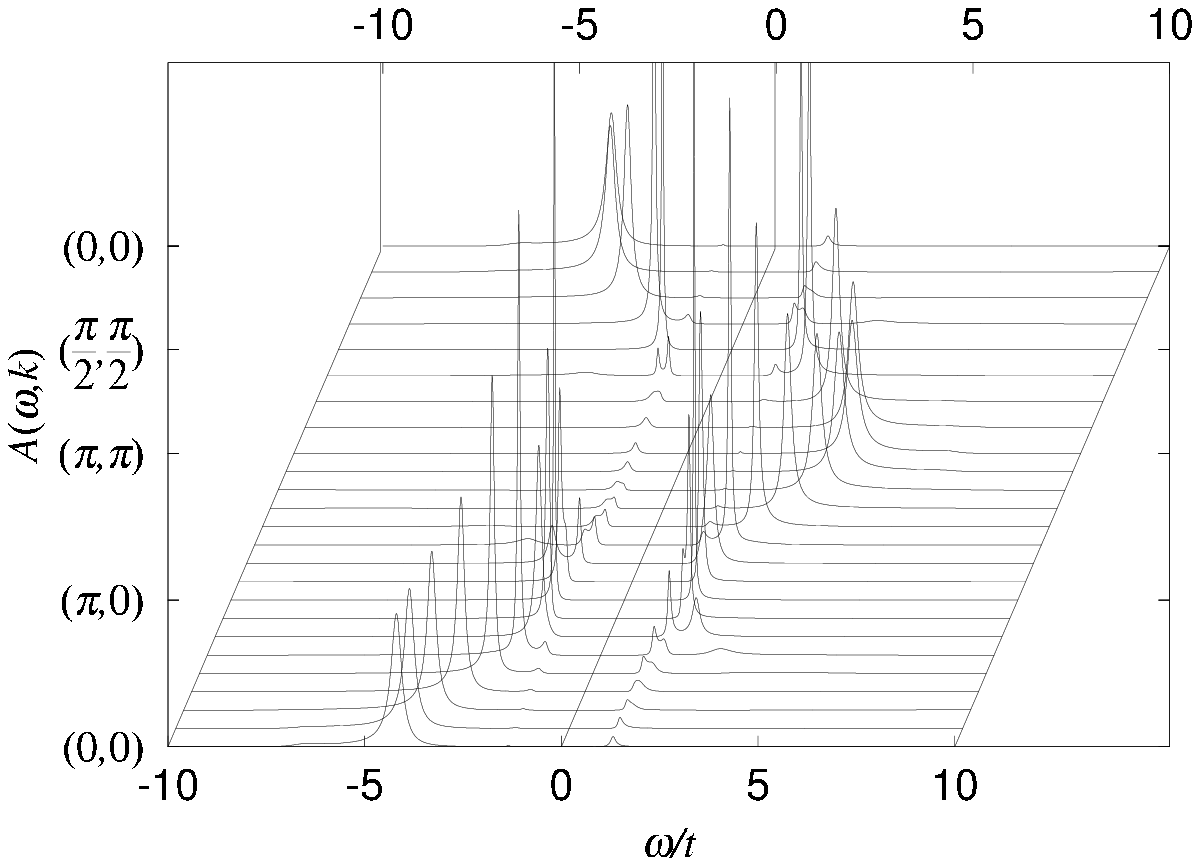}$$
\end{center}
\vspace*{-32pt}(a)
\begin{center}
\epsfxsize=8.0cm
$$\epsffile{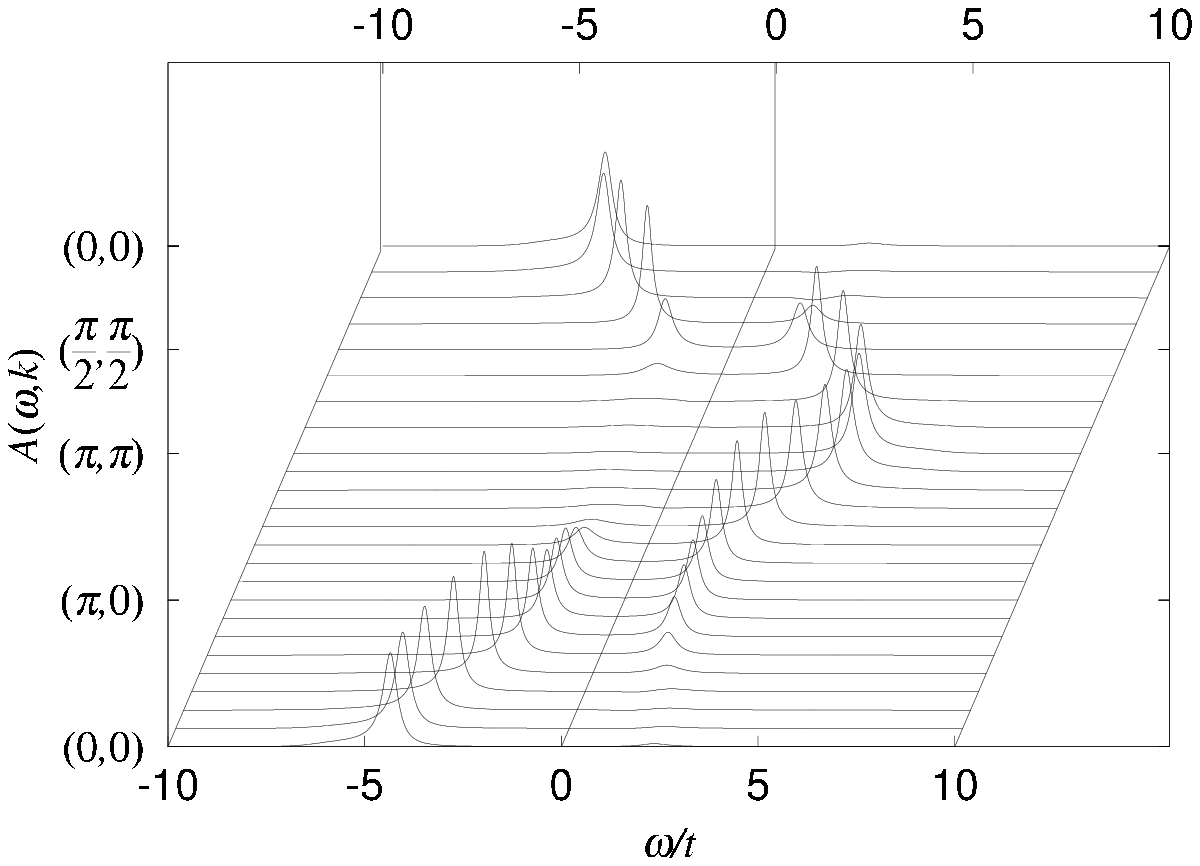}$$
\end{center}
\vspace*{-32pt}(b)
\begin{center}
\epsfxsize=8.0cm
$$\epsffile{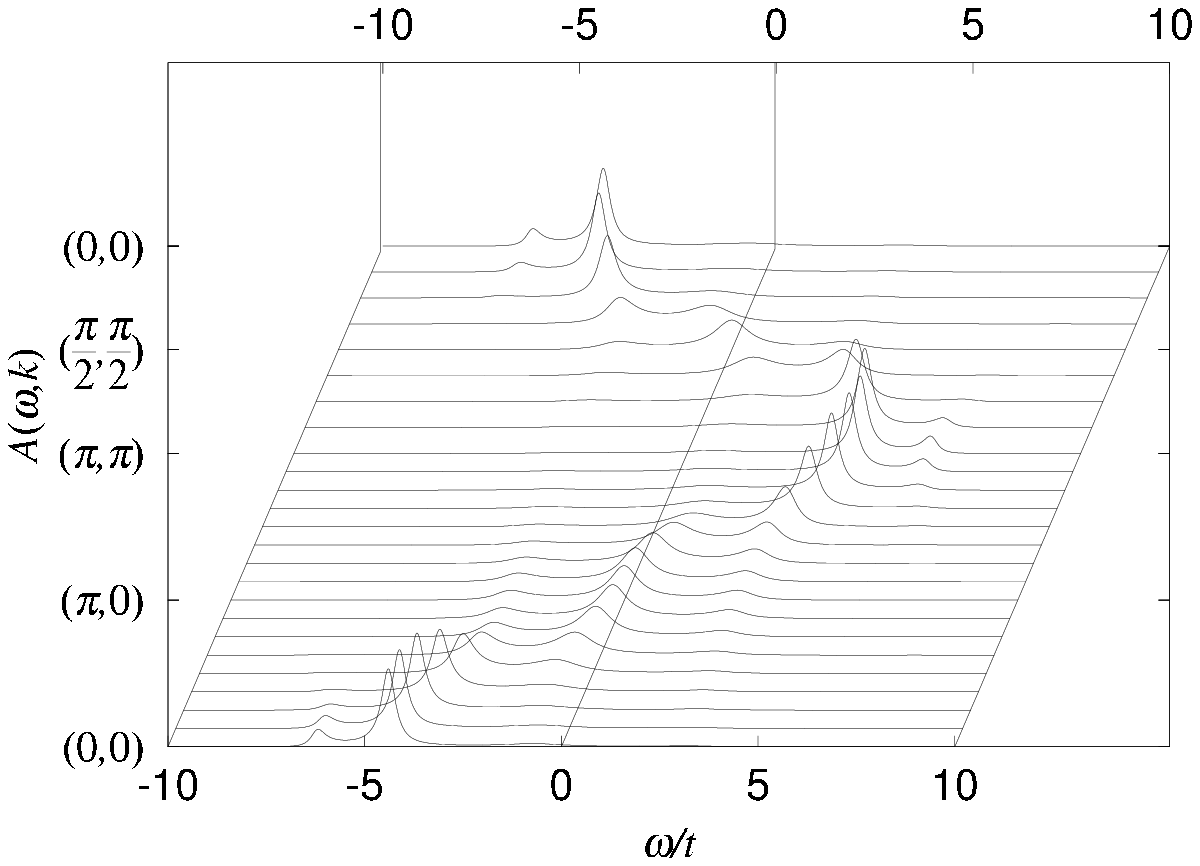}$$
\end{center}
\vspace*{-32pt}(c)
\caption{Single-particle spectral functions $A(\omega,\mib{k})=-{\rm Im}\ G(\omega,\mib{k})/\pi$ along $(0,0)$--$(\pi,0)$--$(\pi,\pi)$--$(0,0)$ for $U/t=4$ at half filling $\ave{n}=1$. The temperatures are $T/t=0.02$, $0.5$ and $4$ in (a), (b) and (c), respectively.}
\label{fig:Akwel1t0U4_T.02_.5_4}
\end{figure}
\begin{figure}[htb]
\begin{center}
\epsfxsize=8.0cm
$$\epsffile{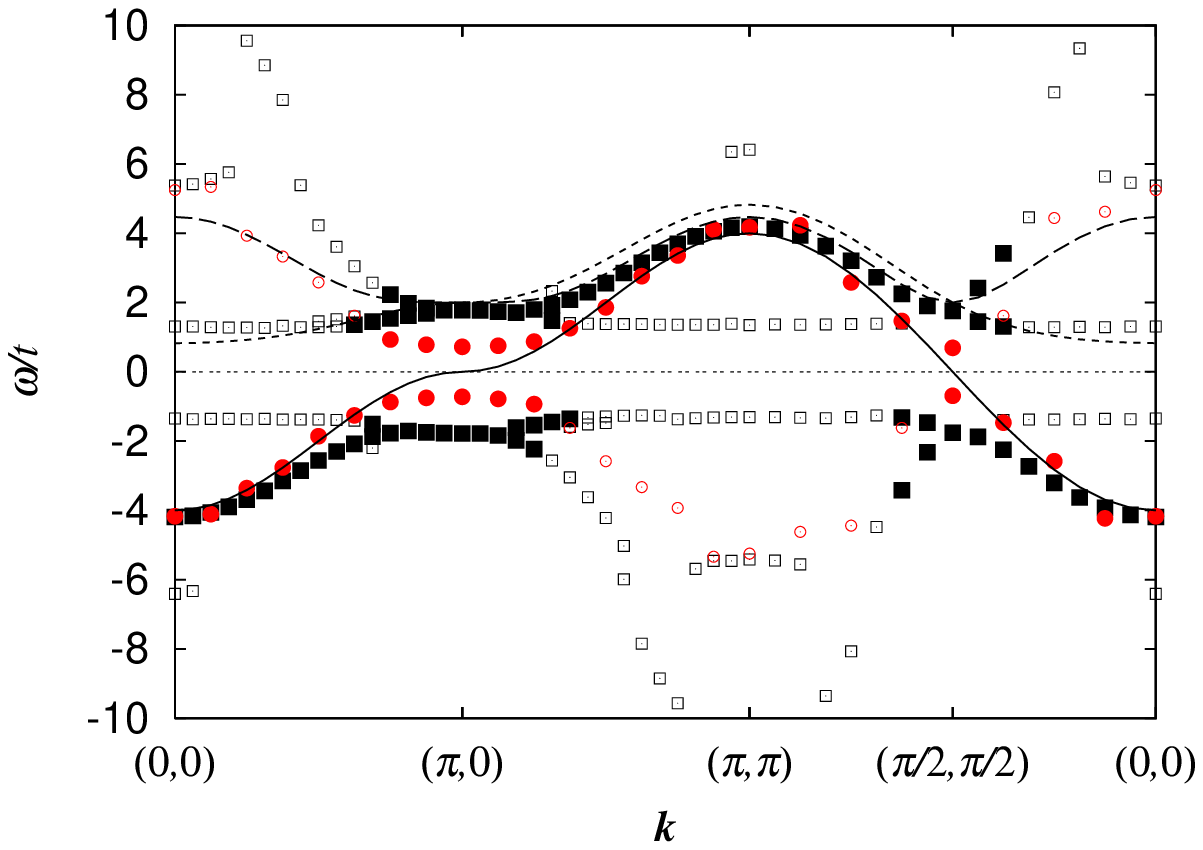}$$
\end{center}
\vspace*{-32pt}(a)
\begin{center}
\epsfxsize=8.0cm
$$\epsffile{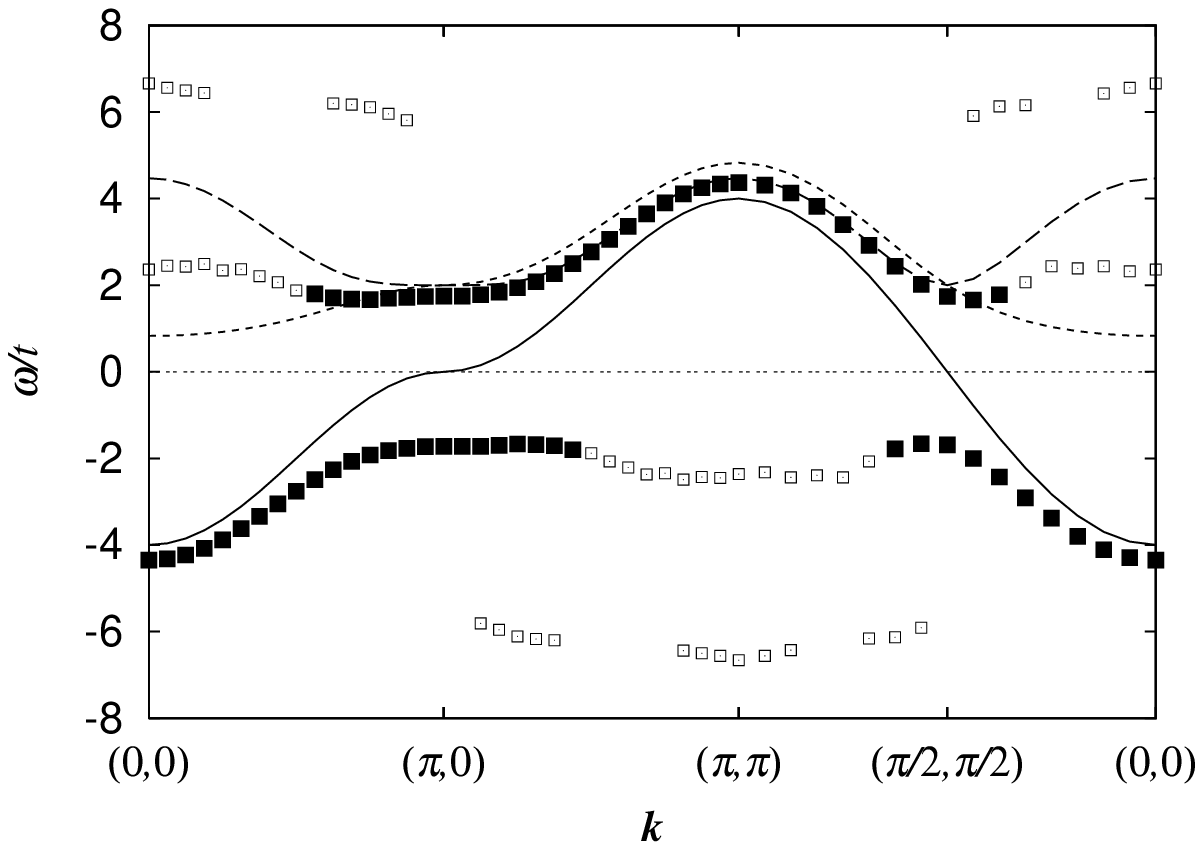}$$
\end{center}
\vspace*{-32pt}(b)
\begin{center}
\epsfxsize=8.0cm
$$\epsffile{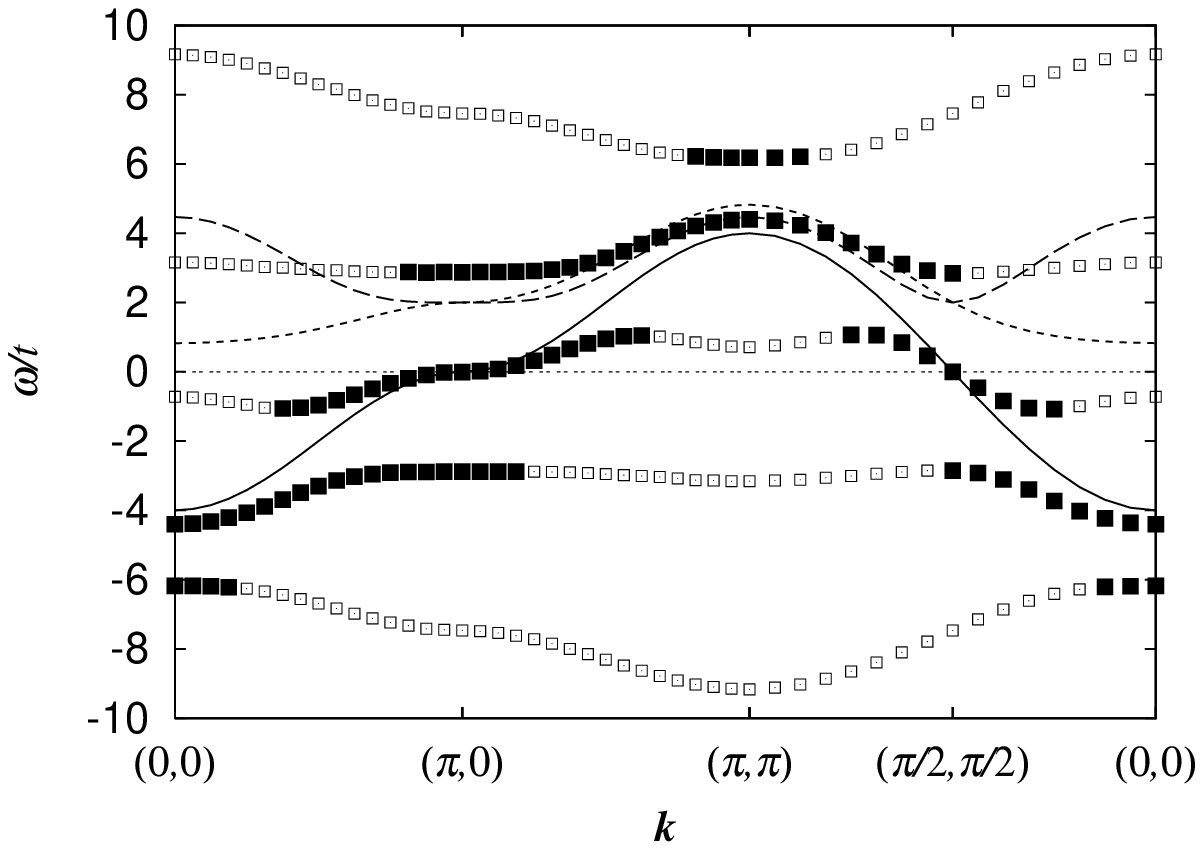}$$
\end{center}
\vspace*{-32pt}(c)
\caption{Single-particle dispersions along $(0,0)$--$(\pi,0)$--$(\pi,\pi)$--$(0,0)$ for $U/t=4$ at $\ave{n}=1$. The temperature is $T/t=0.02$, $0.5$ and $4$ in (a), (b) and (c), respectively. The squares denote the present results and the red circles denote QMC results~\cite{AssaadImada99}. The open and the filled symbols represent the momenta where the peak intensities divided by the largest intensity are less and larger than $0.1$, respectively. Main features of our results marked by the filled symbols are similar to those of QMC results~\cite{AssaadImada99}, except that the Mott-insulating gap is overestimated around the $(\pi,0)$ and $(0,\pi)$ momenta in the present results. The solid lines denote the free-electron dispersions. The dotted and the dashed lines represent the dispersion of the upper Hubbard band within the Hubbard I approximation and that obtained by the two-pole approximations with $\tilde{t}_{\mibs{k}}=t_{\mibs{k}+\mibs{Q}}$.}
\label{fig:disel1t0U4_T.02_.5_4}
\end{figure}

First, we give results at half filling. Results on the local density of states $\rho(\omega)\equiv-(\pi N)^{-1}\sum_{\mibs{k}}{\rm Im} G_{\rme}(\omega, \mib{k})$ for $U/t=1$, $4$ and $8$ are shown in Fig.~\ref{fig:DOSel1t0U1,4,8} (a), (b) and (c), respectively, at several temperatures. At temperatures low enough, a Mott-insulating gap grows. This indicates that the ground states are Mott insulators. At a much smaller coupling $U/t=0.2$, we have also obtained a Mott gap. Then, we expect that the present method reproduces a Mott insulator at any finite value of $U$. This Mott-insulating feature is accompanied by sharp peaks in $\rho(\omega)$ at the low-energy edges of the Hubbard bands in contrast with the Hubbard approximations~\cite{Hubbard1,Hubbard3}. Excitation spectra around $(\pi,0)$ and $(0,\pi)$ mainly contribute to these sharp peaks. These sharp peaks have also been found in QMC calculations~\cite{BulutScalapinoWhite94}. As the temperature increases, the spectral weights fill in the gap and the intensities of the peaks diminish. At high temperatures $T>U/2$, a low-energy incoherent band appears inside the Mott-Hubbard pseudogap separating the Hubbard bands. This is particularly prominent in the strong-coupling regime. The whole band structure becomes similar to the free-electron tight-binding band, particularly in the weak-coupling regime. The excitations in this band are incoherent and can not be described by well-defined quasiparticles. This is confirmed by the smooth momentum distributions $n_{\mibs{k}}$ in the momentum space, as shown in Fig.~\ref{fig:momdisel1t0U1,4,8} (c).

In the Fermi liquids, in general, momentum distributions have jumps along the Fermi surface. On the other hand, in the present case, i.e., the repulsive Hubbard model on the square lattice at half filling, the momentum distribution exhibits neither a jump nor a singularity. It is a smooth function of the momentum, as shown in Fig.~\ref{fig:momdisel1t0U1,4,8}. This means that the effective Fermi temperature vanishes in this case. This is naturally expected at low temperatures, since the system is a Mott insulator at $T=0$. At high temperatures, low-energy excitations are recovered around $\omega=0$. In spite of such metallic features, the momentum dependence of the momentum distributions monotonically decreases. This means that the system approaches the classical limit where $n_{\mibs{k}}$ is independent of $\mib{k}$. Similar situations with a suppressed Fermi temperature also occur in the doped cases, as discussed in \S~\ref{subsec:Doped}. Figure~\ref{fig:momdisel1t0U1,4,8} also shows that the momentum dependence of $n_{\mibs{k}}$ decreases with increasing $U$. The deviations of $n_{\mibs{k}}$ from the line for $U=0$ become smaller for higher temperatures. This is naturally expected, since these states at $T>0$ are adiabatically connected with the free-electron system.

To clarify an accuracy of the present method, the $T=0$ QMC results for $U/t=4$ obtained by Assaad and Imada~\cite{AssaadImada99} are also shown for comparison in Fig.~\ref{fig:momdisel1t0U1,4,8} (b). Those for $U/t=8$ by Assaad~\cite{Assaad} are shown in Fig.~\ref{fig:momdisel1t0U1,4,8} (c). Compared with QMC calculations, the error turns out to be less than $10\%$ for $U/t=4$. The error is even smaller for $U/t=8$ than for $U/t=4$. This small discrepancy originates from an overestimate of the Mott gap for smaller values of $U$. Though the Mott gap is actually reduced from $U$ due to fluctuations, the reduction ratio is still small compared with QMC calculations. The discrepancy in the evaluation of the Mott gap will be examined later by showing the single-particle dispersions as well as by obtaining the chemical potential as a function of the doping concentration in \S~\ref{subsec:Doped}.

The energy and momentum dependences of these single-particle excitations can be understood by calculating the spectral function,
\begin{equation}
A(\omega,\mib{k})\equiv-\frac{1}{\pi}{\rm Im}G_\rme(\omega,\mib{k}).
\label{eq:half-filling:A_kw}
\end{equation}
Figure~\ref{fig:Akwel1t0U4_T.02_.5_4} shows the single-particle spectral function $A(\omega,\mib{k})$ for $U/t=4$ at $T/t=0.02$ (a), $0.5$ (b) and $4$ (c). The upper and lower Hubbard bands nearly separated by the Mott-Hubbard pseudogap exist at $T/t=0.5$ as shown in Fig.~\ref{fig:Akwel1t0U4_T.02_.5_4} (b). At lower temperatures, the growth of AF spin fluctuations produces their shadow bands due to the folding of the Brillouin zone in the AF ground state. This results in a four-band structure, as shown in Fig.~\ref{fig:Akwel1t0U4_T.02_.5_4} (a) for $T/t=0.02$. This evolution of the single-particle dispersions from the Hubbard bands into such four-band structure at low temperatures $T< J$ has also been obtained with QMC calculations by Gr\"{o}ber {\it et al.}~\cite{GroberZacherEder98,GroberEderHanke00}. We will discuss the dispersions in more detail later. On the other hand, at such higher temperatures as $T>U/2$, the Mott-Hubbard pseudogap disappears and the low-energy incoherent band develops, as shown in Fig.~\ref{fig:Akwel1t0U4_T.02_.5_4} (c). It seems that this realizes a similar band structure to a free-electron tight-binding dispersion. However, as we mentioned above, this band does not introduce any abrupt change in the momentum distribution. This reflects that the Fermi temperature vanishes for this half-filled case. Therefore, for arbitrary values of $U>0$ and $T$, it is not justified to describe the electronic structures of the simple two-dimensional half-filled Hubbard model in terms of the Fermi-liquid theory. Though these incoherent metallic states at high temperatures are adiabatically connected with a free-electron ground state, it does not necessarily mean that a simple weak-coupling perturbation theory effectively works for describing Mott-insulating properties at low temperatures. Only those theories which describe both the Fermi liquid and the non-Fermi liquid reproduce correct electronic properties.

Lastly, we discuss an evolution of the electron dispersions at $U/t=4$ as a function of temperature. Figure~\ref{fig:disel1t0U4_T.02_.5_4} shows the peak positions in $A(\omega,\mib{k})$ at the same temperatures as given in Fig.~\ref{fig:Akwel1t0U4_T.02_.5_4}. The QMC results at $T=0$~\cite{AssaadImada99} are also given as the circles in Fig.~\ref{fig:disel1t0U4_T.02_.5_4} (a), while the present results are marked as the squares. This figure indicates the overall similarity of the present results to the QMC results, except for an overestimate of the Mott gap. It should be noted that the dispersion around $(\pi,0)$ and $(0,\pi)$ obtained in the present calculations is much flatter than the Hubbard~\cite{Hubbard1} I approximation, in agreement with the QMC results~\cite{HankePreuss95,GroberZacherEder98,AssaadImada99,GroberEderHanke00} and the two-pole approximation~\cite{HarrisLange67,Roth69,BeenenEdwards95}. Moreover, the four-band structure composed of the upper and lower Hubbard bands and their AF shadows takes the form of a superposition of two SDW-like bands and two low-energy narrow bands. Such structure is consistent with the QMC results~\cite{GroberZacherEder98,GroberEderHanke00}. As the temperature increases above $J$, strong short-ranged AF spin correlations disappear. This smears out the AF shadows of the Hubbard bands. Then, the main structure is characterized by the Hubbard bands, as shown in Fig.~\ref{fig:Akwel1t0U4_T.02_.5_4} (b). At higher temperatures $T>U/2$, an extra low-energy narrow band crossing the chemical potential appears without any coherence, as already mentioned and also shown in Fig.~\ref{fig:disel1t0U4_T.02_.5_4} (c). The main structure shown in Fig.~\ref{fig:Akwel1t0U4_T.02_.5_4} (c) resembles the free-electron tight-binding band.

\subsection{Metallic properties in doped cases}
\label{subsec:Doped}

\begin{figure}[htb]
\begin{center}
\epsfxsize=8.0cm
$$\epsffile{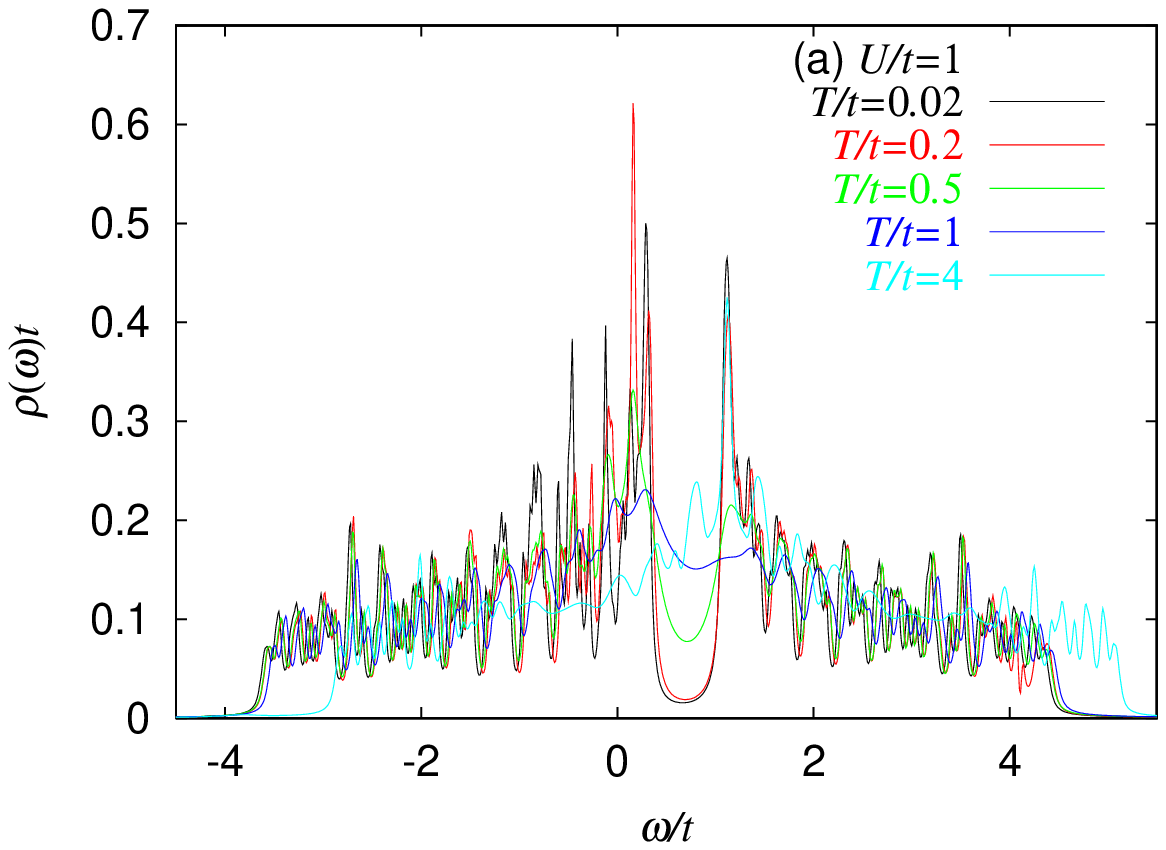}$$
\end{center}
\begin{center}
\epsfxsize=8.0cm
$$\epsffile{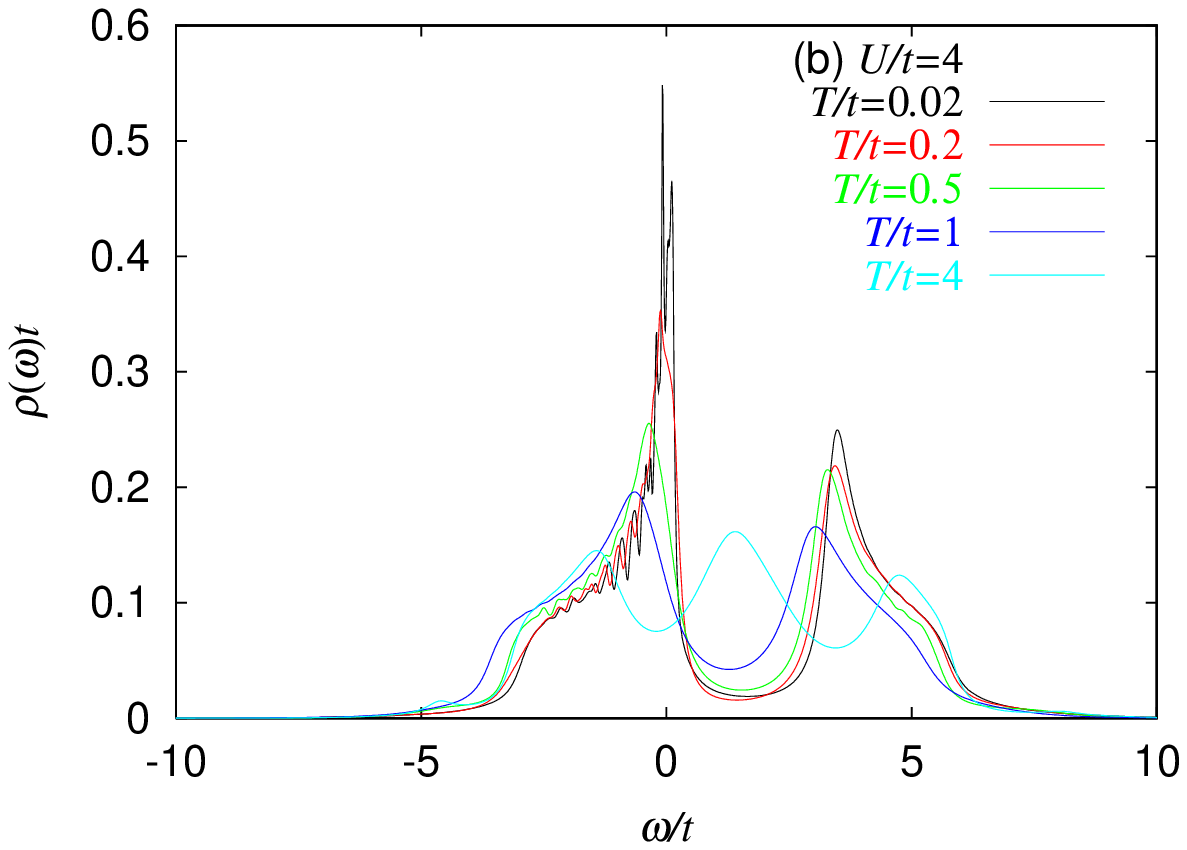}$$
\end{center}
\begin{center}
\epsfxsize=8.0cm
$$\epsffile{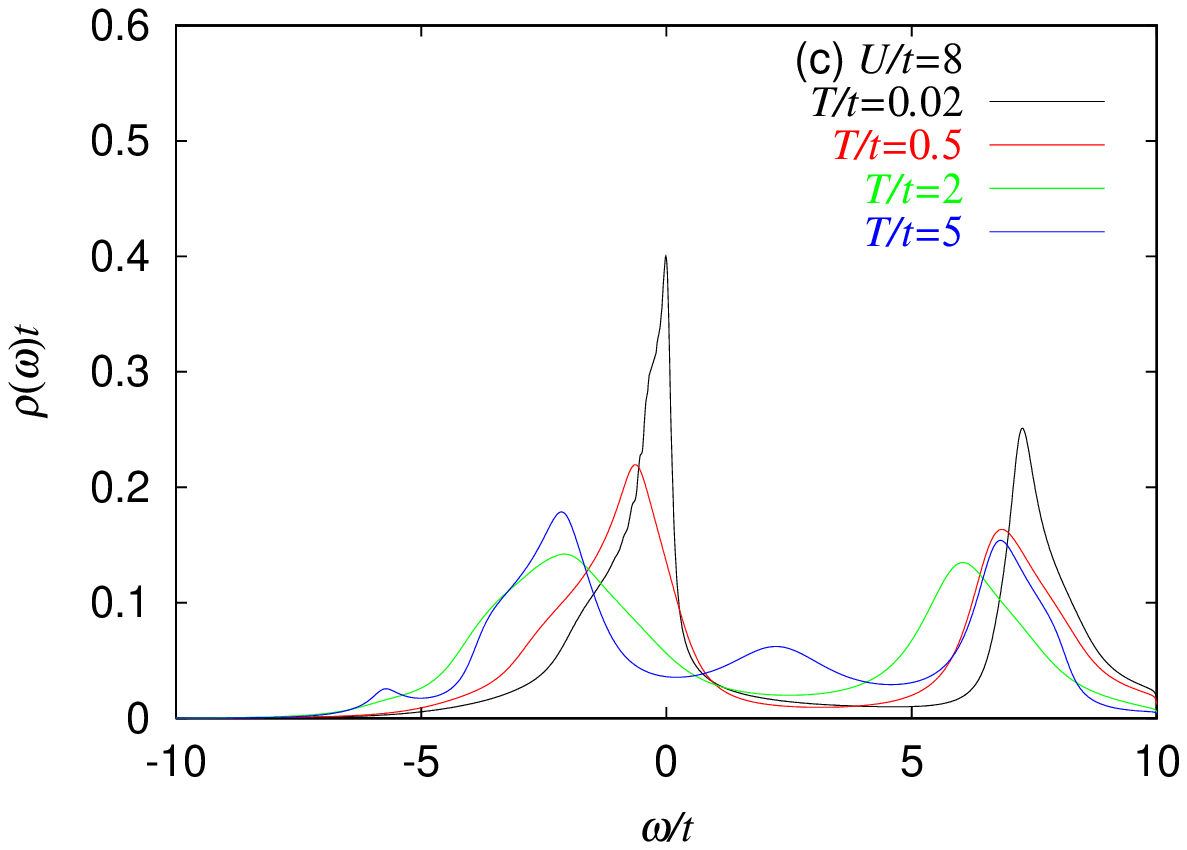}$$
\end{center}
\caption{The temperature dependence of the local density of states $\rho(\omega)$ for $U/t=1$ (a), $4$ (b) and $8$ (c) at $n=0.87$. For (b), $T/t=0.02$, $0.2$, $0.5$, $1$ and $4$. For (c), $T/t=0.02$, $0.5$, $2$, $5$ and $8$. A sharp peak develops near $\omega=0$ at low $T$.}
\label{fig:DOSel.87t0U1,4,8}
\end{figure}
\begin{figure}[htb]
\begin{center}
\epsfxsize=8.0cm
$$\epsffile{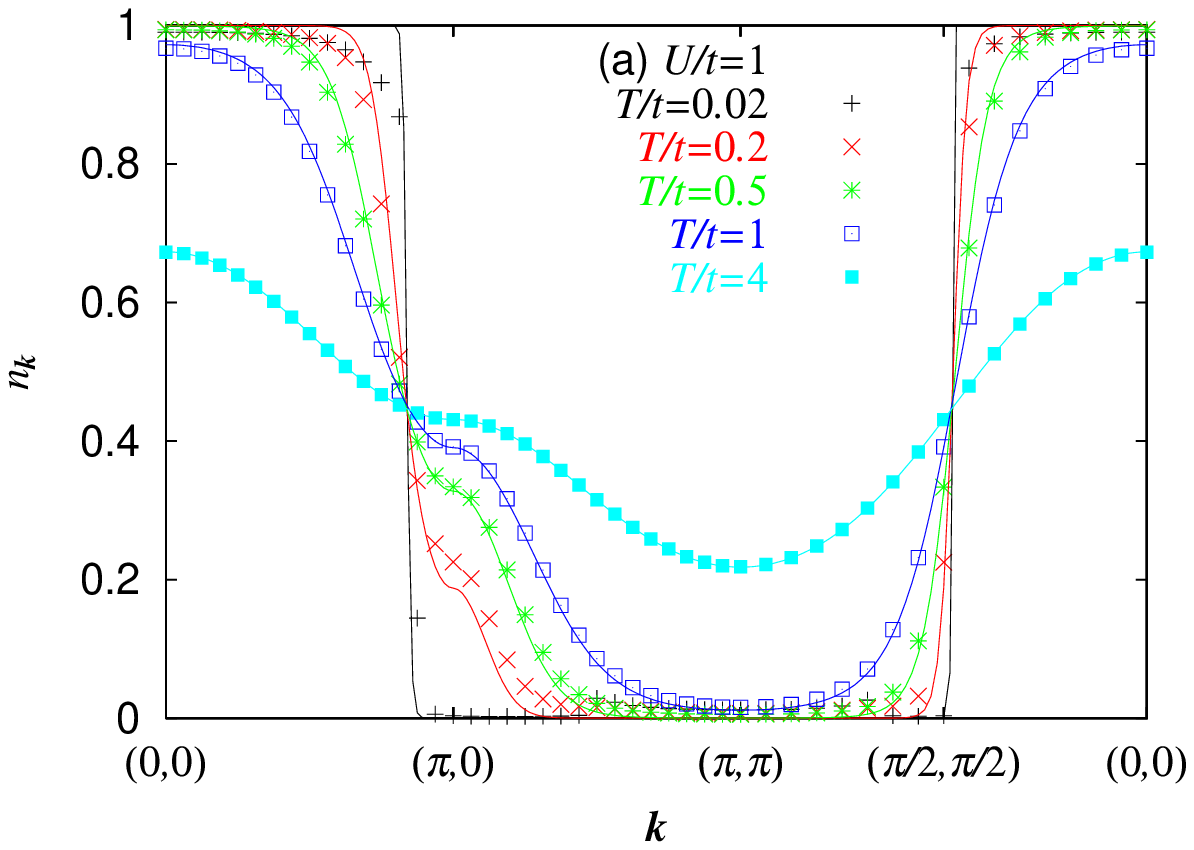}$$
\end{center}
\begin{center}
\epsfxsize=8.0cm
$$\epsffile{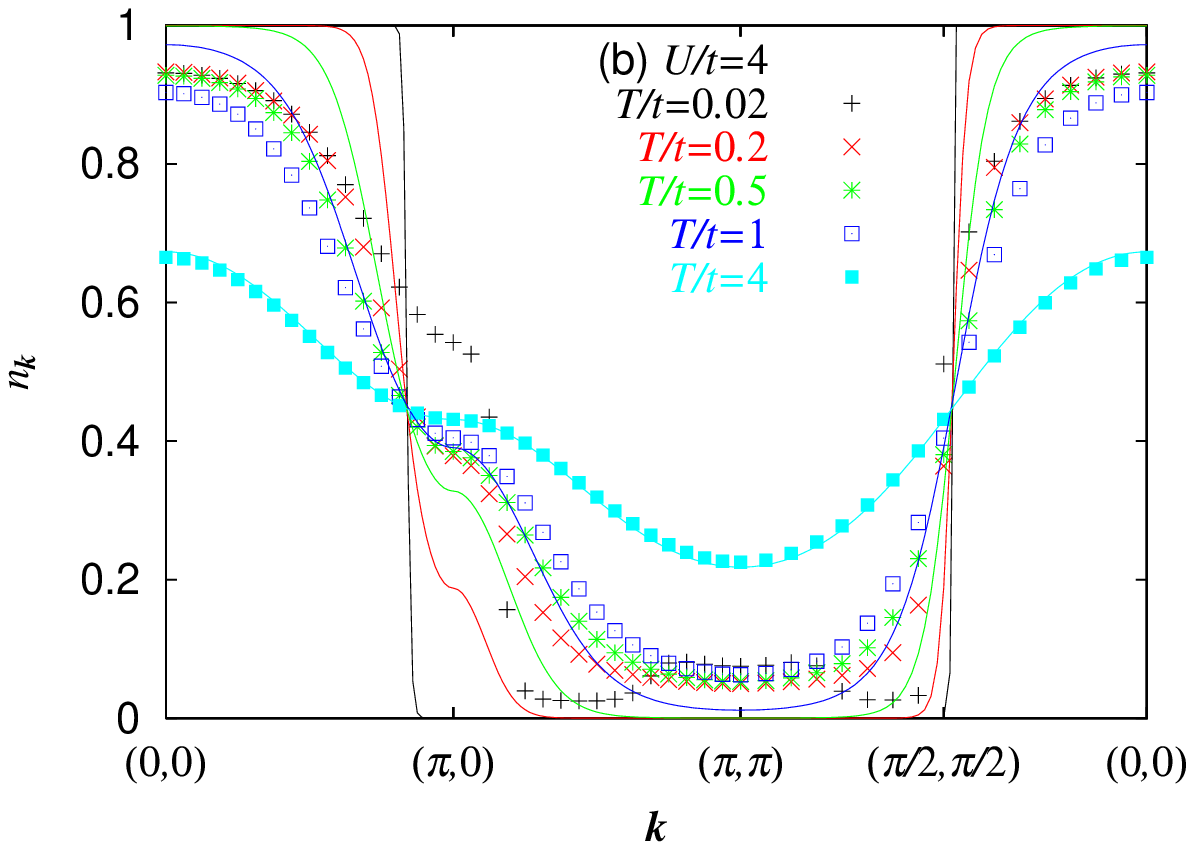}$$
\end{center}
\begin{center}
\epsfxsize=8.0cm
$$\epsffile{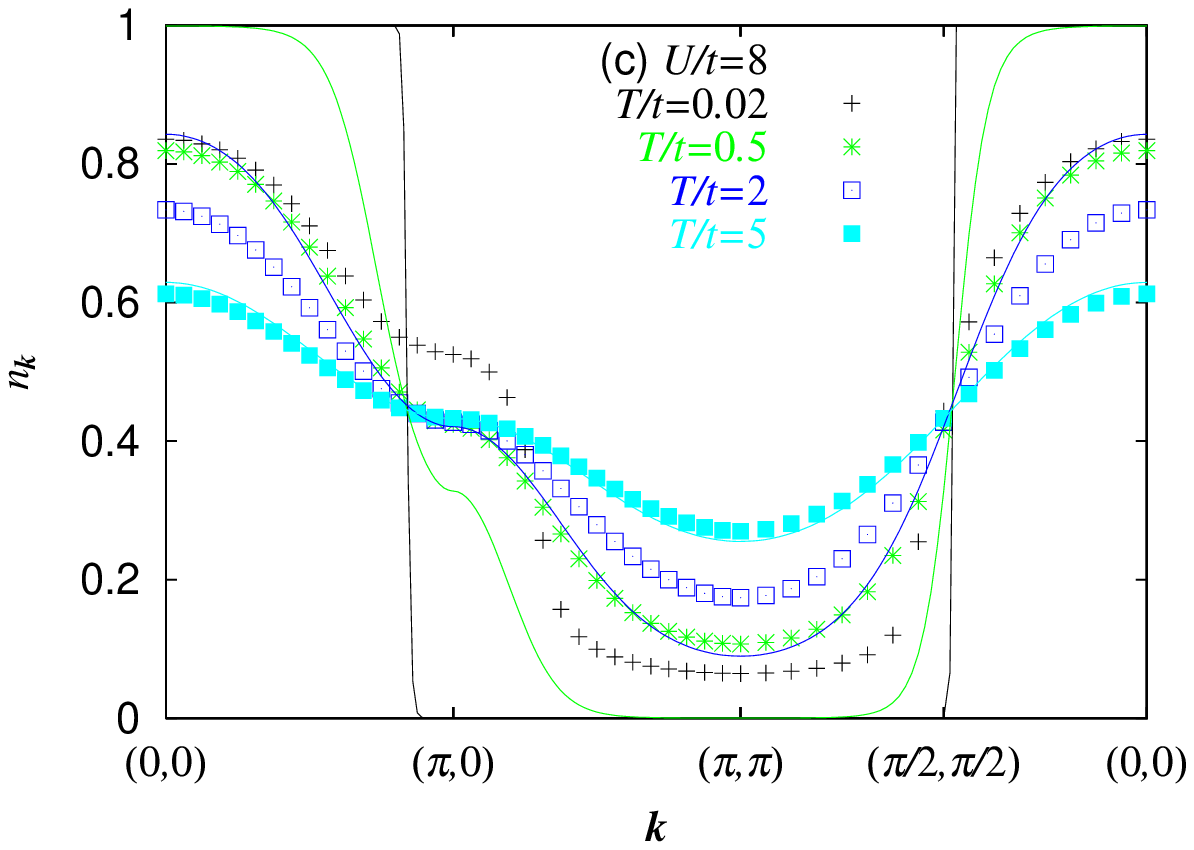}$$
\end{center}
\caption{Momentum distributions along $(0,0)$--$(\pi,0)$--$(\pi,\pi)$--$(0,0)$ for $U/t=1$ (a), $4$ (b) and $8$ (c) at $\ave{n}=0.87$. For comparison, momentum distributions for $U=0$ and $\ave{n}=0.87$ at the same set of temperatures are plotted with solid lines in the corresponding colors in each panel. The Fermi momenta $\mib{k}_{\rm F}$ where $n_\mibs{k}$ most strongly changes move towards $(\pi,\pi)$ with increasing temperature. Thus, the ``Fermi volume'' increases with increasing temperature.}
\label{fig:momdisel.87t0U1,4,8}
\end{figure}
\begin{figure}[htb]
\begin{center}
\epsfxsize=8.0cm
$$\epsffile{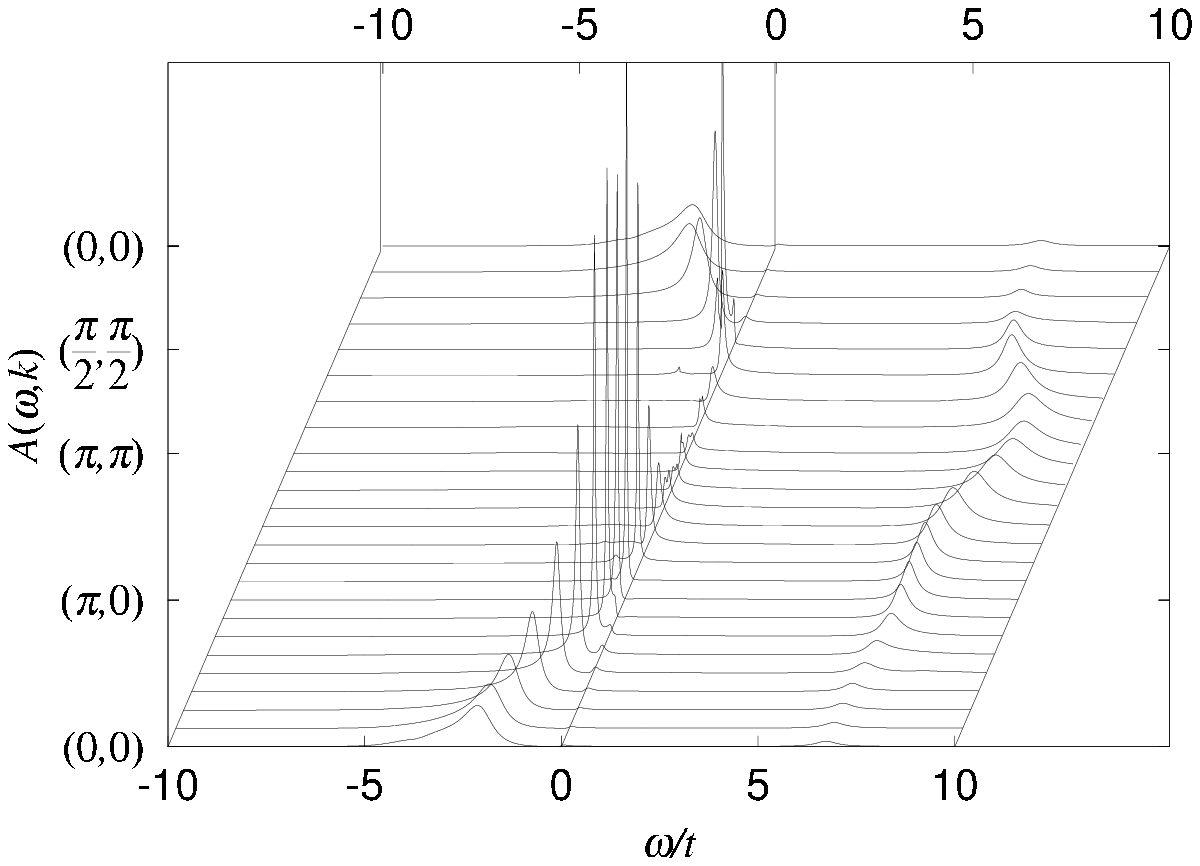}$$
\end{center}
\vspace*{-32pt}(a)
\begin{center}
\epsfxsize=8.0cm
$$\epsffile{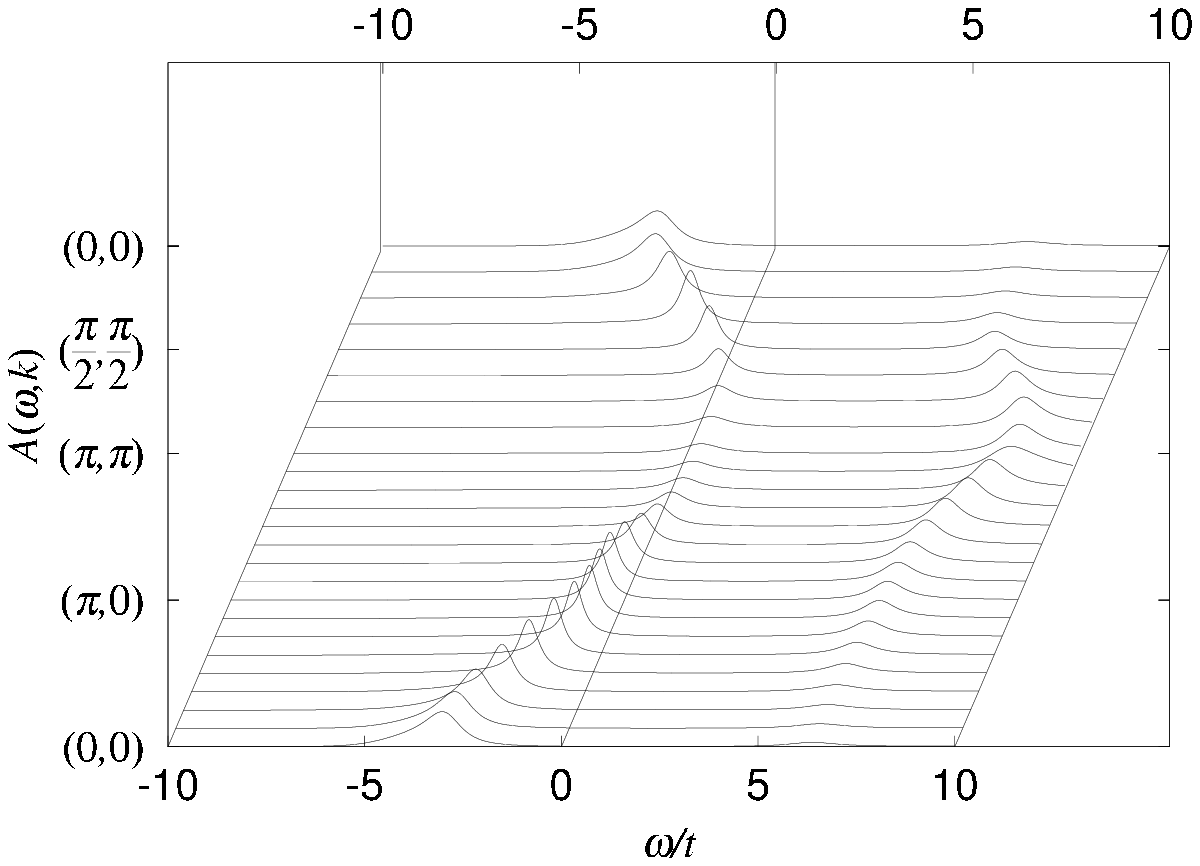}$$
\end{center}
\vspace*{-32pt}(b)
\begin{center}
\epsfxsize=8.0cm
$$\epsffile{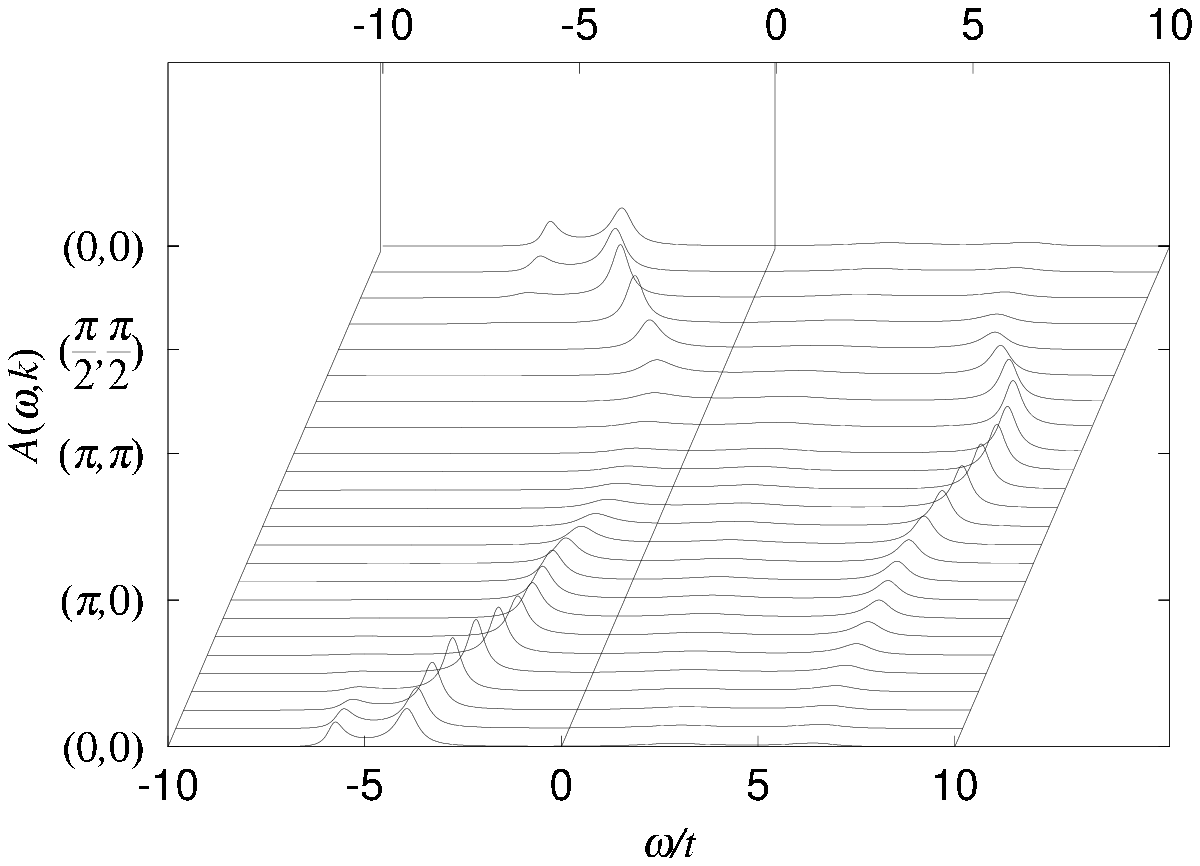}$$
\end{center}
\vspace*{-32pt}(c)
\caption{Single-particle spectral functions $A(\omega,\mib{k})=-{\rm Im}\ G(\omega,\mib{k})/\pi$ along $(0,0)$--$(\pi,0)$--$(\pi,\pi)$--$(0,0)$ for $U/t=8$ at $\ave{n}=0.87$. The temperature is $T/t=0.02$, $0.5$ and $5$ in (a), (b) and (c), respectively. They show that around the $(\pi,\pi)$ momentum, the dispersion of the lower Hubbard band becomes weak and the weights also become small. The dispersion around the $(\pi,0)$ and $(0,\pi)$ momenta are also found to be suppressed.
}
\label{fig:Akwel.87t0U8_T.02_.5_5}
\end{figure}
\begin{figure}[htb]
\begin{center}
\epsfxsize=8.0cm
$$\epsffile{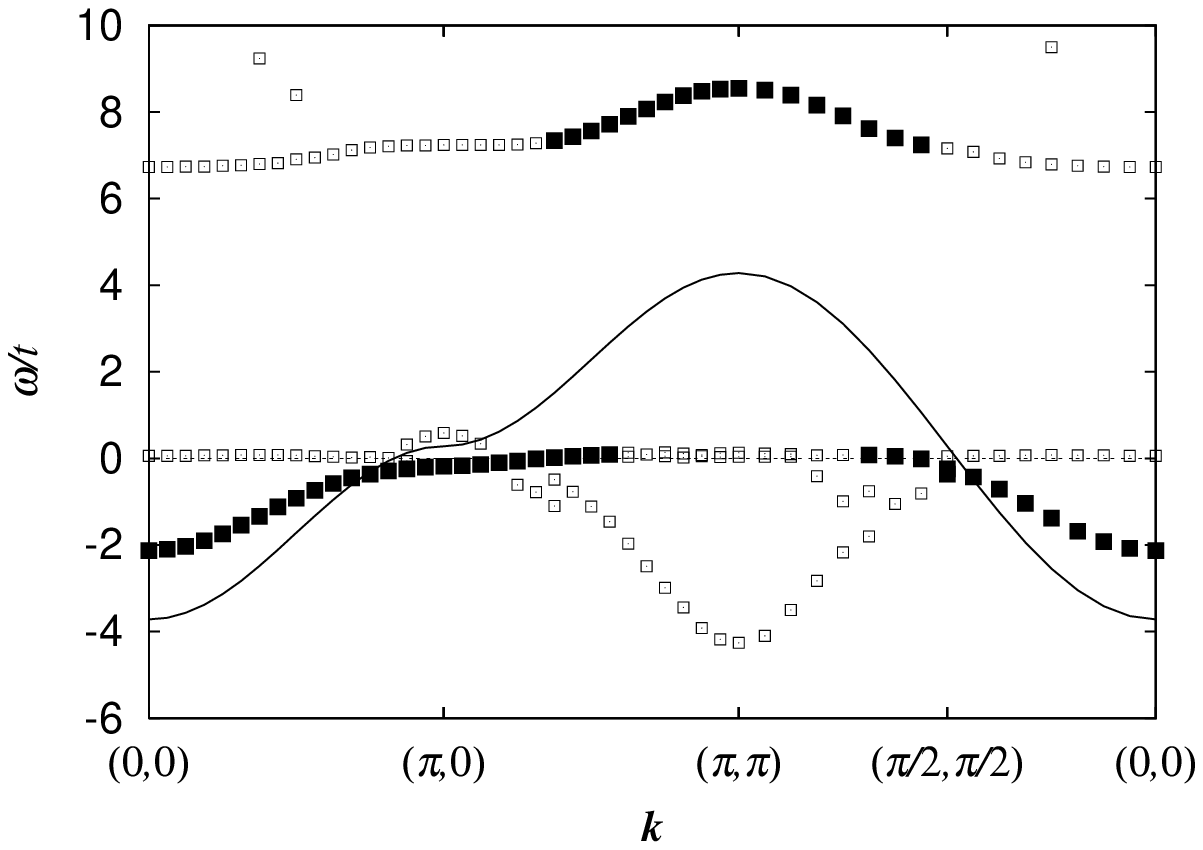}$$
\end{center}
\vspace*{-32pt}(a)
\begin{center}
\epsfxsize=8.0cm
$$\epsffile{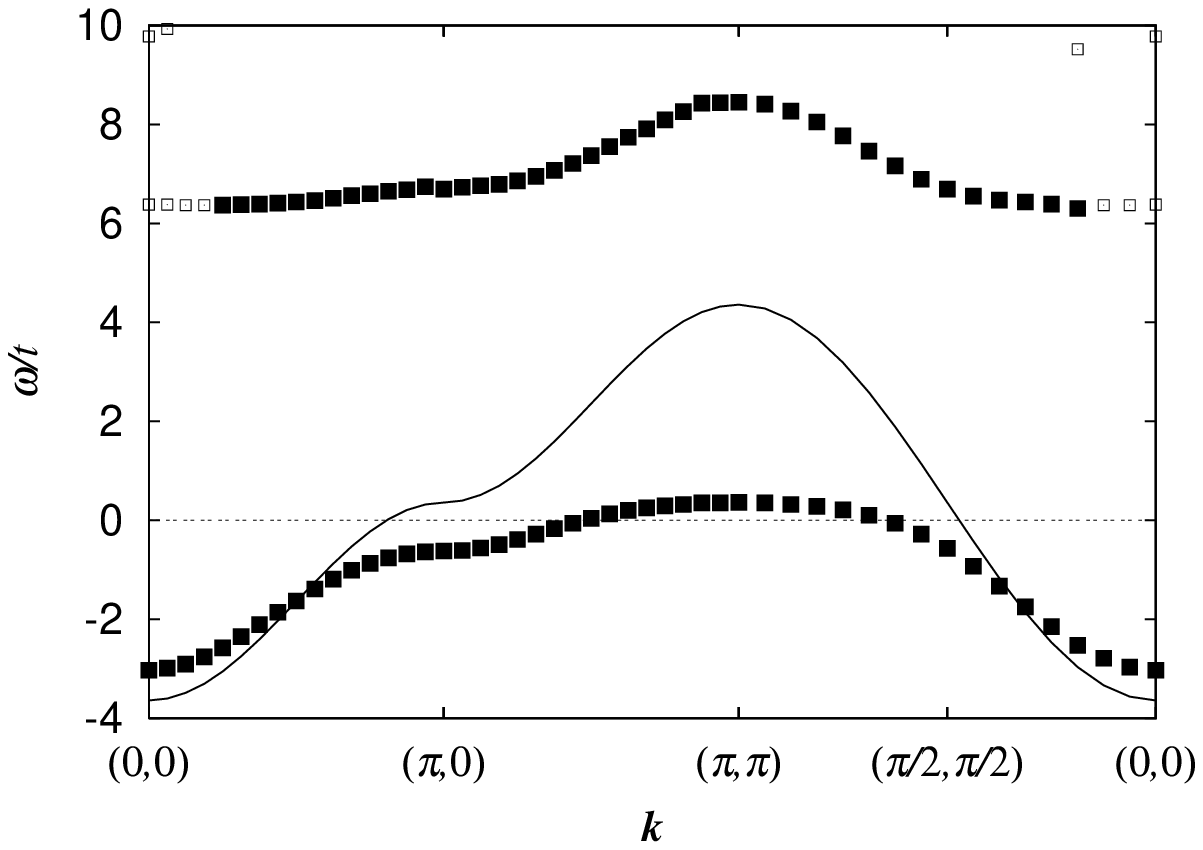}$$
\end{center}
\vspace*{-32pt}(b)
\begin{center}
\epsfxsize=8.0cm
$$\epsffile{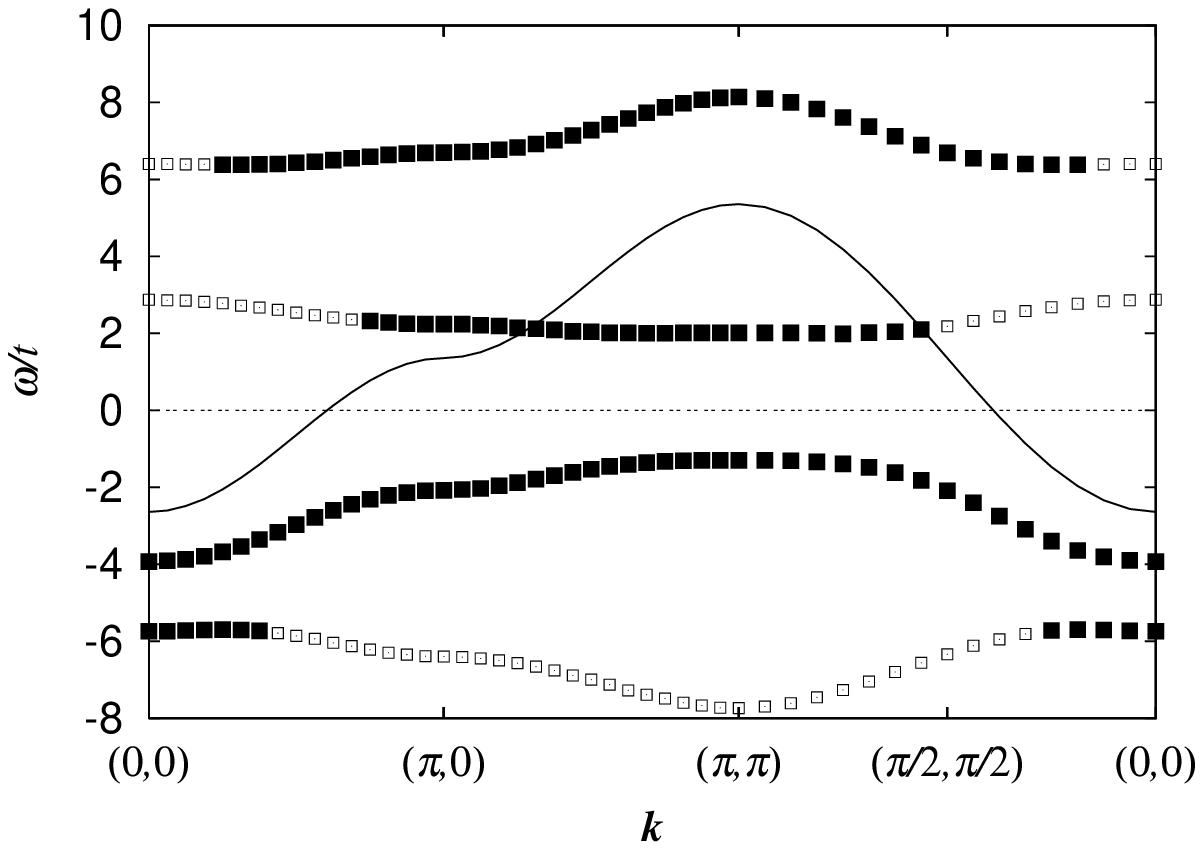}$$
\end{center}
\vspace*{-32pt}(c)
\caption{Single-particle dispersions along $(0,0)$--$(\pi,0)$--$(\pi,\pi)$--$(0,0)$ for $U/t=8$ at $\ave{n}=0.87$. The temperature is $T/t=0.02$, $0.5$ and $5$ in (a), (b) and (c), respectively. The open and the filled symbols represent the momenta where the peak intensities divided by the largest value of $A(\omega,\mib{k})$ are less and larger than $0.1$, respectively. The solid lines denote the free-electron dispersions.}
\label{fig:disel.87t0U8_T.02_.5_5}
\end{figure}

Next, we give the results for doped cases with $\ave{n}=0.87$. Since electron transfers are restricted to the nearest neighbors, the following results also hold for $\ave{n}=1.13$ after particle-hole transformation.

Results of the local density of states $\rho(\omega)$ in the case of $\ave{n}=0.87$ are shown in Fig.~\ref{fig:DOSel.87t0U1,4,8} for $U/t=1$ (a), $4$ (b) and $8$ (c) at several choices of temperatures. As in the two-pole approximations~\cite{Roth69,BeenenEdwards95}, the upper and lower Hubbard bands are nearly separated. In addition, weak incoherent spectra remain between the Hubbard bands even at $T/t=0.02$. For $U/t=8$, these features have also been obtained in QMC calculations~\cite{BulutScalapinoWhite94}. For all the used values of $U$, a sharp peak develops near the Fermi level as the temperature decreases, in agreement with QMC results~\cite{BulutScalapinoWhite94}. This sharp peak emerging at low enough temperatures is also reminiscent of the paramagnetic solution of the doped Hubbard model in infinite dimensions~\cite{RMP_DMFT}. However, an important difference here from the infinite dimensions is that the sharp peak is pinned at the Hubbard gap edge. At high temperatures $T>U/2$, $A(\omega,\mib{k})$ exhibits a rather broad peak between the Hubbard bands, as it does at half filling.

Figure~\ref{fig:momdisel.87t0U1,4,8} shows the momentum distribution $n_{\mibs{k}}$ for the same electron filling as in Fig.~\ref{fig:DOSel.87t0U1,4,8}, $\ave{n}=0.87$ and $U/t=1$ (a), $4$ (b) and $8$ (c) at various temperatures. In the weak-coupling regime, the results for $U/t=1$ and $T/t=0.02$ suggest that the Fermi surface at which the momentum distribution jumps at $T=0$ exists and that it appears to agree with the free-electron case, as shown in Fig.~\ref{fig:momdisel.87t0U1,4,8} (a). This is consistent with the statement that this system belongs to a Fermi liquid. The difference in $n_{\mibs{k}}$ between $U/t=1$ and $0$ cases lies in the value of the Fermi jump. On the other hand, for $U/t=4$ and $8$, Fig.~\ref{fig:momdisel.87t0U1,4,8} (b) and (c) show that the momentum distribution has the maximum slope outside the magnetic Brillouin zone along both directions from $(\pi,0)$ to $(\pi,\pi)$ and from $(0,0)$ to $(\pi,\pi)$. This suggests that the obtained ``Fermi surface'' is larger than the Luttinger volume at least at $T/t=0.02$. The ``Fermi jump'' determined at $T/t=0.02$ decreases with increasing $U$. This leads to the smaller momentum dependence for larger values of $U$. At higher temperatures, the ``Fermi jumps'' are smeared out. At $T/t\le U/2$, the slopes of $n_{\mibs{k}}$ at the ``Fermi surface'' appear to be  smaller than in the free-electron cases. This indicates that the characteristic temperature for smearing the ``jumps'' in the momentum distribution at the ``Fermi surface'' decreases with increase in $U$. This occurs because the coherence in single-particle excitations is suppressed due to correlation effects near the MIT. At higher temperatures, $n_{\mibs{k}}$ approximates the free-electron value. With further increasing temperature, the quantum-mechanical degeneracy of electrons disappears. Then, the momentum dependence of $n_{\mibs{k}}$ gradually disappears.

There is another remarkable property in the momentum distribution $n_{\mibs{k}}$. Figures~\ref{fig:momdisel.87t0U1,4,8} (a) and (b) show that at $T/t=0.02$, after the sudden decrease in $n_{\mibs{k}}$ at the ``Fermi surface'', it exhibits an increase along both $(0,0)$-$(\pi,\pi)$ and $(0,0)$-$(\pi,0)$-$(\pi,\pi)$. If this second large change in $n_{\mibs{k}}$ evolves into another jump at $T=0$, this violates the Luttinger sum rule.

One main reason that we have obtained the Fermi volume inconsistent with the Luttinger volume at $T/t=0.02$ is the downward shift of the main dispersion in the lower Hubbard band around the momentum $(\pi,\pi)$. As shown for $U/t=8$ in Figs.~\ref{fig:Akwel.87t0U8_T.02_.5_5} and \ref{fig:disel.87t0U8_T.02_.5_5} and discussed later, the low-energy single-particle dispersion in the lower Hubbard band is significantly weak outside the magnetic Brillouin zone boundary as well as around the $(\pi,0)$ and $(0,\pi)$ momenta. Figure~\ref{fig:Akwel.87t0U8_T.02_.5_5} also shows that spectral weights in the lower Hubbard band outside the magnetic Brillouin zone boundary are strongly reduced. This makes it difficult to speculate the $T=0$ properties from the present results. More detailed and rigorous considerations on the Luttinger sum rule requires results at $T=0$. In addition, the dispersion and the momentum distribution are still sensitive to a choice of approximations: Matsumoto and Mancini obtained a large Fermi volume for similar choices of parameters by employing a different decoupling scheme based on a local picture~\cite{MatsumotoMancini97}. It is not clear whether their method satisfies the Luttinger sum rule as the system approaches the half filling. Particularly, detailed low-energy structures depend on the spin fluctuation properties. Further studies are required for this problem.

\begin{figure}[htb]
\begin{center}
\epsfxsize=8.0cm
$$\epsffile{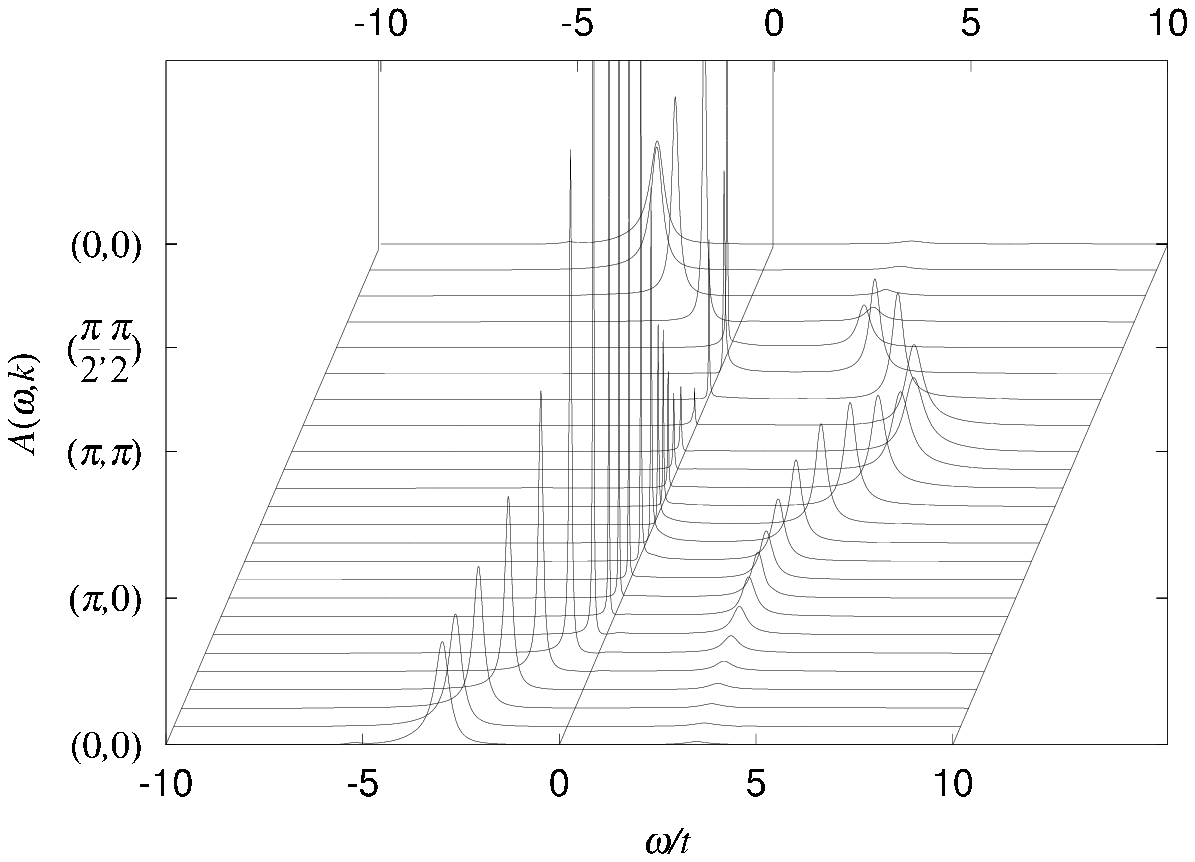}$$
\end{center}
\vspace*{-32pt}(a)
\begin{center}
\epsfxsize=8.0cm
$$\epsffile{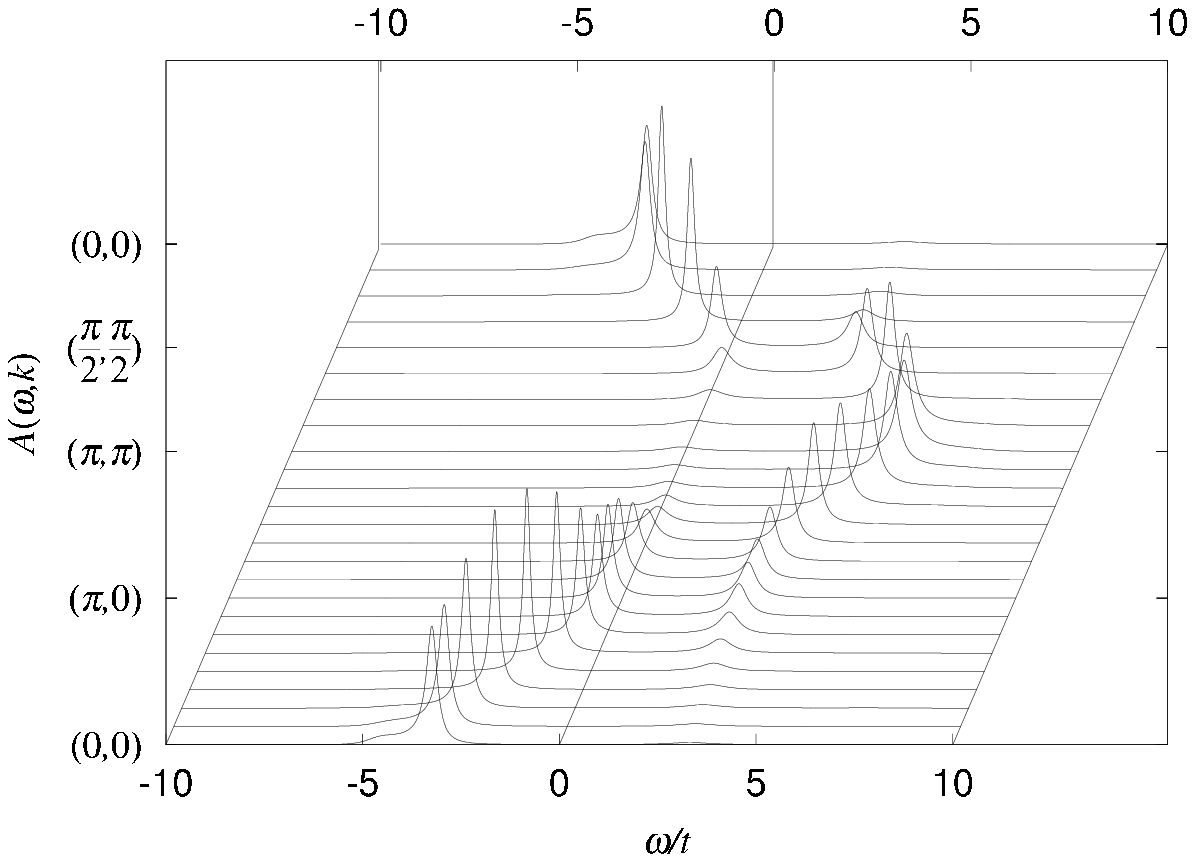}$$
\end{center}
\vspace*{-32pt}(b)
\begin{center}
\epsfxsize=8.0cm
$$\epsffile{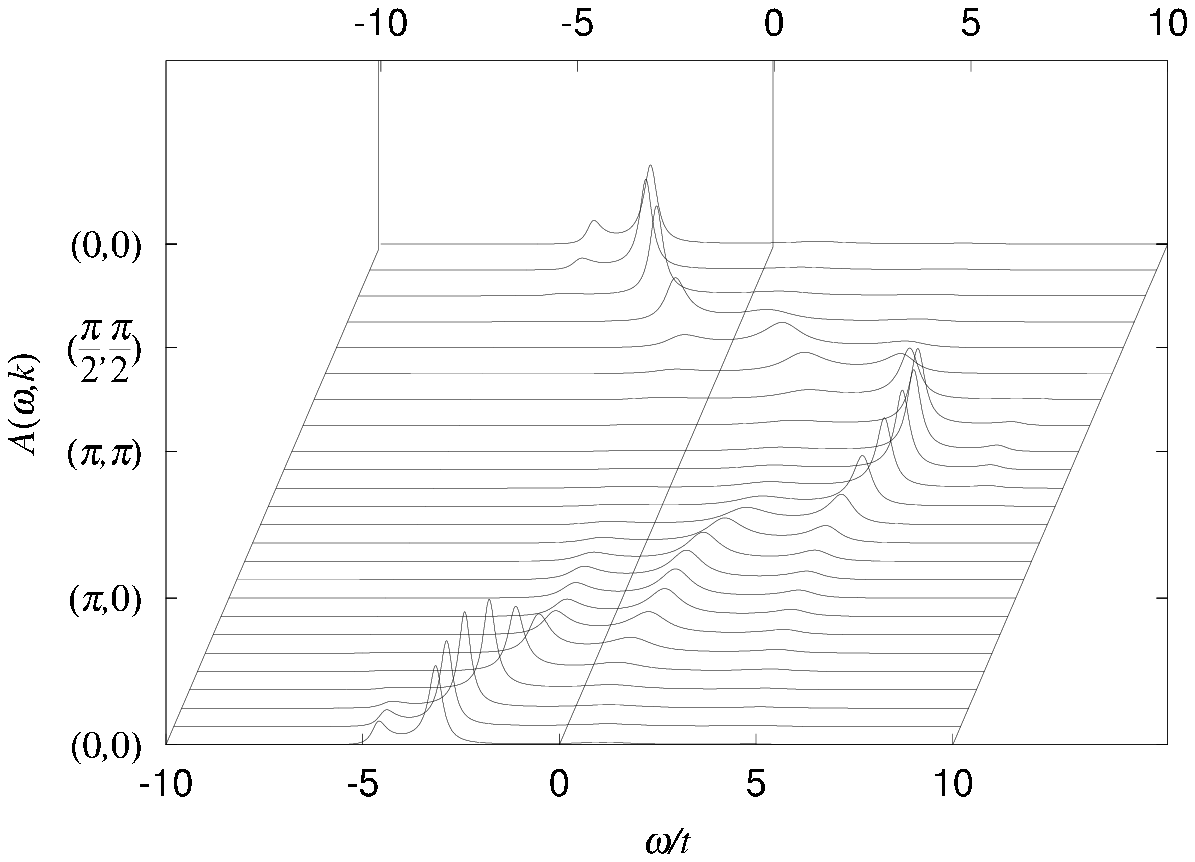}$$
\end{center}
\vspace*{-32pt}(c)
\caption{Single-particle spectral functions $A(\omega,\mib{k})=-{\rm Im}\ G(\omega,\mib{k})/\pi$ along $(0,0)$--$(\pi,0)$--$(\pi,\pi)$--$(0,0)$ for $U/t=4$ at $\ave{n}=0.87$. The temperature is $T/t=0.02$, $0.5$ and $4$ in (a), (b) and (c), respectively.}
\label{fig:Akwel.87t0U4_T.02_.5_4}
\end{figure}

The single-particle spectral functions and the corresponding dispersions for the filling $\ave{n}=0.87$ and $U/t=8$ are given in Figs.~\ref{fig:Akwel.87t0U8_T.02_.5_5} and \ref{fig:disel.87t0U8_T.02_.5_5}. The energetically separated Hubbard bands clearly appear even in this doped case at all the values of temperature we have used. The main structures are similar to the QMC results~\cite{BulutScalapinoWhite94}. Particularly, the single-particle dispersion in the lower Hubbard band becomes strongly weak not only outside the magnetic Brillouin zone boundary but also around the $(\pi,0)$ and $(0,\pi)$ momenta. This agrees with QMC calculations~\cite{BulutScalapinoWhite94} and results of the two-pole approximation~\cite{Roth69, BeenenEdwards95}. At $T/t=0.02$, an extra low-energy band appears at the top of the lower Hubbard band and is pinned at the top. This seems to evolve into the low-energy band obtained as an AF shadow of the lower Hubbard band at half filling. Another important feature is, as mentioned above, that the deviation from the Luttinger sum rule is found at least for the temperatures investigated here. Actually, QMC calculations show that for $T/t\sim0.5$, the $(\pi,0)$ level is below the Fermi level~\cite{BulutScalapinoWhite94}. In Fig.~\ref{fig:Akwel.87t0U4_T.02_.5_4}, we also show the results of $A(\omega,\mib{k})$ for the same filling $\ave{n}=0.87$ but with $U/t=4$ at $T/t=0.02$, $0.5$ and $4$. They appears to be similar to those with $U/t=8$, except a smaller value of the Hubbard gap.

\begin{figure}[htb]
\begin{center}
\epsfxsize=8.0cm
$$\epsffile{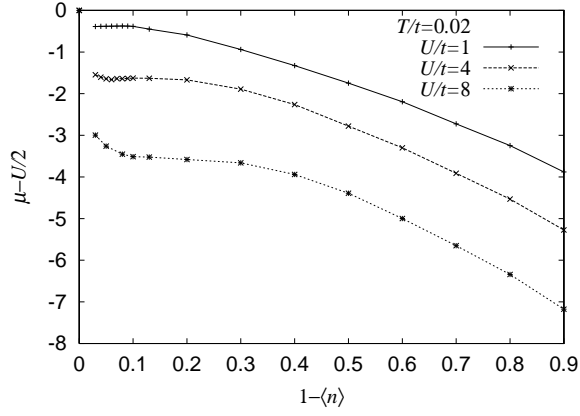}$$
\end{center}
\caption{The chemical potential measured from the particle-hole symmetric energy $\mu-U/2$ versus the hole concentration $1-\ave{n}$ for the two-dimensional Hubbard model with $U/t=1$, $4$ and $8$ at $T/t=0.02$.}
\label{fig:ef-elt0U1,4,8}
\end{figure}
Next, we consider the chemical potential compared with the particle-hole symmetric position $\mu-U/2$ versus the electron filling $\ave{n}$ shown in Fig.~\ref{fig:ef-elt0U1,4,8}. Near half filling, the $\mu$-$\ave{n}$ curve shows that $\kappa=\del\ave{n}/\del\mu$ appears to diverge as $\ave{n}\to1$ in agreement with the QMC results~\cite{FurukawaImada91}. For $U/t=4$ and $8$, the chemical potential shows strong $\ave{n}$ dependence for $\ave{n}$ very close to $1$, namely for $\ave{n}\ge 0.95$ and $0.9$, respectively. This occurs because slight but finite spectral weights still remain inside the zero-temperature Mott-Hubbard gap of order of $U$ due to finite-temperature effects. Because of the smaller residual states for larger $U$, $\mu$ dependence of $n$ becomes larger for larger values of $U$. In addition, with increase in $U$, the difference in momentum distributions between the momenta $(0,0)$ and $(\pi,\pi)$ decreases, due to larger low-energy spectral weights around $(\pi,\pi)$ and $(0,0)$ in the lower and the upper Hubbard bands, respectively. They enlarge the range of the hole concentrations where the finite-temperature effect appears. At strictly $T=0$, $\del\ave{n}/\del\mu\to\infty$ is expected with $\ave{n}\to1$. We note that the divergence of the compressibility is in contrast with the prediction of the Brinkman-Rice picture where the compressibility remains finite. This important difference is more clearly seen in the density of states at half filling in Fig.~\ref{fig:DOSel1t0U1,4,8}. The nearly diverging density of states near the edge of the lower Hubbard band appears due to the low-energy band with extremely weak dispersion especially around $(\pi,0)$ and $(0,\pi)$. Such low-energy band near the level of $(\pi,0)$ and $(0,\pi)$ momenta is beyond the Brinkman-Rice picture. As we will argue in \S~\ref{sec:TotalView}, strong AF spin correlations and their contributions to the strong momentum dependence of the self-energy part produce such sharp contrast with the Brinkman-Rice picture and the dynamical mean-field theory.

\begin{figure}[htb]
\begin{center}
\epsfxsize=8.cm
$$\epsffile{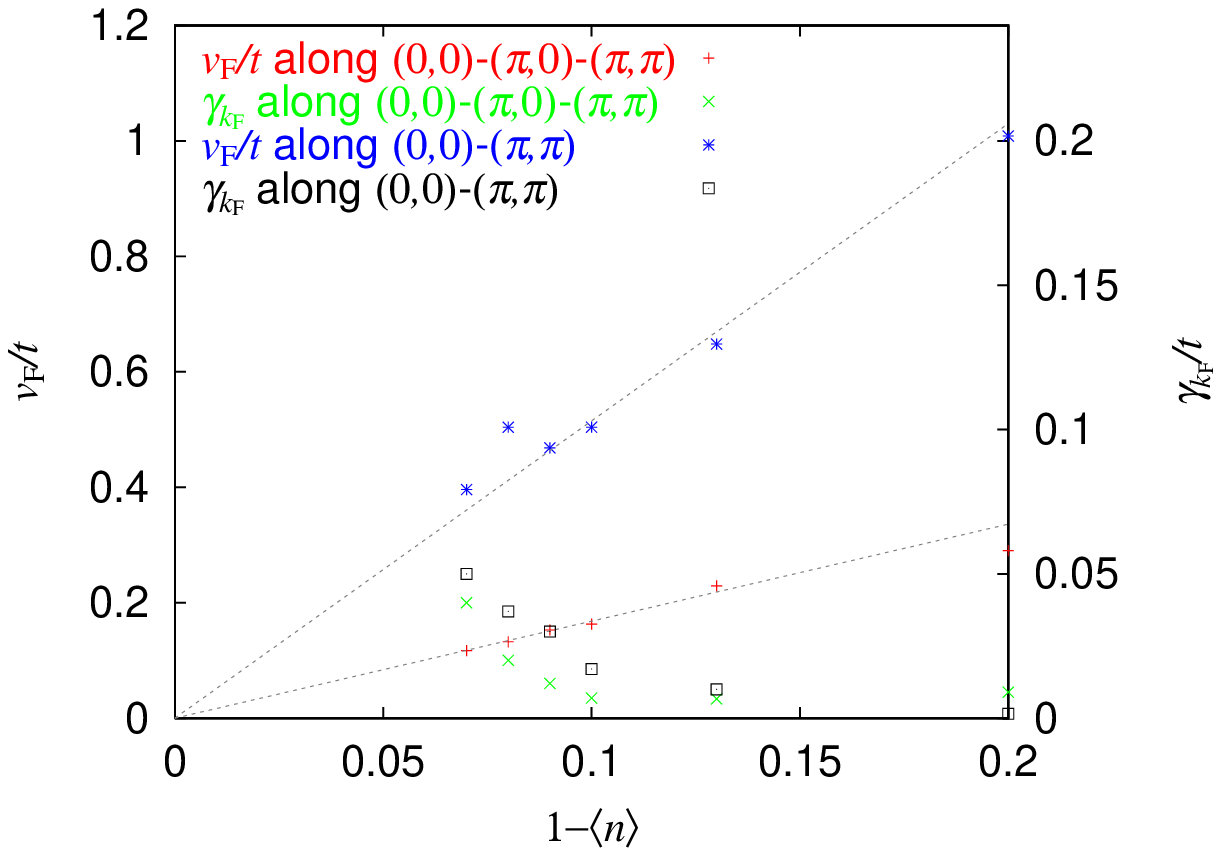}$$
\end{center}
\caption{Fermi velocities $v_{\rm F}$ and damping rates $\gamma_{{\mibs k}_{\rm F}}$ are given as a function of the hole concentration $1-\ave{n}$. They are given for the Fermi momenta in directions along both $(0,0)$-$(\pi,0)$-$(\pi,\pi)$ and $(0,0)$-$(\pi,\pi)$. $\gamma_{{\mibs k}_{\rm F}}$ is determined as the width of the lorentzian fit to $A(\omega,\mib{k})$ at $T/t=0.02$. The figure shows that in both momentum directions, the quasiparticle mass diverges towards the MIT, as in the Brinkman-Rice picture. However, $v_{{\mibs k}_{\rm F}}$ exhibits a momentum dependence so that that around $(\pi,0)$ and $(0,\pi)$ is less than that along $(0,0)$-$(\pi,\pi)$. It also shows the enhancement of the quasiparticle damping rates as $\ave{n}\to1$. The lines in this figure are guide to the eye. In the case of $U=0$, $v_{\rm F}/t=2$ and $1.998$ along the momentum direction $(0,0)$-$(\pi,\pi)$ at $\ave{n}=1$ and $0.87$, respectively. Along $(0,0)$-$(\pi,0)$-$(\pi,\pi)$, $v_{\rm F}/t=0$ and $0.942$ at $\ave{n}=1$ and $0.87$. For $0.94\le\ave{n}\le1$, low-energy excitations near the Fermi level and momenta in $A(\omega,{\mib k})$ exhibit a too broad structure to recognize quasiparticles with well-defined velocities and damping rates. Therefore, the Fermi surface is also ill-defined in this filling region at least at this temperature $T/t=0.02$.}
\label{fig:quasit0U4}
\end{figure}

Lastly, we discuss the Fermi velocity and damping rate that characterizes the low-energy single-particle exciations. Figure~\ref{fig:quasit0U4} shows the Fermi velocities $v_{\rm F}=[|\nabla\veps^*(\mib{k})|]_{{\mibs k}={\mibs k}_{\rm F}}$ and the damping rates $\gamma_{{\mibs k}_{\rm F}}$ at the Fermi momenta along both $(0,0)$-$(\pi,0)$-$(\pi,\pi)$ and $(0,0)$-$(\pi,\pi)$. It shows the divergence of the quasiparticle mass in both momentum regions near $(\pi,0)$ and $(\pi/2,\pi/2)$, although the anisotropy is large. The diverging mass itself is consistent with the Brinkman-Rice picture~\cite{BrinkmanRice70}. These results reflect the generic suppression of the Fermi degeneracy near the MIT. However, the enhanced density of states due to the low-energy band is pinned at the top of the lower Hubbard band. This is the origin of new physics not contained in the Brinkman-Rice picture. Furthermore, quasiparticle damping rates given as $|{\rm Im} \Sigma_{1{\rm e}}(\omega=0,\mib{k}_{\rm F})|$ exhibit a prominent increase as $\ave{n}\to0$. Thefore, quasiparticles become ill-defined at least at $T=0.02t$, as one approaches the MIT. In fact, for $0.94\le\ave{n}\le1$, low-energy excitations near the Fermi level and momenta in $A(\omega,{\mib k})$ exhibit a too broad structure to recognize quasiparticles with well-defined velocities and damping rates. Therefore, the Fermi surface is also ill-defined in this filling region at least at this temperature $T/t=0.02$. Further studies on the low-energy single-particle excitations are required in order to judge the validity of the Fermi-liquid theory.

\subsection{Mott insulator in one dimension}
\label{subsec:1D}

As mentioned in \S~\ref{sec:Introduction}, one-electron excitation spectra of the Hubbard model in one dimension have been understood in detail only for $U\to\infty$~\cite{SorellaParola92}, where spinon and holons are decoupled at an arbitrary energy scale. This produces spinon and holon branches in one-electron excitations. These branches do not have poles due to the one-dimensionality, leading to the Luttinger liquid instead of the Fermi liquid. In the case of $U\to\infty$ at half filling, it is only at $|k|<\pi/2$ that the spinon branch has finite spectral weights below the chemical potential because of the conservation laws for energy and momentum in creating a holon and annihilating a spinon.
For the half-filled Hubbard model with a finite value of $U$, their one-electron spectral properties are still not clear, since the spinons and the holons can not be decoupled at all energies. In this case, QMC calculations~\cite{PreussMuramatsuLindenDieterichAssaadHanke94} give a similarity to the SBMF prediction~\cite{KotliarRuckenstein86} and predict that at low but finite temperatures without any AF long-ranged order, the shadow structures develop. Here, we apply the present formalism to the one-dimensional Hubbard model and compare the results in two dimensions with those in one dimension.

\begin{figure}[htb]
\begin{center}
\epsfxsize=8.0cm
$$\epsffile{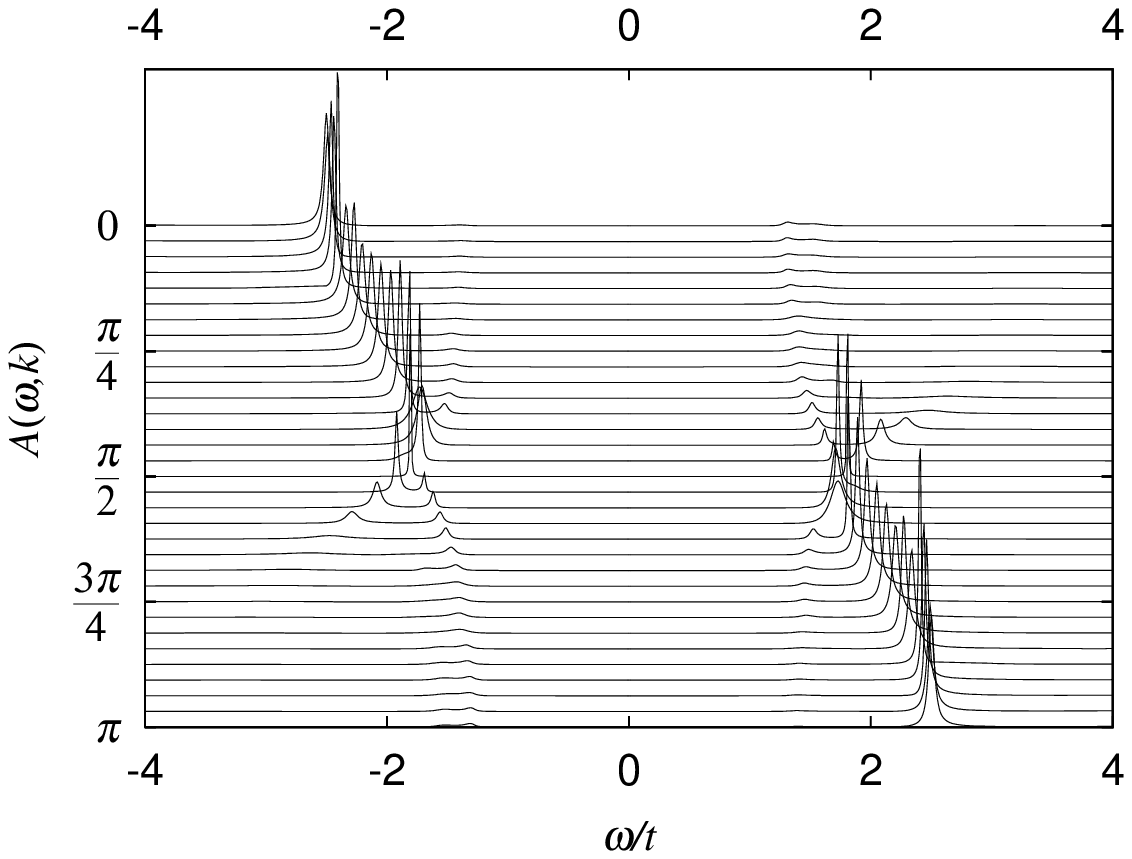}$$
\end{center}
\vspace*{-32pt}(a)
\begin{center}
\epsfxsize=8.0cm
$$\epsffile{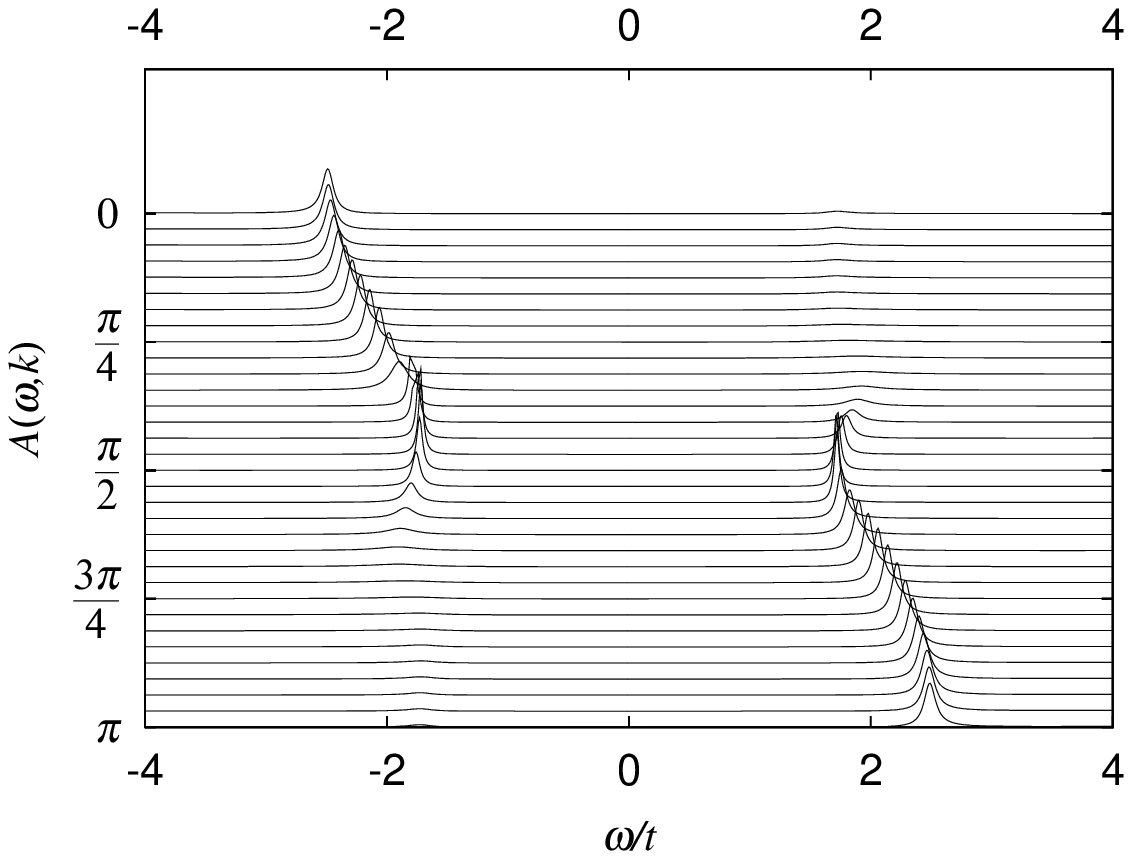}$$
\end{center}
\vspace*{-32pt}(b)
\caption{Spectral function $A(\omega,k)$ as a function of the momentum in the one-dimensional Hubbard model for $U/t=4$ at half filling. The temperature is $T/t=0.01$ (a) and $0.2$ (b).}
\label{fig:Akw1Del1t0U4_T.01_.2}
\end{figure}
Figure~\ref{fig:Akw1Del1t0U4_T.01_.2} shows our results for the spectral function $A(\omega,k)$ of the one-dimensional Hubbard model with $U/t=4$ at half filling. Similarly to the two-dimensional case, a Mott gap opens and separates the Hubbard bands. At $T/t=0.2$, single-particle excitations at the momenta $k=\pm\pi/2$ become the closest to the Fermi level. This reflects that a negative value of $\tilde{t}/t$ provided by AF short-ranged spin correlations yields SDW-like bands instead of the simple Hubbard bands, in terms of the two-pole approximation.
Apart from the overestimate of the Mott gap, these main structures are similar to the QMC results~\cite{PreussMuramatsuLindenDieterichAssaadHanke94}.

At lower temperature $T/t=0.01$, the four-band structure emerges in the spectral function, as in two dimensions. The origin of the four-band structure is the same as that in the two-dimensional case: The AF shadows of the Hubbard bands develop due to the growth of AF short-ranged spin fluctuations. The structure composed of the Hubbard bands and their shadows yields the four bands. In the low-energy band emerging at the top (bottom) of the lower (upper) Hubbard band, spectral weights appear not only for $|k|\le\pi/2$ ($|k|\ge\pi/2$) but also for $|k|\ge\pi/2$ ($|k|\le\pi/2$).
This contrasts with the result at $U\to\infty$~\cite{SorellaParola92} or exact diagonalization results in the $t$-$J$ model~\cite{TohyamaMaekawa96, KimShenMotoyamaEisakiUchidaTohyamaMaekawa97}, which show that the spinon branch appearing at the top of the lower Hubbard band has finite spectral weights only for $|k|\le\pi/2$. The present results for an intermediate coupling can not be well understood from the arguments at $U\to\infty$ in terms of the decoupled spinons and holons. This indicates strong mixing effects between spinons and holons. However, these low-energy bands may still be sensitive to the choice of an approximation to $\Sigma_{2\rme}(\omega,\mib{k})$, as previously mentioned. Therefore, further studies are required to clarify these low-energy bands in more detail.

\section{Total View of the Mott Transition in Two Dimensions}
\label{sec:TotalView}

In this section, we discuss the route to the Mott metal-insulator transition (MIT) in the two-dimensional Hubbard model. We provide a unified view of the Mott-Hubbard, Brinkman-Rice and Slater pictures. Then, we clarify to what extent these old pictures hold and how they are modified.

First, we note that as one approaches the MIT, the local spin moments develops as expressed in \eq{eq:higher:sum_chis}. It increases with increasing $U$ through the reduction of the double occupancy. This gives rise to the splitting of the upper and lower Hubbard bands as in the Mott-Hubbard picture~\cite{Mott,Hubbard1,Hubbard3}. At low temperatures, this large local spin moments tend to align antiferromagnetically to gain a superexchange energy, as in the Slater picture~\cite{Slater}. Then, these local AF spin correlations contribute to formation of the SDW-like dispersions in the Hubbard bands through the momentum dependence of the self-energy part as in the two-pole approximation at half filling~\cite{Roth69}. We note that not only the simple Hubbard approximations~\cite{Hubbard1,Hubbard2,Hubbard3} but also the paramagnetic solution of the dynamical mean-field theory~\cite{RMP_DMFT} gives a simple cosine band in each Hubbard band but not the SDW-like dispersion. This is because the momentum dependence of the self-energy part is oversimplified or neglected. At low temperatures, a narrow band appears at the top of the lower Hubbard band for hole-doped cases, as in the Brinkman-Rice picture~\cite{BrinkmanRice70}. However, the origin of the narrow band at the gap edge is clearly different from the Brinkman-Rice and dynamical mean-field theories, because the enhanced density of states already exists at the gap edge even in the insulator phase.  Since the dispersions in the Hubbard bands take the form of SDW-like dispersions, the level of the narrow band is naturally located near the magnetic Brillouin zone but not $(\pi,\pi)$ in contrast with the Brinkman-Rice picture. The chemical potential shifts within the narrow band at low doping concentrations. The velocity of these low-energy single-particle excitations increases with increasing the doping concentration. But the level of the narrow band compared with the SDW-like dispersion in the lower Hubbard band does not change and pinned at the edge of the Hubbard band, in contrast with the results of the dynamical mean-field theory. These lead to the divergence of the charge compressibility.

Further studies on the low-energy narrow band are necessary not only to discuss the Luttinger sum rule but also to clarify the mechanism of the high-$\Tc$ cuprate superconductivity, since it seems to play a significant role in recovering the quantum-mechanical coherence.

Next, we summarize the modifications of the three old pictures of the MIT clarified by the present theory.

The Mott-Hubbard picture holds at high temperatures where the spin correlations are random. However, when the spin correlations develop at low temperatures, this picture breaks down. Instead, equal-time local AF spin correlations yield the SDW-like dispersions in the Hubbard bands, as in the two-pole approximation at half filling. Simultaneously, the original disperion of the lower (upper) Hubbard band shrinks outside (inside) the magnetic Brillouin zone. Then, the narrow nearly localized band appears at the Hubbard gap edge. This narrow band crosses the chemical potential for underdoped cases. Towards half filling, the velocity characterizing the narrow band appears to vanish.

The Brinkman-Rice picture holds in that the quasiparticle mass diverges towards the MIT. As is well known, this is realized in the dynamical mean-field theory~\cite{RMP_DMFT} which incorporates the Brinkman-Rice and Mott-Hubbard pictures. The theory gives nearly localized single-particle excitations at the chemical potential for doped cases. Since the momentum dependence of the self-energy is neglected, the dispersion of the Hubbard band takes basically simple one given by the Hubbard I approximation~\cite{Hubbard1}. Then, it is natural that the resonant peak appears first near the $(\pi,\pi)$ level for one-hole doping. However, this prediction does not reproduce the divergence of the charge compressibility. This is because in the dynamical mean-field theory, the peak of the local density of states shifts in the Hubbard band as the doping concentration changes. In contrast, as already mentioned, the sharp peak at the gap edge exists already in the insulating phase in our results and it pins the chemical potential. In addition, this one-hole state around $(\pi,\pi)$ in the dynamical mean-field theory clearly violates the Luttinger sum rule. These problems are solved after AF spin correlations are properly included through the momentum dependence of the self-energy part. As we have already noted, AF spin correlations suppresses the dispersion outside (inside) the magnetic Brillouin zone in the lower (upper) Hubbard band and gives rise to formation of the SDW-like dispersions.

The Slater picture holds only in the following two aspects; one, the ground state is antiferromagnetically ordered and the other, the SDW-like dispersions appear at low enough temperatures. This picture is improved when in addition to AF spin correlations, the growth of the local spin moment and thereby the splitting of the Hubbard bands are taken into considerations in the paramagnetic phase. This enables one to describe insulating state without the AF long-ranged order at finite temperatures as well as doped metallic states. Under strong AF spin correlations, the high-energy behaviour of the self-energy contributes to formation of the upper and lower Hubbard bands with the SDW-like dispersions. However, an extra narrow band appears at low energies as resonant single-particle excitations, in contrast with the Slater picture.

\section{Comparison with high temperature superconductors}
\label{sec:ARPES}

Lastly, we discuss a strongly weak dispersion relation both around the $(\pi,0)$ and $(0,\pi)$ momenta and along $(\pi,0)$ to $(\pi,\pi)$ observed with the angle-resolved photoemission spectroscopy (ARPES) in \BSCCO{2}{2}{}{2}{8+\delta}~\cite{Dessau_prl93,Ding_bilayer}. In the high-$T_{\rm c}$ compound, the dispersion around the $(\pi,0)$ and $(0,\pi)$ momenta is significantly reduced compared with the band structure calculations. This flatness depends only weakly on both the temperature and the hole concentration; it persists below and above the superconducting critical temperature $T_{\rm c}$. This suppression of the dispersion is attributed to strong correlation effects related to the MIT, as discussed previously based on QMC calculations~\cite{BulutScalapinoWhite94, AssaadImada99}. The present calculations of the OPM gives an intuitive explanation of such enhancement of the electron mass: Local AF or singlet correlations increase $|\tilde{t}_{\mibs{x},\mibs{x}'}|$ between the nearest-neighbor sets of $\mibs{x}$ and $\mibs{x}'$. This affects the dispersion in the level of the two-pole approximation so that the $(\pi,\pi)$ level in the lower Hubbard band shifts downwards and the dispersion around the $(\pi,0)$ and $(0,\pi)$ momenta becomes flatter. Such strong correlation effects on the single-particle dispersions in the Hubbard bands can not be obtained if one introduces a decoupling approximation to the self-energy part as done in the FLEX approximation~\cite{FLEX}. The similarity of the dispersions between the two-dimensional Hubbard model and high-$\Tc$ cuprates above the pseudogap temperature indicates that cuprates are actually located in the proximity of the MIT described by the Hubbard model. It also suggests that to capture essential features of the superconductivity, it is crucial to describe the correlation effects with strongly momentum-dependent renormalization obtained in this paper. We speculate that the low-energy narrow band obtained in this paper play an important role in realizing the superconductivity. Further studies are required to clarify its contribution to the superconductivity.

\section{Summary}
\label{sec:Summary}

We have developed the operator projection method (OPM) to study strong electron correlations in the Hubbard model. The OPM is a non-perturbative analytic theoretical framework based on the continued-fraction or Dyson-equation expression of response functions~\cite{OPM_jpsjletter01}. It is remarkable that the method systematically improves the Hartree-Fock theory, approximations to the self-energy part based on the perturbation expansion or decoupling approximations, Hubbard approximations and two-pole approximations. In fact, the present theory captures three crucial ingredients of the MIT derived from (i) Mott-Hubbard, (ii) Brinkman-Rice and (iii) Slater pictures. Furthermore, our present theory incorporates the strong momentum dependence of the single-particle spectra, which emerges as a crucial aspect of the strong correlation and a source of electron differentiation.

This method reproduces the Mott-insulating phase as well as the filling-control metal-insulator transition (MIT) in the Hubbard model in both one and two dimensions. A transition from metals to a Mott insulator occurs at half filling as a function of the filling. We have obtained the following characteristic features in the vicinity of the MIT in two dimensions:

At half filling, the Mott gap separates two Hubbard bands. At low temperatures, the growth of AF spin correlations produces the AF shadows of the simple Hubbard bands. The formation of the Hubbard bands and their AF shadows result in a four-band structures at half filling. This four-band structure can be regarded as the superposition of two low-energy narrow bands and two SDW-like bands. The sharp peaks of the density of states at the gap edges are associated with an unusually flattened dispersion around $(\pi,0)$ and $(0,\pi)$. At moderate temperatures, the AF shadows disappear and the single-particle excitations are well described by the Hubbard I approximation~\cite{Hubbard1}. At much higher temperatures $T>U/2$, incoherent low-energy spectral weights appear in the Mott gap, reflecting the loss of the quantum-mechanical features of electrons.

In the hole-doped case, a narrow band appears at the top of the lower Hubbard band energetically separated from the upper Hubbard band by the Mott-Hubbard pseudogap. At low temperatures, a sharp peak of the local density of states develops near the Fermi level due to the narrow band. This low-energy narrow band has a particularly flat dispersion and dominant spectral weights around the $(\pi,0)$ and $(0,\pi)$ momenta. This low-energy band is pinned at the edge of the incoherent lower Hubbard bands and generates low-energy hierarchy structure. Such extremely weak single-particle dispersion obtained in the present results agrees with the results observed in the high-$\Tc$ cuprates with the angle-resolved photoemission spectroscopy. The velocities characterizing the low-energy single-particle excitations vanish towards the MIT. This suggests that the quasiparticle mass diverges towards the MIT. The diverging compressibility is generated from the mass divergence of the pinned low-energy band at the Hubbard gap edge. This novel feature is certainly beyond the Brinkman-Rice picture. The damping rates for these low-energy excitations are also enhanced near half filling. Then, quasiparticles become ill-defined at least at $T=0.02t$, since the damping rate remains comparable to $v_{\rm F}$.

In the one-dimensional Hubbard model at half filling, the results for the single-particle spectra show similar Mott-insulating features for a Mott gap separating two Hubbard bands. As in two dimensions, strong AF spin correlations at low enough temperature produce AF shadows of the Hubbard bands in the plane of energy and momentum, in spite of the absence of the AF long-ranged order even at $T=0$.
Then, single-particle spectra have two branches in each Hubbard band. The bandwidth of the low-energy branch is narrower than the high-energy branch. This low-energy narrow branch is, however, not described as a simple spinon branch in the $U=\infty$ case. This is because the low-energy branch has finite weights for the momenta $|k|>\pi/2$ in the lower Hubbard band in the present results in contrast with the $U=\infty$ case. A mixing between spinons and holons is considered to yield such deviation from the $U=\infty$ results.

The above main features obtained in the present study well reproduce the results of the QMC calculations~\cite{FurukawaImada91,PreussMuramatsuLindenDieterichAssaadHanke94}.
For further improvements of the present OPM, it is required to impose constraints on the consistency between the single-particle and two-particle properties. Furthermore, the resulting spin degrees of freedom at half filling in the strong-coupling regime must agree with predictions for the Heisenberg model. The higher-order projections are also important to extract low-energy properties more clearly. Studies on these improvements are currently in progress.

\acknowledgement

The authors acknowledge F. Mancini, A. Avella and H. Matsumoto for sending us their paper on the present subject. We thank F. F. Assaad for valuable discussions and providing results of Monte-Carlo calculations. S. O. also thanks M. Kohno, Y. Saiga and Y. Motome for useful conversations. The work was supported by the ``Research for the Future" program of the Japan Society for the Promotion of Science under grant number JSPS-RFTF97P01103.


\end{document}